\documentclass[
10pt,
 amsmath,amssymb,
 aps,
 prd,nofootinbib,
]{revtex4-2}
\usepackage[hhmmss]{datetime}
\usepackage[english]{babel}
\usepackage{amssymb,amsmath,amsfonts,graphicx,mathrsfs,color,bm}
\usepackage{graphicx}
\usepackage{dcolumn}
\usepackage{hyperref}
\hypersetup{colorlinks=true,linkcolor=magenta,anchorcolor=green,citecolor=cyan,filecolor=black,menucolor=black,urlcolor=blue}
\usepackage[table]{xcolor}
\usepackage{slashbox}
\allowdisplaybreaks
\usepackage{epstopdf}
\usepackage[utf8]{inputenc} 
\usepackage{pgf}
\usepackage{tikz}
\usepackage{tikz-cd}
\usepackage{tikz-feynman}
\usepackage{float}
\usepackage{subcaption}
\usepackage{ulem}
\usetikzlibrary{shapes.arrows,calc,decorations.pathreplacing,decorations.markings,bending,arrows,positioning}
\tikzset{
    >=stealth',
    punkt/.style={
           rectangle,
           rounded corners,
           draw=black, very very thick,
           text width=6.5em,
           minimum height=2em,
           text centered},
    pil/.style={
           ->,
           very thick,
           shorten <=2pt,
           shorten >=2pt,}
}

\newcommand{\be}{\begin{equation}}
\newcommand{\ee}{\end{equation}}
\newcommand{\sbe}{\begin{subequations}}
\newcommand{\see}{\end{subequations}}
\newcommand{\bea}{\begin{eqnarray}}
\newcommand{\eea}{\end{eqnarray}}
\newcommand{\p}{\partial}
\newcommand{\nn}{\nonumber}
\newcommand{\ud}{\mathrm{d}}

\newcommand{\bdm}{\begin{displaymath}}
\newcommand{\edm}{\end{displaymath}}
\def\switch{[1 \leftrightarrow 2]}

\definecolor{darkred}{rgb}{0.8,0,0}
\definecolor{darkcerulean}{rgb}{0.02, 0.29, 0.69}
\definecolor{vertcem}{cmyk}{0.45 , 0 , 0.90, 0}

\usepackage[force]{feynmp-auto}

\begin{document}

\title{Tidal effects up to next-to-next-to leading post-Newtonian order in massless scalar-tensor theories}

\author{Laura Bernard}
 \email{laura.bernard@obspm.fr}
 \affiliation{%
 Laboratoire Univers et Théories, Observatoire de Paris, Université PSL, Université Paris Cité, CNRS, F-92190 Meudon, France
}%

\author{Eve Dones}
 \email{eve.dones@obspm.fr}
 \affiliation{%
 Laboratoire Univers et Théories, Observatoire de Paris, Université PSL, Université Paris Cité, CNRS, F-92190 Meudon, France
}%

\author{Stavros Mougiakakos}
 \email{stavros.mougiakakos@obspm.fr}
 \affiliation{%
 Laboratoire Univers et Théories, Observatoire de Paris, Université PSL, Université Paris Cité, CNRS, F-92190 Meudon, France
}%

\date{\today}

\begin{abstract}
In this article, we study the tidal effects in the gravitationally bound two-body system at next-to-next-to leading post-Newtonian order for spin-less sources in massless scalar-tensor theories. We compute the conservative dynamics, using both a Fokker Lagrangian approach and effective field theory with the PN-EFT formalism. We also compute the ten conserved quantities at the same NNLO order. Finally, we extend our results from simple ST theories to Einstein-scalar-Gauss-Bonnet gravity. Such results are important in preparation of the science case of the next generation of gravitational wave detectors.
\end{abstract}



\maketitle

\newpage

\section{Introduction} 

Gravitational wave astronomy is becoming a mature field, nourished by the current and future gravitational wave experiments such as LIGO-Virgo-KAGRA, LISA or Einstein Telescope. To accompany the development of the field, it is crucial to provide the community with all the necessary tools to achieve a very high standard science interpretation. Among this, one will explore the strong-field and highly dynamical regime of gravity which will allow us to test fundamental physics.

Compact binary systems are the most common sources of gravitational waves and their detection and parameter estimation heavily rely on our ability to model their waveform at a very high precision. In order to test our gravitational paradigm, such a program should be done not only in general relativity (GR) but also in a representative selection of alternative theories of gravity. In this work, we focus on the scalar-tensor (ST) class of theories in which a single massless scalar field is introduced in addition to the gravitational field~\cite{DeFelice:2010aj}. Although such a class is wide, with many different models that can usually be classified in the DHOST theories~\cite{Langlois:2018dxi}, we will focus here on the simplest one, namely the generalized Brans-Dicke theories~\cite{Damour:1992we}. However, we will see that our results can easily be extended to other theories, such as Einstein-scalar-Gauss Bonnet gravity~\cite{Moura:2006pz}.

Gravitational wave modelling in scalar-tensor theories has been developed for the different phases of the coalescence since several years. The merger part is being tackled using different numerical relativity approaches~\cite{Barausse:2012da,Palenzuela:2013hsa,Shibata:2013pra,Corman:2022xqg,Figueras:2021abd,Witek:2020uzz}. On the analytical side, which is used to model the inspiral phase, results have now reached a high post-Newtonian (PN) accuracy. Hence, the dynamics is now known at 3PN order~\footnote{We call nPN order the $2n^{\rm th}$ order in an expansion in 1/c, namely nPN$={\cal O}\left(\frac{1}{c^{2n}}\right)$.}, while the waveform and flux have been obtained at 1.5PN order~\footnote{In scalar-tensor theories, the leading order flux is at -1PN order compared to GR due to the presence of dipolar emission. In this article, we choose to refer PN orders with respect to the leading GR contribution. For example 1PN order in the flux corresponds to 2PN order beyond the leading order ST contribution.}~\cite{Bernard:2018hta,Bernard:2018ivi,Bernard:2022noq}. In particular, tidal effects have also been investigated and derived at leading order~\cite{Bernard:2019yfz}. Note that all these results were first obtained in the generalized BD framework and then extended to other theories like Einstein-Maxwell-Dilaton or EsGB for which only the leading order correction was necessary to achieve the same accuracy~\cite{Julie:2017rpw,Julie:2018lfp,Julie:2019sab,Shiralilou:2020gah,Shiralilou:2021mfl,vanGemeren:2023rhh}. 

The purpose of this work is to compute the tidal effects at the next-to-next-to-leading order (NNLO). Tidal effects are particularly interesting in ST theories as they start at 3PN order compared to 5PN order in GR. This is due to the presence of a time-varying dipole moment that generates a scalar-induced tidal deformation of compact objects. The motivation to go to the NNLO is to reach a level where the gravitationally-induced tidal deformations start contributing~\cite{Flanagan:2007ix}.

After a short sub-section on the notations, the rest of the article is organized as follows.  First in Sec.~\ref{sec:STtheory}, we present the massless-scalar-tensor theories that will be studied through this work. Then in Sec.~\ref{Sec:PNform}, we explain how the post-Newtonian formalism is adapted to the treatment of tidal effects in ST theories and we present in Sec.~\ref{Sec:PNEFT} an alternative calculation based on the PN-EFT formalism that we used to check our results. In Secs~\ref{Sec:NNLOlag} and~\ref{Sec:noether}, the NNLO Lagrangian and the conserved quantities are respectively presented. Before concluding, the short section~\ref{Sec:EsGB} explains why our results are also valid for EsGB gravity. The article ends with some appendices presenting technical details. We have relegated most of the lengthy results to the auxiliary file that accompanies the article.

\subsection{Notations}
\label{notations}

In this section, we present the notation that will be used throughout the article. Some quantities are related to ST theories and their generalized PPN parameters, while others are linked to compact objects and binary systems.
\begin{itemize}	
%
\item[-] We adopt the convention that the leading order contribution due to tidal effect, which is formally at 3PN order, is noted as LO. The higher order corrections will be called next-to-leading-order (NLO) and next-to-next-to-leading-order (NNLO) and they respectively correspond to 4PN and 5PN orders.
\item[-] The two masses are indicated by $m_{1}$ and $m_{2}$. We denote by $\bm{y}_a(t)$ the two ordinary coordinate trajectories in a harmonic coordinate system $\left\{t,\mathbf{x}\right\}$, by $\bm{v}_a(t)=\ud\bm{y}_a/\ud t$ the two ordinary velocities and by $\bm{a}_a(t)=\ud\bm{v}_a/\ud t$ the two ordinary accelerations. The ordinary separation vector reads $\bm{n}_{12}=(\bm{y}_{1}-\bm{y}_{2})/r_{12}$, where $r_{12}=\left\vert\bm{y}_{1}-\bm{y}_{2}\right\vert$; ordinary scalar products are denoted by parentheses, \textit{e.g.} $\left(n_{12}v_{1}\right)=\bm{n}_{12}\cdot\bm{v}_{1}$, while 3-dimensional Dirac function is denoted $\delta^{(3)}(\mathbf{x})$, and its value at the position $\bm{y}_a$ is written $\delta_a \equiv \delta^{(3)} (\mathbf{x} - \bm{y}_a) $. We denote by $L=i_1\cdots i_\ell$ a multi-index with $\ell$ spatial indices; $\nabla_L = \nabla_{i_1}\cdots\nabla_{i_\ell}$ and so on;  similarly,  $n_{L} = n_{i_1}\cdots n_{i_\ell}$.
\item[-] To express quantities in the center of mass (CM) frame, we introduce the notations $\bm{n}=\bm{n_{12}}$, $r=r_{12}$ and define the relative position $\bm{x}=\bm{y_{1}}-\bm{y_{2}}$, velocity $\bm{v}=\bm{v_{1}}-\bm{v_{2}}$, and acceleration $\bm{a}=\bm{a_{1}}-\bm{a_{2}}$; we pose $v^2=(vv)=\bm{v}\cdot\bm{v}$ and $\dot{r}=(nv)=\bm{n}\cdot\bm{v}$. In the CM frame we use the total mass $m=m_1+m_2$, the reduced mass $\mu = m_1 m_2 / m$, the symmetric mass ratio $\nu= \mu/ m \in \ ]0,1/4]$ and the relative mass difference $\delta = (m_1-m_2)/m \in \ [0,1[ $. Note that the symmetric mass ratio and the relative mass difference are linked by the relation $\delta^2 = 1-4\nu$. To reduce our expressions in the CM frame, we define convenient combinations of the tidal deformabilities, namely : 
\begin{align}\label{tidal_deformabilities_COM}
& \lambda^{(n)}_{\pm} = \frac{m_2}{m_1} \bar{\delta}_2 \, \lambda^{(n)}_1 \pm \frac{m_1}{m_2} \bar{\delta}_1 \, \lambda^{(n)}_2 \,, \qquad \Lambda^{(n)}_{\pm} = \frac{m_2}{m_1}  (1-2 s_2) \bar{\delta}_2 \, \lambda^{(n)}_1 \pm \frac{m_1}{m_2} (1-2 s_1) \bar{\delta}_1 \, \lambda^{(n)}_2  \,, \nn \\
& \mu^{(n)}_{\pm} = \frac{m_2}{m_1} \bar{\delta}_2 \, \mu^{(n)}_1 \pm \frac{m_1}{m_2} \bar{\delta}_1 \, \mu^{(n)}_2 \,, \nn \\
& c^{(n)}_{\pm} = \frac{m_2}{m_1} \frac{1-\zeta}{\zeta} \bigl(1-\zeta +\zeta(1-2 s_2)\bigr)^2 c^{(n)}_1 \pm \frac{m_1}{m_2} \frac{1-\zeta}{\zeta} \bigl(1-\zeta +\zeta(1-2 s_1)\bigr)^2 \, c^{(n)}_2\,, \nn \\
& {\nu}^{(n)}_{\pm}= \frac{m_{2}{}}{m_{1}{}}(1 -  \zeta) (1 - 2 s_2) \bigl(1-\zeta +\zeta(1-2 s_2)\bigr)\, \nu_1^{(n)} \pm \frac{m_{1}{}}{m_{2}{}}(1 -  \zeta)  (1 - 2 s_1) \bigl(1-\zeta +\zeta(1-2 s_1)\bigr) \, \nu_2^{(n)} \,.
\end{align}
\item[-] Finally, in order to later present our results, following~\cite{Bernard:2018hta}, we introduce a number of ST and post-Newtonian parameters. The ST parameters are defined based on the value $\phi_0$ of the scalar field $\phi$ at spatial infinity, on the Brans-Dicke-like scalar function $\omega(\phi)$ and on the mass-functions $m_a(\phi)$. We pose $\varphi\equiv\phi/\phi_{0}$. The post-Newtonian parameters naturally extend and generalize the usual PPN parameters to the case of a general ST theory~\cite{Will:1972zz,Will:2018bme}. All these parameters are given and summarized in the following Table~\ref{table1}.
\hspace{0.5cm}\begin{small}
\begin{center}
\begin{tabular}{|c||cc|}
	\hline
	& \multicolumn{2}{|c|}{\textbf{ST parameters}} \\[2pt]
	\hline &&\\[-10pt]
	general & \multicolumn{2}{c|}{$\omega_0=\omega(\phi_0),\qquad\omega_0'=\left.\frac{\ud\omega}{\ud\phi}\right\vert_{\phi=\phi_0}, \qquad\omega_0''=\left.\frac{\ud^2\omega}{\ud\phi^2}\right\vert_{\phi=\phi_0},\qquad\varphi = \frac{\phi}{\phi_{0}},\qquad\tilde{g}_{\mu\nu}=\varphi\,g_{\mu\nu},$} \\[12pt]
	& \multicolumn{2}{|c|}{$\tilde{G} = \frac{G(4+2\omega_{0})}{\phi_{0}(3+2\omega_{0})},\qquad \zeta = \frac{1}{4+2\omega_{0}},$} \\[8pt]
	& \multicolumn{2}{|c|}{$\lambda_{1} = \frac{\zeta^{2}}{(1-\zeta)}\left.\frac{\ud\omega}{\ud\varphi}\right\vert_{\varphi=1},\qquad \lambda_{2} = \frac{\zeta^{3}}{(1-\zeta)}\left.\frac{\ud^{2}\omega}{\ud\varphi^{2}}\right\vert_{\varphi=1}, \qquad \lambda_{3} = \frac{\zeta^{4}}{(1-\zeta)}\left.\frac{\ud^{3}\omega}{\ud\varphi^{3}}\right\vert_{\varphi=1}.$} \\[8pt]
	& \multicolumn{2}{|c|}{$\switch$ switches the particle's labels (note the index on the $\lambda_i$'s in not a particle label)} \\[7pt]
	\hline &&\\[-7pt]
	~sensitivities~ & \multicolumn{2}{|c|}{$s_a = \left.\frac{\ud \ln{m_a(\phi)}}{\ud\ln{\phi}}\right\vert_{\phi=\phi_0},\qquad s_a^{(k)} = \left.\frac{\ud^{k+1}\ln{m_a(\phi)}}{\ud(\ln{\phi})^{k+1}}\right\vert_{\phi=\phi_0},\qquad(a=1,2)$} \\[9pt]
    & \multicolumn{2}{|c|}{$s'_a = s_a^{(1)},\qquad s''_a = s_a^{(2)},\qquad s'''_a = s_a^{(3)},$} \\[5pt]
	& \multicolumn{2}{|c|}{$\mathcal{S}_+ = \frac{1-s_1 - s_2}{\sqrt{\alpha}}\,,\qquad \mathcal{S}_- = \frac{s_2 - s_1}{\sqrt{\alpha}}.$} \\[7pt]	\hline\hline 
	Order & \multicolumn{2}{|c|}{\textbf{PN parameters}} \\[2pt]
	\hline &&\\[-10pt]
	N & \multicolumn{2}{|c|}{$\alpha= 1-\zeta+\zeta\left(1-2s_{1}\right)\left(1-2s_{2}\right)$}   \\[5pt]
	\hline &&\\[-10pt]
	1PN & $\overline{\gamma} = -\frac{2\zeta}{\alpha}\left(1-2s_{1}\right)\left(1-2s_{2}\right),$ & Degeneracy \\[5pt]
	& ~~$\overline{\beta}_{1} = \frac{\zeta}{\alpha^{2}}\left(1-2s_{2}\right)^{2}\left(\lambda_{1}\left(1-2s_{1}\right)+2\zeta s'_{1}\right),$~~~~&  $\alpha(2+\overline{\gamma})=2(1-\zeta)$ \\[5pt]
	& $\overline{\beta}_{2} = \frac{\zeta}{\alpha^{2}}\left(1-2s_{1}\right)^{2}\left(\lambda_{1}\left(1-2s_{2}\right)+2\zeta s'_{2}\right),$~~~~& \\[5pt]
	&  $\overline{\beta}_+ = \frac{\overline{\beta}_1+\overline{\beta}_2}{2}, \qquad \overline{\beta}_- = \frac{\overline{\beta}_1-\overline{\beta}_2}{2}.$ &  \\[5pt]
	\hline &\\[-10pt]
	2PN & $\overline{\delta}_{1} = \frac{\zeta\left(1-\zeta\right)}{\alpha^{2}}\left(1-2s_{1}\right)^{2}\,,\qquad \overline{\delta}_{2} = \frac{\zeta\left(1-\zeta\right)}{\alpha^{2}}\left(1-2s_{2}\right)^{2},$ & Degeneracy \\[5pt]
	&  $\overline{\delta}_+ = \frac{\overline{\delta}_1+\overline{\delta}_2}{2}, \qquad \overline{\delta}_- = \frac{\overline{\delta}_1-\overline{\delta}_2}{2},$ &  $16\overline{\delta}_{1}\overline{\delta}_{2} = \overline{\gamma}^{2}(2+\overline{\gamma})^{2}$\\[5pt]
	& $~~\overline{\chi}_{1} = \frac{\zeta}{\alpha^{3}}\left(1-2s_{2}\right)^{3}\left[\left(\lambda_{2}-4\lambda_{1}^{2}+\zeta\lambda_{1}\right)\left(1-2s_{1}\right)-6\zeta\lambda_{1}s'_{1}+2\zeta^{2}s''_{1}\right],~~$  &  \\[5pt]
	& $\overline{\chi}_{2} = \frac{\zeta}{\alpha^{3}}\left(1-2s_{1}\right)^{3}\left[\left(\lambda_{2}-4\lambda_{1}^{2}+\zeta\lambda_{1}\right)\left(1-2s_{2}\right)-6\zeta\lambda_{1}s'_{2}+2\zeta^{2}s''_{2}\right],$ &  \\[5pt]
	&  $\overline{\chi}_+ = \frac{\overline{\chi}_1+\overline{\chi}_2}{2}, \qquad \overline{\chi}_- = \frac{\overline{\chi}_1-\overline{\chi}_2}{2}.$ &  \\[5pt]
\hline
\end{tabular}
\captionof{table}{Parameters for the general ST theory and our notation for PN parameters. \label{table1}}
\end{center}
\end{small}

\end{itemize}

\section{Massless scalar-tensor theories}
\label{sec:STtheory}

\subsection{The gravitational action}
\label{gravact}

We consider a generic class of scalar-tensor theories in which a single massless scalar field $\phi$ minimally couples to the metric $g_{\mu\nu}$. It is described by the action
\be\label{STactionJF}
S_{\mathrm{ST}} = \frac{c^{3}}{16\pi G} \int\ud^{4}x\,\sqrt{-g}\left[\phi R - \frac{\omega(\phi)}{\phi}g^{\alpha\beta}\p_{\alpha}\phi\p_{\beta}\phi\right] +S_{\mathrm{m}}\left(\mathfrak{M},g_{\alpha\beta}\right)\,,
\ee
where $R$ and $g$ are respectively the Ricci scalar and the determinant of the metric, $\omega$ is a function of the scalar field and $\mathfrak{M}$ stands generically for the matter fields. The action for the matter $S_{\mathrm{m}}$ is a function only of the matter fields and the metric. 

Equivalently, one can consider the same theory in another frame, the Einstein frame, which is more practical to perform the computations. We consider the following conformal transformation for the metric and the redefinition of the scalar field as
\be\label{conf}
\tilde{g}_{\mu\nu}=\varphi\,g_{\mu\nu}\,, \qquad \varphi=\frac{\phi}{\phi_0}\,.
\ee
After some integration by part, the action becomes
\be\label{STactionEF}
S^\text{GF}_{\mathrm{ST}} = \frac{c^{3}\phi_{0}}{16\pi G} \int\ud^{4}x\,\sqrt{-\tilde{g}}\left[ \tilde{R} -\frac{1}{2}\tilde{g}_{\mu\nu}\tilde{\Gamma}^{\mu}\tilde{\Gamma}^{\nu} - \frac{3+2\omega(\phi)}{2\varphi^{2}}\tilde{g}^{\alpha\beta}\p_{\alpha}\varphi\p_{\beta}\varphi\right] +S_{\mathrm{m}}\left(\mathfrak{m},g_{\alpha\beta}\right)\,,
\ee
where we have introduced a gauge fixing term, $\propto\tilde{g}_{\mu\nu}\tilde{\Gamma}^{\mu}\tilde{\Gamma}^{\nu}$ with $\tilde{\Gamma}^{\nu}\equiv\tilde{g}^{\rho\sigma}\tilde{\Gamma}^{\nu}_{\rho\sigma}$, to enforce working in the harmonic gauge.

\subsection{The matter action}
\label{matteract}

In order to study the tidal effects, we will consider finite-size objects and go beyond the point-particle approximation that is commonly used. Following the post-Newtonian effective field theory (EFT) approach, we model the coupling to matter by a worldline action describing the coupling of the two objects to gravity. First, as it is usual in the PN formalism, we start with the following point-particle action
\begin{align}\label{Spp}
    S_{\rm pp} = -c\sum_{a=1,2}\,\int\ud\tau_a\,m_a\left(\phi\right)\,,
\end{align}
where $\ud\tau_a$ is the proper time of particle $a$ along its worldline $y_a^{\mu}$, defined as $\ud\tau_a\equiv c\,\ud t\,\sqrt{-(g_{\mu\nu})_a\frac{v_a^{\mu}v_a^{\nu}}{c^2}}$ . We have also introduced a dependence of the masses on the scalar field, $m_a\left(\phi\right)$, in order to take into account the internal self-gravity of each object w.r.t. the scalar field~\cite{1975ApJ...196L..59E}.

Then, going beyond the point-particle action, we construct a tidal action, decomposed as
\begin{align}\label{Stidal}
     S_{\rm tidal} = S_{\rm fs}^{(s)}+S_{\rm fs}^{(g)}+S_{\rm fs}^{(g-s)}
\end{align}

The first piece of Eq.~\eqref{Stidal} encodes scalar-induced tidal effects, \textit{i.e.} the response of each object w.r.t. an external scalar field. Still using the EFT approach, we consider the following action for the scalar tidal contribution
\begin{align}\label{Stidscal}
    S_{\rm fs}^{(s)} = -c\sum_{a=1,2}\,\int\ud\tau_a\,\ \sum_{l=1}^{\infty} \frac{1}{2 l !} \,\ \lambda_a^{l}\left(\phi\right) \left(\nabla_{L}^{\perp}\varphi\right)_a\,\left(\nabla^{L}_{\perp}\varphi \right)_a \,,
 \end{align}
where $L=\mu_1\cdots\mu_l$ is a multi-index. We have introduced the projection onto the hypersurface orthogonal to the four velocity, namely $\nabla_{\mu}^{\perp}\equiv\left(\delta_{\mu}^{\nu}+u_{\mu}u^{\nu}\right)\nabla_{\nu}$ and $\nabla^{\perp}_L=\nabla^{\perp}_{\mu_1}\cdots\nabla^{\perp}_{\mu_l}$.
The coefficients $\lambda^{l}_a$ are the $l^{th}$-order scalar tidal deformability parameters. As for the masses, we have added an explicit dependence in $\phi$ to describe their internal response to the scalar field. In the following, we set $\lambda_a(\phi)\equiv\lambda_a^{1}(\phi)$ and $\mu_a(\phi)\equiv\lambda_a^{2}(\phi)$.

Through the second piece of Eq.~\eqref{Stidal}, we also model the gravitational tidal effects, \textit{i.e.} the usual gravitational response of each object w.r.t. the companion body. Following the approach \cite{Henry:2019xhg} pioneered by \cite{Damour:2009wj}, the corresponding action is given by
\begin{align}\label{Stidgrav}
S_{\rm fs}^{(g)} = \sum_{a=1,2}\,\int\frac{\ud\tau_a}{c}\sum_{l=2}^{\infty} \frac{1}{2l!} \left[ c_a^{l}(\phi)  \,  G^a_{L} \, G_a^{L} \; + \;  \frac{l}{(l+1)c^2} \, d_a^{l}(\phi)  \, H^a_{L} \, H_a^{L}\right]\,.
\end{align}
$G_L$ and $H_L$ are the tidal moments whose expressions read
\begin{subequations}\label{tidal_moment}
\begin{align}
& G_{\mu_1 ...\mu_l}^a = - c^2 \left[ \nabla^{\perp}_{<\mu_1}...\nabla^{\perp}_{\mu_l-2} C_{\mu_{l-1} \underline{\rho} \mu_l > \sigma} \right] \, u_a^{\rho} \, u_a^{\sigma} \,, \\
& H_{\mu_1 ...\mu_l}^a = 2\, c^3 \left[ \nabla^{\perp}_{<\mu_1}...\nabla^{\perp}_{\mu_l-2} C^{*}_{\mu_{l-1} \underline{\rho} \mu_l > \sigma} \right] \, u_a^{\rho} \, u_a^{\sigma} \,, 
\end{align}
\end{subequations}
with $C_{\mu\nu\rho\sigma}$ the Weyl tensor and $C^{*}_{\mu\nu\rho\sigma}=\frac{1}{2}\varepsilon_{\mu\nu \lambda \kappa}{C^{\lambda \kappa}}_{\rho\sigma}$, where $\varepsilon_{\mu\nu \lambda \kappa}$ denotes the completely anti-symmetric Levi-Civita tensor. This action is formally the same as GR, with $\{c_a^{(l)},d_a^{(l)} \}$ being respectively the $l^{th}$-order mass-type and current-type tidal deformability parameters and where, once again, we have introduced an explicit dependence on the scalar field.
In our case, as we are interested in the NNLO tidal effects w.r.t. the leading order scalar contribution, it will be sufficient to consider the mass quadrupole tidal moment only, defined as 
\begin{align}
(G_{\mu \nu})_a= - c^2\,{\left(R_{\mu \rho \nu \sigma}\right)}_a \, u_a^{\rho} \, u_a^{\sigma}\,.
\end{align}
Note that in this definition, the Weyl tensor in Eqs. (\ref{tidal_moment}) has been replaced by the Riemann tensor. This is due to the fact that the traces of the Riemann tensor do not impact the dynamics at the accuracy we consider. In other words, using both definitions yields the same equations of motion for the system. As only the mass-type quadrupolar deformation will contribute at the NNLO, we set $c_a(\phi)\equiv c_a^{2}(\phi)$.

Finally, by adding the third piece of Eq.~\eqref{Stidal},  we also introduce gravito-scalar tidal effects, \textit{i.e.} the possibility to have a mixing between scalar and gravitational tidal effects. Such effects were already mentioned in~\cite{Creci:2023cfx}. The action is
\begin{align}\label{Stidscalgrav}
S_{\rm fs}^{(g-s)} = -c \sum_{a=1,2}\,\int\ud\tau_a\sum_{l=2}^{\infty} \frac{1}{l!} \nu_a^{l}(\phi)  \,  G_a^{L} \, \left(\nabla_{L}^{\perp}\varphi\right)_a \,,
\end{align}
where $\nu_a^l(\phi)$ is the $l^{th}$-order gravito-scalar tidal deformability parameter and possesses, as the other parameters, an explicit dependence on the scalar field. As for the purely gravitational case, only the $l=2$ mode contributes to the NNLO dynamics so we will restrict to this case in the following and use the loose notation  $\nu_a(\phi)\equiv\nu_a^{2}(\phi)$.

At this point, a clarification on the different frames is needed. In our formalism, we have coupled the matter fields directly to the (Jordan-frame) physical metric; the possible matter interaction being relegated to the dependence of the mass and tidal parameters on the scalar field. However, as will be clearer in Sec.~\ref{Sec:PNEFT}, the definition of scalar and gravitational perturbations and the rest of the calculation are done in the conformal (Einstein) frame. As a consequence, it will introduce additional couplings between the matter and scalar fields, as well as a mixing of the different type of tidal deformability parameters. In other work, such as Ref.~\cite{Creci:2023cfx}, the coupling to matter is directly performed in the Einstein frame, hence no such mixing is appearing. We will come back on this point in Sec.~\ref{Sec:EsGB}.

\section{The post-Newtonian formalism}
\label{Sec:PNform}

To derive the equations of motion for each object, we follow a Lagrangian approach, often referred to as the Fokker Lagrangian approach~\cite{Bernard:2015njp}. We remind here the main steps for such a construction:
\begin{itemize}
    \item[i)] Starting from the action~\eqref{STactionEF}, we derive the field equations for the metric and the scalar field perturbations, displayed in Eqs. (\ref{rEFE});
    \item[ii)] We solve iteratively these equations up to a certain PN order, determined by the Fokker approach. Here, as we are interested only in the correction due to tidal effects, it is sufficient to know the point-particle contributions to the metric and scalar field perturbations, see App.~\ref{App:Fokker}.
    \item[iii)] We inject these solutions in the total action up to the required order. It results in a generalized (Fokker) Lagrangian that depends not only on the positions and velocities of the particles but also on their higher order derivatives. The result is presented in Sec.~\ref{Sec:NNLOlag}. 
    \item[iv)] Varying the generalized action w.r.t. to the position of the particles, we obtain the equations of motion for each particle. In Sec.~\ref{Sec:EOMinCOM}, we display the resulting tidal correction to the equations of motion after reduction to the center of mass frame at the LO and NLO, and we relegate the NNLO result to App.~\ref{App:CMeom}.
\end{itemize}
In the following subsections, we give more details on the implementation of the PN formalism in ST theories.

\subsection{The field equations}
\label{fieldeqs}

First, after introducing the conformal gothic metric $\mathfrak{g}^{\mu\nu} \equiv \sqrt{-\tilde{g}}\,\tilde{g}^{\mu\nu}$, we define the scalar perturbation $\psi$ and  the metric perturbation $h^{\mu\nu}$ as
\begin{equation}
\psi\equiv\varphi-1 \,,\qquad \text{ and } h^{\mu\nu}\equiv  \mathfrak{g}^{\mu\nu} -\eta^{\mu\nu} \,,
\end{equation}
where $\eta^{\mu\nu}$ is the Minkowski metric. Then, from the harmonic gauge-fixed action~\eqref{STactionEF}, we get the field equations
\begin{subequations}\label{rEFE}
\begin{align}
& \Box_{\eta}\,h^{\mu\nu} = \frac{16\pi G}{c^{4}}\tau^{\mu\nu}\,,\label{EFE_V1_1}\\
& \Box_{\eta}\,\psi = -\frac{8\pi G}{c^{4}}\tau_{s}\label{EFE_V1_2}\,,
\end{align}
\end{subequations}
where $\Box_{\eta}$ denotes the ordinary flat space-time d'Alembertian operator. The source terms read
\begin{subequations}\label{tau}
\begin{align}\label{taumunu}
& \tau^{\mu\nu} = \frac{\varphi}{\phi_{0}} \vert g\vert (T^{\mu\nu}+ \Delta T^{\mu\nu}) +\frac{c^{4}}{16\pi G} \Lambda^{\mu\nu} \,,\\
\label{taus} & \tau_{s} = -\frac{\varphi}{\phi_{0}(3+2\omega)}\sqrt{-g}\left(T - 2\phi(\frac{\p T}{\p \phi}+ \Delta S)\right) -\frac{c^{4}}{8\pi G}\Lambda_s\,.
\end{align}
\end{subequations} The non-linearities in the source terms are encoded by $\Lambda^{\mu\nu}$ and $\Lambda_s$ whose explicit expressions can be found in Ref.~\cite{Bernard:2022noq}. In Eq.~\eqref{tau}, we have introduced the classical matter stress-energy tensor $T^{\mu\nu} \equiv \frac{2}{\sqrt{-g}}\frac{\delta S_{\mathrm{pp}}}{\delta g_{\mu\nu}}$, the tidal correction to this tensor $\Delta T^{\mu\nu} \equiv \frac{2}{\sqrt{-g}}\frac{\delta S_{\mathrm{tidal}}}{\delta g_{\mu\nu}}$, as well as their respective contractions with the metric  $T\equiv g_{\mu\nu}T^{\mu\nu}$. For convenience, we have also defined $\frac{\p T}{\p \phi}\equiv \left.\frac{\p T(g_{\mu\nu}, \phi)}{\p\phi}\right|_{g\;\mathrm{fixed}}$ and $\Delta S = \frac{1}{\sqrt{-g}}\frac{\delta S_{\mathrm{tidal}}}{\delta \phi}$.

The point-particle contribution to the stress energy tensor $T^{\mu\nu}$ reads
\begin{align}\label{Tmunupp}
    T^{\mu\nu}\left(t,\mathbf{x}\right) = \sum_{a=1,2}\,\dfrac{m_a\left(\phi\right)v_a^{\mu}v_a^{\nu}}{\sqrt{-g_{\mu\nu}v_a^{\mu}v_a^{\nu}/c^2}}\frac{\delta_a^{(3)}(\mathbf{x}-\mathbf{y_a(t)})}{\sqrt{-g}}\,.
\end{align}
Then, we expand the mass $m_a\left(\phi\right)$ around the asymptotic value of the scalar field at infinity, $\phi_0$. Using the definition of the sensitivities and higher order sensitivities of \cite{Mirshekari:2013vb},
\begin{align}
s_a^{(n)}\equiv\frac{d^{n+1}\ln\ m_a(\phi)}{d \ln\phi^{n+1}}\Big|_{\phi=\phi_0}\,,
\end{align}
we obtain, at the minimal order required for this work,
\begin{align}
m_a\left(\phi\right)=m_a \Biggl[ 1 + s_a \psi + \frac{1}{2} \Bigl( s_a^2 + s_a' - s_a\Bigr) \psi^2 + \frac{1}{6}\Bigl(s_a'' + 3 s_a's_a -3s_a' + s_a^3-3s_a^2 +2s_a\Bigr) \psi^3 + \mathcal{O}\left(\psi^4\right)\Biggr]\,.
\end{align}
As explained at the beginning of Sec.~\ref{Sec:PNform} and in the App.~\ref{App:Fokker}, only the point-particle solution is required to compute the NNLO tidal correction to the dynamics. Hence, we do not display here the explicit expressions for $\Delta T^{\mu\nu}$ and $\Delta S$. They can be found in the forthcoming companion article~\cite{Dones:2023} in which we compute the NNLO gravitational and scalar fluxes and waveforms.
Similarly to the mass, we expand the tidal deformation parameters as a function of the scalar perturbation $\psi $ as
\begin{align}\label{tiddeform}
& \lambda_a(\phi) = \sum_{n=0}^{\infty} \frac{\lambda_a^{(n)}}{n!} \,{\phi_0}^n \, {\psi}^n\,, \qquad \mu_a(\phi) = \sum_{n=0}^{\infty} \frac{\mu_a^{(n)}}{n!} \,{\phi_0}^n \, {\psi}^n\,, \qquad \nu_a(\phi) = \sum_{n=0}^{\infty} \frac{\nu_a^{(n)}}{n!} \,{\phi_0}^n \, {\psi}^n\,,\qquad c_a(\phi) = \sum_{n=0}^{\infty} \frac{c_a^{(n)}}{n!} \,{\phi_0}^n \, {\psi}^n\,.
\end{align}

\subsection{The metric and scalar-field decomposition}
\label{decomp}

To solve the field equations, we further decompose the metric and the scalar field in terms of PN potentials that obey flat d'Alembertian equations~\cite{Blanchet:2013haa}. We start by decomposing the metric perturbation in its component $h^{\mu\nu}=\left(h^{00ii}, h^{0i}, h^{ij}\right)$, where $h^{00ii}\equiv h^{00}+h^{ii}$, which are in turn decomposed in terms of some PN potentials. The same applies for the scalar perturbation $\psi$, see Eqs.~\eqref{metric_decomp}. The next step is to determine the minimal order that is required in order to get the dynamics at the next-to-next-to leading order  (NNLO) in the tidal effects. Following~\cite{Henry:2019xhg}, we start by noticing that it is enough to solve the point-particle field equations (i.e. neglecting the tidal corrections to the potentials) in order to get the corrections to the Lagrangian due to tidal effects. See App.~\ref{App:Fokker} for the full reasoning. Hence, the potentials will be sourced by the point-particle matter action~\eqref{Spp} only and the tidal corrections will come from the injection of these field equation solutions into $S_{\mathrm{tidal}}=S_{\mathrm{fs}}^{(s)} + S_{\mathrm{fs}}^{(g)}+S_{\mathrm{fs}}^{(g-s)}$. 
Furthermore, to get the NNLO corrections for tidal effects, we formally need to know the metric and scalar field up to 2PN beyond the leading order, i.e. to order $\left(h^{00ii}, h^{0i}, h^{ij};\,\psi\right) = \mathcal{O}\left(\frac{1}{c^6},\frac{1}{c^5},\frac{1}{c^6};\frac{1}{c^6}\right)$. However, as the leading term is sourced only by the scalar field, see Ref.~\cite{Bernard:2019yfz}, such a high order is only required for the scalar field, while one can go to one lower PN order for the metric components. At the end, we get that we should know the metric and the scalar field at the orders
\begin{equation}
    \left(h^{00ii}, h^{0i}, h^{ij};\,\psi\right) = \mathcal{O}\left(\frac{1}{c^4},\frac{1}{c^3},\frac{1}{c^4};\frac{1}{c^6}\right).
\end{equation}

Using the standard PN decomposition, we introduce the potentials $\left(V,\,V^i,\,\hat{W}^{ij}\right)$ and  $\left(\psi_{(0)},\,\psi_{(1)}\right)$ to parametrize the metric and scalar perturbations~\cite{Blanchet:2013haa,Bernard:2018hta}:
\begin{subequations}
\label{metric_decomp}
\begin{align}
& h^{00ii}=- \frac{4}{c^2}V- \frac{8}{c^4} V^2 + \mathcal{O}\left(\frac{1}{c^6}\right) \,, \\
& h^{0i}=- \frac{4}{c^3}V_i+ \mathcal{O}\left(\frac{1}{c^5}\right) \,, \\
& h^{ij}=- \frac{4}{c^4}(\hat{W}_{ij}-\frac{1}{2} \delta_{ij} \hat{W})+\mathcal{O}\left(\frac{1}{c^6}\right) \,, \\
& \psi=-\frac{2}{c^2}\psi_{(0)}+\frac{2}{c^4}\left(1-\frac{\phi_0 \omega_0^{'}}{3+2 \omega_0}\right)\psi_{(0)}^2 +\frac{1}{c^6}\left[-\frac{4}{3}\Biggl(1-\frac{4\phi_0 \omega_0^{'}}{3+2 \omega_0}-\frac{\phi_0^2(-4(\omega_0^{'})^2+ (3+2 \omega_0)\omega_0^{''})}{(3+2 \omega_0)^2}\right)\psi_{(0)}^3 +\psi_{(1)}\Biggr] +\mathcal{O}\left(\frac{1}{c^8}\right) \,, 
\end{align}
\end{subequations}
with $\hat{W}\equiv\hat{W}_{ii}$. Each PN potential satisfies a flat space-time wave equation
\begin{subequations}
\label{PotentialWaveEq}
\begin{align}
&\Box V = -4 \pi G \sigma \,, \\
&\Box V_i= -4 \pi G \sigma_i \,, \\
&\Box \hat{W}_{ij}=-4 \pi G \left( \sigma_{ij}-\delta_{ij}\sigma_{kk}\right) -\p_i V\p_iV - (3+2 \omega_0)\p_i \psi_{(0)} \p_j \psi_{(0)} \,, \\
&\Box \psi_{(0)}= -4 \pi G \sigma_s \,, \\
&\Box \psi_{(1)}= -16 \pi G \sigma_s \hat{W} -4 \biggl( 4 V_i\p_{ti}\psi_{(0)} + 2 \hat{W}_{ij}\p_{ij}\psi_{(0)}+2V \p_t^2 \psi_{(0)} + 2 \bigl( \p_t V_i + \p_j \hat{W}_{ij} -\frac{1}{2} \p_i \hat{W}\bigr)\p_i \psi_{(0)} \biggr) \label{PotentialWaveEq_4}\,,
\end{align}
\end{subequations}
where we have introduced the compact-support matter source densities in the point-particle approximation
\begin{subequations}
\begin{align}
&	\sigma = \frac{1}{\phi_0 \varphi^3}\frac{T^{00} + T^{ii}}{c^2}\,,\qquad	\sigma_i = \frac{1}{\phi_0 \varphi^3}\frac{T^{0i}}{c} \,,\qquad	\sigma_{ij} = \frac{1}{\phi_0 \varphi^3}T^{ij} \,,\\[5pt]
&	\sigma_s = - \frac{1}{c^2\phi_0}\frac{\sqrt{-g}}{\sqrt{\left(3+2\omega_0\right)(3+2\omega)}} \left( T - 2\varphi \frac{\partial T}{\partial \varphi} \right) \,.\label{sigmas}	\end{align}
\end{subequations}
 Among these potentials, only $\psi_{(1)}$, which has a non-compact support source, is new to this paper. Note that the derivatives in Eqs.~\eqref{PotentialWaveEq_4} should be understood as Schwartz distributional derivatives~\cite{Schwartz:1978}.
Solving these equations does not introduce new difficulties and it was achieved using the techniques already introduced in the literature~\cite{Blanchet:2013haa}. In Table~\ref{table_PN_Potentials}, we summarize the orders at which each potential is needed in order to derive the Fokker Lagrangian up to the NNLO in the tidal effects. 
\begin{center}
\begin{tabular}{ |c||c|c|c|c|c| } 
\hline 
 & $\quad$ V $\quad$& $\quad V_i \quad $ & $ \quad W_{ij} \quad $ & $\quad \psi_{(0)} \quad $ & $\quad \psi_{(1)} \quad $\\ [4pt]
\hline \hline
$\quad$ N $\quad$ & $\times$ & $\times$ & $\times$ & $\times$ & $\times$ \\ [4pt]
\hline
$\quad$ 1PN $\quad $ & $\times$ &  &  & $\times$ &  \\ [4pt]
\hline
$\quad$ 2PN $\quad $ &  &  &  & $\times$ & \\ [4pt]
\hline
\end{tabular}
\captionof{table}{Summary of the PN potentials and their required order for the computation of the NNLO tidal corrections in the Lagrangian \label{table_PN_Potentials}}
\end{center}

We then inject the gravitational and scalar solutions in the total action in order to get a generalized Lagrangian that depends on the positions and their successive derivatives. As we are interested in the tidal corrections only, it is sufficient to incorporate the p.p. solutions in the finite-size actions~(\ref{Stidscal},\ref{Stidgrav},\ref{Stidscalgrav}). We present the result of the calculations in Sec.~\ref{Sec:NNLOlag}.

\section{Alternative derivation: PN-EFT formalism}
\label{Sec:PNEFT}

In parallel to the traditional post-Newtonian calculation presented in Sec.~\ref{Sec:PNform}, we have also computed the Lagrangian using an effective field theory (EFT) method~\cite{Goldberger:2004jt}. In addition to paving the way to perform other heavy calculations with this method, it provided us with an additional check of our PN results. We recall here the main steps of the EFT framework and explain how to adapt it to a ST theory. The core of the calculation, namely the calculations of Feynman rules and diagrams, are respectively put in the App.~\ref{App:Feyn} and in the supplementary file.

Our goal is to work with a worldline theory and canonically normalized bulk fields so that we can exploit the PN-EFT machinery~\cite{Kuntz:2019zef,Brax:2021qqo,Porto:2016pyg}. Specifically, in addition to the conformal transformation~\eqref{conf}, we perform a redefinition of the scalar field,
\begin{align}\label{redefphi}
&\frac{\psi}{\tilde{M}_{\rm pl}}=\sqrt{\frac{3+2\omega_0}{2}}\ln \frac{\phi}{\phi_0}\,,
\end{align}
where we have introduced the effective Planck mass $\tilde{M}_{\rm pl}^{2}\equiv\frac{c^3\phi_0}{8\pi G}$ following closely the EFT vocabulary but replacing the gravitational constant by an effective one, $\frac{G}{\phi_0}$. As before, we expand all the field-dependent couplings around the asymptotic value $\phi_0$  with respect to the canonically normalized field $\psi$ ending up with the following action,
\begin{align}\label{Left}
S_{\rm EFT} = & \int\ud^4 x\sqrt{-\tilde{g}}\,\biggl[\frac{\tilde{M}_{pl}^2}{2}\tilde{R} - \frac{1}{2}\tilde{g}_{\mu\nu}\partial_{\mu}\psi\partial_{\nu}\psi\sum_{n=0}^{\infty}\frac{x_n}{n!}\left(\frac{\psi}{\tilde{M}_{pl}}\right)^n\biggr] - \sum_{a=1,2} \int \ud t_a\ m_a\sqrt{-(\tilde{g}_{\mu\nu})_a v^{\mu}_av^{\nu}_a}\sum_{n=0}^{\infty}\frac{\tilde{d}_n^{(a)}}{n!}\left(\frac{\psi}{\tilde{M}_{pl}}\right)^n \nn\\
& -\frac{1}{2}\sum_{a=1,2} \int \ud t_a\ \lambda^{(0)}_a(\phi_0)\sqrt{-(\tilde{g}_{\mu\nu})_a v^{\mu}_av^{\nu}_a}\Big(\tilde{g}^{\mu\nu}-\frac{v_a^{\mu}v_a^{\nu}}{\tilde{g}_{\mu\nu}v^{\mu}_av^{\nu}_a}\Big)\frac{\partial_{\mu}\psi\partial_{\nu}\psi}{\tilde{M}_{pl}^2}\sum_{n=0}^{\infty}\frac{\tilde{f}_n^{(a)}}{n!}\left(\frac{\psi}{\tilde{M}_{pl}}\right)^n\nn\\
&-\frac{1}{4}\sum_{a=1,2}\int \ud t_a\ \mu_a^{(0)}\tilde{f}_0^{(a)}\  \frac{\nabla_{\mu\nu}^{\perp}\psi\ \nabla^{\mu\nu}_{\perp}\psi}{\tilde{M}_{pl}^2} +\frac{1}{4}\sum_{a=1,2}\int \ud t_a\ \nu_a^{(0)}\ \frac{\sqrt{2}}{\sqrt{3+2\omega_0}} \frac{\partial_{\mu}\partial_{\nu}(e^{-\sqrt{\frac{2}{3+2\omega_0}}\frac{\psi}{\tilde{M}_{pl}}}\tilde{g}_{00})\ \nabla^{\mu\nu}_{\perp}\psi}{\tilde{M}_{pl}\ (e^{-\sqrt{\frac{2}{3+2\omega_0}}\frac{\psi}{\tilde{M}_{pl}}}\tilde{g}_{00})}\nn\\ 
& +\sum_{a=1,2}\int \ud t_a\ \frac{c_a^{(0)}}{16}\  \frac{\partial_{\mu}\partial_{\nu}(e^{-\sqrt{\frac{2}{3+2\omega_0}}\frac{\psi}{\tilde{M}_{pl}}}\tilde{g}_{00})\ \partial^{\mu}\partial^{\nu}(e^{-\sqrt{\frac{2}{3+2\omega_0}}\frac{\psi}{\tilde{M}_{pl}}}\tilde{g}_{00})}{(e^{-\sqrt{\frac{2}{3+2\omega_0}}\frac{\psi}{\tilde{M}_{pl}}}\tilde{g}_{00})^2}\,,\nn\\
\end{align}
where, only in this section, we have set $c=1$. Note that, for the last three terms we are displaying only the leading order contribution needed for this work. From the two last terms in Eq.~\eqref{Left}, we directly observe the mixing between the different kind of tidal deformability parameters. In particular, the purely gravitational contribution will also contain terms corresponding to purely scalar and gravito-scalar interactions. The coupling constants, $x_n$, $\tilde{d}_n$ and $\tilde{f}_n$, are given up to the order needed for our calculations by the following expressions
\begin{eqnarray}
&x_n=\big(\frac{2}{3+2\omega_0}\big)^{1+n/2}\sum_{m=1}^n(-1)^m\omega_0^{(m)}\phi_0^m\Big(\sum_{l=1}^m\frac{l^{n-1}(-1)^l}{\Gamma(l)\Gamma(1+m-l)}\Big)\, \quad \text{ for}\ n\ge 1,& \qquad\qquad x_0=1\,,
\end{eqnarray}
and
\begin{eqnarray}
&\tilde{d}_n^{(a)}=\frac{d_n^{(a)}}{2^n}\big(\frac{2}{3+2\omega_0}\big)^{n/2}\,,\qquad\qquad\qquad\qquad\qquad\qquad &d_0^{(a)}=1\,,\nn\\
& &d_1^{(a)} = 2s_a - 1 \,,\nn\\
& &d_2^{(a)} = 4s'_a + (2s_a-1)^2\,,\nn\\
& &d_3^{(a)} = 8s''_a+12s'_a(2s_a-1)+(2s_a-1)^3 \,,\\[7pt]
&\tilde{f}_n^{(a)}=f_n^{(a)}\big(\frac{2}{3+2\omega_0}\big)^{1+n/2}\,,\qquad\qquad\qquad\qquad\qquad\qquad  &f_0^{(a)}=1\,,\nn\\
& &f_1^{(a)}=\frac{\lambda^{(1)}_a}{\lambda^{(0)}_a}\phi_0+5/2\,,\nn\\
& &f_2^{(a)}=\frac{\lambda^{(2)}_a}{\lambda^{(0)}_a}\phi_0^2+6\frac{\lambda^{(1)}_a}{\lambda^{(0)}_a}\phi_0+25/4\,.\nn\\
\end{eqnarray}

The action~\eqref{Left} is the starting point to develop the machinery of the PN-EFT formalism. To do so, we also perform a Kaluza-Klein decomposition of the metric as
\begin{equation}\label{KKdecomp}
\tilde{g}_{\mu\nu}=e^{4\frac{\phi_g}{\tilde{M}_{pl}}}
\begin{pmatrix} -1 & 2A_j/\tilde{M}_{pl}  \\ 2A_i/\tilde{M}_{pl}\ \ \ \ & e^{-8\frac{\phi_g}{\tilde{M}_{pl}}}\gamma_{ij}-4A_iA_j/\tilde{M}_{pl}^2
 \end{pmatrix},
\end{equation}
where $\gamma_{ij}=\delta_{ij}+\sigma_{ij}/\Lambda$. In the following, we will work with the Kaluza-Klein gravitational fields, $\left(\phi_g,\,A_i,\,\sigma_{ij}\right)$, instead of the perturbation $h^{\mu\nu}$. We derive all the Feynman rules with respect to the scalar, vector and tensor modes of the Kaluza-Klein decomposition and the canonically normalized massless scalar field $\psi$, that we introduced earlier, which can be found in App.~\ref{App:Feyn}. For completeness, we have also put all the Feynman diagrams and their values in the supplementary file (Section III).

Summing together all the diagrams, we get the NNLO tidal Lagrangian computed from EFT techniques. First in order to absorb terms non-linear in the accelerations, we add a double-zero term of the form~\cite{Damour:1985mt}
\begin{equation}
-\frac{G^2m_1^2\lambda_2^{(0)}}{r^2}\tilde{f}_0^{(2)}(\tilde{d}_1^{(1)})^2\Big(2\ na_1^{LO}(2\ na_2^{LO}-na_1^{LO})-a_1^{LO}\cdot a_2^{LO}\Big)+[1\leftrightarrow 2]\,,
\end{equation}
where $a_i^{LO}$ denotes the replacement of the LO equations of motion (eom). 

Then, in order to compare directly our results at the level of the Lagrangian we need to make another operation. Indeed, in a separate computation within PNEFT, we have computed the 2PN order Lagrangian in the point-particle approximation in ST theories~\cite{Mougiakakos:2023}. We found explicitly, that in order to bring the 2PN ST point-particle Lagrangian to the form displayed in~\cite{Bernard:2018ivi}, we have to add a double-zero term of the form
\begin{equation}
\delta L^{2PN}=\frac{\alpha \tilde{G}m_1 m_2\ r}{8}na_1^{LO}\cdot na_2^{LO}-(15+8\overline{\gamma})\frac{\alpha \tilde{G}m_1 m_2\ r}{8}a_1^{LO}\cdot a_2^{LO}\,, 
\end{equation}
where the constants are defined according to Table~\ref{table1}. Such a double-zero term, when we include the tidal effects to the LO eoms, contributes to the NNLO tidal Lagrangian as
\begin{equation}
\delta L^{2PN}_{{\rm tidal}}=-2\Big(7-2\tilde{d}_1^{(1)}\tilde{d}_1^{(2)}\Big)\frac{G^3}{r^4}\Big[m_2^2\lambda_{1}^{(0)}\tilde{f}_0^{(1)}(\tilde{d}_1^{(2)})^2+m_1^2\lambda_{2}^{(0)}\tilde{f}_0^{(2)}(\tilde{d}_1^{(1)})^2\Big]\Big[m_1na_1-m_2na_2+2\frac{Gm_1m_2}{r^2}(1+2\tilde{d}_1^{(1)}\tilde{d}_1^{(2)})\Big]\,.
\end{equation}
Finally, after taking into account the above contribution, we find complete agreement for the Lagrangian computed with the traditional PN formalism.

\section{NNLO conservative Lagrangian}
\label{Sec:NNLOlag}


The Lagrangian describing the conservative dynamics of compact binary systems in ST theories at 3PN order in harmonic coordinates and the leading order scalar tidal correction have been previously derived in~\cite{Bernard:2018hta} and~\cite{Bernard:2019yfz}. In the present work, we have pushed the computation of the tidal correction up to the next-to-next-to-leading order (NNLO) associated to the 2PN dynamics. The resulting Lagrangian is displayed below. As explained in the previous section, it has been obtained both with the traditional PN formalism and the most recent PN-EFT one.
The resulting Fokker Lagrangian is a generalized one, meaning that it is a function of the positions $y_a^i(t)$, the velocities $v_a^i(t)$, but also of the accelerations $a_a^i(t)$ and their successive derivatives. In order to get a Lagrangian that is at most linear in the acceleration, we have applied a reduction procedure which consists in iteratively adding total time derivatives, to remove the dependence in higher-order derivative of the acceleration, and double-zero terms, to remove terms non-linear in acceleration~\cite{Damour:1985mt}. After this procedure, the Fokker Lagrangian in its reduced form is completely equivalent to the original one, meaning that they both yield to the same equations of motion in a harmonic gauge. 
We write the total Lagrangian as
\be\label{Ltot}
L=L_{\rm pp}^{\rm 2PN}+L_{\rm tidal}\,,\qquad  L_{\rm tidal}= L_{\rm LO}+ L_{\rm NLO}+L_{\rm NNLO}\,,
\ee
where $L_{\rm pp}^{\rm 2PN}$ is the 2PN order Lagrangian in the point-particle approximation in ST theories displayed in Ref.~\cite{Bernard:2018ivi}. The NNLO result is further split in increasing powers of $\tilde{G}$ as 
\be
L_{\rm NNLO}=\tilde{G}^2 L_{\rm NNLO}^{(2)} + \tilde{G}^3 L_{\rm NNLO}^{(3)} + \tilde{G}^4 L_{\rm NNLO}^{(4)}\,.
\ee
The complete result for each components is then
\begin{subequations}
\begin{align}
&L_{\rm LO}=\frac{\alpha ^2 \tilde{G} ^2 }{c^2 r_{12}^4} m_1 m_2\frac{-2 \zeta}{1-\zeta}\frac{m_2}{m_1} \bar{\delta}_2\lambda_1^{(0)}  + [1\leftrightarrow 2] \,,\label{LtidLO}\\
&L_{\rm NLO}=\frac{\alpha^3 \tilde{G}^3}{c^4 r_{12}{}^5} \frac{\zeta}{1 -  \zeta} (2 + \bar{\gamma}) \Bigg[\bar{\delta}_2 \lambda_1^{(0)} \Bigl(\frac{4 m_{1}{} m_{2}{}^2 (\bar{\gamma} - 4 \bar{\beta}_2)}{\bar{\gamma} (2 + \bar{\gamma})} + m_{2}{}^3 \bigl(3 + \frac{(5 \zeta - 4 \lambda_1) (1 - 2 s_2)}{-1 + \zeta}\bigr)\Bigr) + \frac{2 \zeta m_{2}{}^3 {} \bar{\delta}_2 \phi_{0} \lambda_1^{(1)} (1 - 2 s_2)}{-1 + \zeta} \Biggr] \nn\\
& \hspace{1.1cm}+\frac{\alpha^2 \tilde{G}^2}{c^4 r_{12}{}^4} \frac{\zeta}{1 -  \zeta} \; m_{2}{}^2 \bar{\delta}_2 \lambda_1^{(0)} \Bigl(-2 (n_{12} v_{1})^2 + v_{1}^2 + 4 (n_{12} v_{1}) (n_{12} v_{2}) + 2 (n_{12} v_{2})^2\Bigr) + [1\leftrightarrow 2] \,,\label{LtidNLO}\\
&L_{\rm NNLO}^{(2)}= \alpha^2  m_{2}{}^2 \Biggl[ \frac{1}{r_{12}{}^6}\left(c_1^{(0)} -4 \nu_1^{(0)}  \frac{\zeta (1 - 2 s_2)}{1 - 2 \zeta  s_2}-4 \frac{\mu_1^{(0)}}{c^2} \Big(\frac{\zeta(1 - 2 s_2)}{1 - 2 \zeta  s_2}\Big)^2\right) \frac{3}{8}(2 + \bar{\gamma})^2\Bigl(1+\frac{\zeta}{1-\zeta}(1-2s_2)\Bigr)^2 \nn\\
& \hspace{1cm}+ \frac{1}{c^6 r_{12}{}^3} \frac{- \zeta}{ 1 -  \zeta}\bar{\delta}_2 \lambda_1^{(0)} \Biggl(2 \Bigl(-2 (a_{2} v_{1}) (n_{12} v_{1}) + 7 (a_{2} n_{12}) (n_{12} v_{1})^2 - 2 (a_{2} n_{12}) v_{1}^2+ 2 (n_{12} v_{1}) (a_{2} v_{2})  \nn\\
& \hspace{1cm}+ 2 (a_{2} v_{1}) (n_{12} v_{2})- 12 (a_{2} n_{12}) (n_{12} v_{1}) (n_{12} v_{2}) - 2 (a_{2} v_{2}) (n_{12} v_{2}) + 7 (a_{2} n_{12}) (n_{12} v_{2})^2 + 4 (a_{2} n_{12}) (v_{1} v_{2})- 2 (a_{2} n_{12}) v_{2}^2\Bigr) \nn\\
& \hspace{1cm}+ \frac{1}{4r_{12}}\Biggr\{ 4(n_{12} v_{1})^2 v_{1}^2 -  v_{1}^4 - 8 (n_{12} v_{1}) (n_{12} v_{2}) v_{1}^2 - 24 (n_{12} v_{1})^2 (n_{12} v_{2})^2 + 12 (n_{12} v_{2})^2v_{1}^2 + 48 (n_{12} v_{1}) (n_{12} v_{2})^3 \nn\\
& \hspace{1cm}+ 16 (n_{12} v_{1}) (n_{12} v_{2}) (v_{1} v_{2}) - 16 (n_{12} v_{2})^2 (v_{1} v_{2})- 16 (n_{12} v_{1}) (n_{12} v_{2}) v_{2}^2\Biggr\}\Biggr)\Biggr] + [1\leftrightarrow 2] \,,\label{LtidNNLO2}\\
&L_{\rm NNLO}^{(3)}= \frac{\alpha^3 }{c^6 r_{12}{}^5} \frac{\zeta}{ 1 -  \zeta} (2 + \bar{\gamma}) \bar{\delta}_2 \Biggl[ \lambda_1^{(0)} \Biggl( m_{1}{} m_{2}{}^2 \Bigl(- \frac{(43 \bar{\gamma} + 44 \bar{\gamma}^2 + 80 \bar{\beta}_2) }{\bar{\gamma} (2 + \bar{\gamma})}(n_{12} v_{1})^2 + \frac{(7 \bar{\gamma} + 8 \bar{\gamma}^2 + 16 \bar{\beta}_2) }{\bar{\gamma} (2 + \bar{\gamma})}v_{1}^2 \nn\\
& \hspace{1cm}+ \frac{3 (7 \bar{\gamma} + 16 \bar{\gamma}^2 + 40 \bar{\beta}_2) }{\bar{\gamma} (2 + \bar{\gamma})}(n_{12} v_{1}) (n_{12} v_{2}) -  \frac{4 (-3 + \bar{\gamma}) }{2 + \bar{\gamma}}(n_{12} v_{2})^2 -  \frac{(5 \bar{\gamma} + 8 \bar{\gamma}^2 + 8 \bar{\beta}_2) }{\bar{\gamma} (2 + \bar{\gamma})}(v_{1} v_{2}) -  \frac{4 v_{2}^2}{2 + \bar{\gamma}}\Bigr) \nn\\
& \hspace{1cm}+ m_{2}{}^3 \Bigl(- (n_{12} v_{1})^2 + \tfrac{1}{2} v_{1}^2 + \frac{(88 + 69 \bar{\gamma}) }{2 (2 + \bar{\gamma})}(n_{12} v_{1}) (n_{12} v_{2}) -  \frac{(57 + 41 \bar{\gamma}) }{2 + \bar{\gamma}}(n_{12} v_{2})^2 -  \frac{(32 + 21 \bar{\gamma}) }{2 (2 + \bar{\gamma})}(v_{1} v_{2}) + \frac{(24 + 17 \bar{\gamma}) }{2 (2 + \bar{\gamma})}v_{2}^2 \nn\\
& \hspace{1cm}- (1 - 2 s_2)\frac{(5 \zeta - 4 \lambda_1)}{1 - \zeta} \Bigl\{(n_{12} v_{1})^2 -  \frac{ v_{1}^2}{2} -  \frac{9 (n_{12} v_{1}) (n_{12} v_{2})}{2}+ (n_{12} v_{2})^2 + \frac{ (v_{1} v_{2})}{2} -  \frac{ v_{2}^2}{2}\Bigr\}\Bigr)\Biggr) \nn\\
& \hspace{1cm}+ \frac{\zeta}{1 -  \zeta} m_{2}{}^3 \phi_{0}{}\lambda_1^{(1)} (1 - 2 s_2)\Bigl(-2 (n_{12} v_{1})^2 + v_{1}^2 + 9 (n_{12} v_{1}) (n_{12} v_{2}) - 2 (n_{12} v_{2})^2 -  (v_{1} v_{2}) + v_{2}^2\Bigr)\Biggr]  + [1\leftrightarrow 2] \,,\label{LtidNNLO3}\\
&L_{\rm NNLO}^{(4)}= \frac{\alpha^4 }{c^6 r_{12}{}^6} \frac{\zeta}{1 -  \zeta} \Biggl[ \lambda_1^{(0)} \Bigg( m_{1}{}^2 m_{2}{}^2 \Bigl(  \bar{\gamma} (2 + \bar{\gamma})^2 (
\tfrac{4}{20} - \tfrac{3}{20} \bar{\gamma}) -  \frac{\bar{\delta}_2 \bigl(22 \bar{\gamma}^2 + 12 \bar{\gamma}^3 + 3 \bar{\gamma}^4 - 160 \bar{\gamma} \bar{\beta}_2 + 20 \bar{\gamma}^2 \bar{\beta}_2 + 160 (\bar{\beta}_2)^2 + 80 \bar{\gamma} \bar{\chi}_2\bigr)}{5 \bar{\gamma}^2}\Bigr)  \nn\\
& \hspace{1cm}+ m_{2}{}^4 \Bigl(-(\bar{\delta}_2)^2\frac{1}{(1 - \zeta) \zeta} \bigl(4 \zeta + 21 \zeta^2 - 58 \zeta \lambda_1 + 40 (\lambda_1)^2 - 8 \lambda_2\bigr) + \bar{\delta}_2 (2 + \bar{\gamma})^2 \bigl(- \tfrac{13}{4}  +  \frac{3}{2(1 - \zeta)}(5 \zeta - 4 \lambda_1)(1 - 2 s_2)\bigr)\Bigr)  \nn\\
& \hspace{1cm}+ m_{1}{} m_{2}{}^3 \Bigl((\bar{\delta}_2)^2 \frac{1}{(1 -\zeta) \bar{\gamma}^2}6 (\bar{\gamma}^2 + 8 \bar{\beta}_2)  (5 \zeta - 4 \lambda_1) (1 - 2 s_1) + \bar{\delta}_2 \bigl(- \frac{79 \bar{\gamma}^2 + 41 \bar{\gamma}^3 + 8 \bar{\gamma}^2 \bar{\beta}_1 - 64 \bar{\gamma} \bar{\beta}_2 - 24 \bar{\gamma}^2 \bar{\beta}_2 + 64 \bar{\beta}_1 \bar{\beta}_2}{\bar{\gamma}^2}  \nn\\
& \hspace{1cm}+  \frac{3}{2 (1 - \zeta)} (2 + \bar{\gamma})^2 (5 \zeta - 4 \lambda_1) (1 - 2 s_2)\bigr)\Bigr)\Biggr)  \nn\\
& \hspace{1cm}+  \frac{\zeta}{1 -  \zeta} \bar{\delta}_2 \phi_{0}{} \lambda_1^{(1)} \Bigg( - \frac{4 m_{2}{}^4 \bar{\delta}_2 (6 \zeta - 5 \lambda_1)}{\zeta} + 3 (2 + \bar{\gamma})^2 m_{2}{}^4 (1 - 2 s_2) + \frac{6 (2 + \bar{\gamma}) m_{1}{} m_{2}{}^3 (\bar{\gamma} - 4 \bar{\beta}_2) (1 - 2 s_2)}{\bar{\gamma}}\Biggr)  \nn\\
& \hspace{1cm}+ \frac{-4\zeta}{1 -  \zeta} m_{2}{}^4  (\bar{\delta}_2)^2 \phi_{0}{}^2\lambda_1^{(2)}\Biggr]+ [1\leftrightarrow 2]\,.\label{LtidNNLO3}
\end{align}
\end{subequations}
Note that the tidal corrections in the Lagrangian do not show directly the PN order to which they contribute. This is a well-known feature already in GR that comes from the fact that the tidal parameters $\left(\lambda_a,\mu_a,\nu_a,c_a\right)$ are dimension-full parameters. It can be better understood by a simple scaling argument. The couplings $\left(\lambda_a,\mu_a,\nu_a,c_a\right)$ respectively scale as $\left(ML^2,ML^4,ML^2T^2,ML^2T^2\right)$. It leads us to define the dimensionless parameters,
\begin{align}
    k_{a,\lambda}^{(n)}\equiv \lambda_a^{(n)}\cdot\frac{\tilde{G}\alpha }{c^2R_a^3}\cdot\left(\frac{\tilde{G}\alpha M_a}{c^2R_a}\right)^n\,,\qquad
    k_{a,\mu}^{(0)}\equiv \mu_a^{(0)}\cdot\frac{\tilde{G}\alpha }{c^2R_a^5}\,,\qquad
    k_{a,\nu}^{(0)}\equiv \nu_a^{(0)}\cdot\frac{\tilde{G}\alpha }{R_a^5}\,,\qquad
    k_{a,c}^{(0)}\equiv c_a^{(0)}\cdot\frac{\tilde{G}\alpha }{R_a^5}\,.
\end{align}
Substituting these parameters in the Lagrangian and using the compacity argument, namely that $\frac{\tilde{G}\alpha M_a}{c^2R_a}\sim 1$, we see that the leading order contribution of each family of tidal parameters is indeed $\mathcal{O}\left(\frac{1}{c^6},\frac{1}{c^{10}},\frac{1}{c^{10}},\frac{1}{c^{10}}\right)$ as expected. Hence, we have computed the correction to the p.p. Lagrangian up to the NNLO, \textit{i.e.} 5PN order.

We have performed a number of consistency checks on our results. First, the leading order Lagrangian~\eqref{LtidLO} is in total agreement with the leading order result from~\cite{Bernard:2019yfz}. Then by taking the GR limit, \textit{i.e.} when $\omega_0 \rightarrow \infty$ and $\phi_0 \rightarrow 1$, we have verified that we retrieve the LO tidal correction in GR computed in~\cite{Henry:2019xhg}. Finally, we have checked that the equations of motion for each particle, derived from the Lagrangian~\eqref{Ltot}, are indeed Lorentz invariant. In order to keep this article short and readable, we have relegated the result for the equations of motion in harmonic coordinates up to the NNLO in the supplementary material.

\section{Noetherian quantities and center of mass frame}
\label{Sec:noether}

In this section, we aim at computing the ten conserved Noetherian quantities, namely the energy $E$, the linear and angular momenta, $P^i$ and $J^i$ and the boost $K^i$. We then use them to define the center of mass frame by solving $G^i=0$ where $G^i$ is the center of mass position. We finally reduce the expressions of the relative acceleration and the conserved quantities to the center of mass frame. We start by briefly recalling the reasoning to derive such quantities.

In Sec.~\ref{Sec:NNLOlag}, we have obtained the Lagrangian up to the NNLO as a function of only the positions $y_a^i(t)$, velocities $v_a^i(t)$ and accelerations $a_a^i(t)$. If we consider an infinitesimal transformation for the body A at some time t, namely $ \delta y_a(t) = y_a'(t)-y_a(t)$, the Lagrangian should transform at linear order as
\begin{align}
\delta L = \frac{dQ}{dt} + \sum \frac{\delta L}{\delta y_a}\delta y_a + \mathcal{O}(\delta y_a^2)\,,
\end{align}
where the function derivative $\delta L/ \delta y_a $  is zero ``on-shell". The quantity $Q$ is defined as
\begin{align}
Q = \sum \Bigl[(p_a\,\delta y_a) + (q_a\,\delta v_a)\Bigr]\,,
\end{align}
where $p_a^i$ and $q_a^i$ are respectively the conjugate momenta to the positions and velocities, namely
\begin{align}
p_a^i & \equiv \frac{\delta L}{\delta v_a^i} = \frac{\p L}{\p v_a^i}-\frac{d}{dt}\left( \frac{\p L}{\p  a_a^i}\right) \,,\\
q_a^i & \equiv \frac{\delta L}{\delta a_a^i} = \frac{\p L}{\p  a_a^i}\,.
\end{align}
The Lagrangian is invariant under the Poincaré group. Hence, under arbitrary infinitesimal time translation $\delta t = \tau $, spatial translation $\delta y_a^i = \epsilon^i $ and spatial rotation $\delta y_a^i = {w^i}_j y_a^j$, we have $ \delta L = 0$. This yields to the conservation on-shell of the energy, and the linear and angular momenta, defined as \cite{deAndrade:2000gf}
\begin{align}
    E &= \sum_a \Bigl[ (p_a v_a) + (q_a a_a)\Bigr] - L \,, \\
    P^i &= \sum_a p_a^i \,, \label{linear_momentum}\\
    J^i &= \varepsilon_{ijk} \sum_a \Bigl[ (p_a y_a) + (q_a v_a)\Bigr] \,.
\end{align}
In addition, the Lagrangian is also invariant under an infinitesimal Lorentz boost. At the linear order in the boost velocity $W^i$, the transformation of the body trajectories reads
\begin{align}\label{Lorentz_boost}
    \delta y_a^i = - W^i t  -\frac{1}{c^2} (W r_a)v_a^i+ \mathcal{O}(W^2) \,,
\end{align}
where $r_a$ is the distance between the field point and the body a. Following \cite{deAndrade:2000gf}, under such a linear transformation there should exist a functional $Z^i$ such that $\delta L = (W dZ/dt) + \mathcal{O}(W^2) $ plus some ``double-zero" terms which give zero ``on-shell" by the Noether theorem. The transformation (\ref{Lorentz_boost}) is associated with the conservation of the Noetherian integral $K^i = G^i - P^i t$, where $P^i$ is the linear momentum (\ref{linear_momentum}) and $G^i $ stands for the center of mass position,
\begin{align}\label{formGi}
    G^i = - Z^i + \sum_a \Biggl( -q_a^i + \frac{1}{c^2} \Bigl[ (p_a v_a)y_a^i +(q_a a_a) y_a^i +(q_a v_a) v_a^i \Bigr]\Biggr)\,.
\end{align}
The existence of such a boost symmetry of the Lagrangian is confirmed since our equations of motion at the NNLO in the tidal effects are Lorentz invariant, as mentioned in section~\ref{Sec:NNLOlag}.

In the following, we present the results for the NNLO tidal correction to all the conserved quantities. In order to stay concise, we are only displaying the center of mass position $G^i$ in generic harmonic coordinates while the relative acceleration $a^i$, and the conserved quantities $(E ,J^i)$, will be given only in the center of mass frame. The results in the generic coordinates are displayed in the auxiliary file.

\subsection{Center of mass frame}

First, using Eq.~\eqref{formGi}, we have obtained the position of the center of mass at NNLO and in a generic harmonic frame,
\begin{subequations}
\begin{align}
&G_{\rm LO}^i=0\,,\\
&G_{\rm NLO}^i= \frac{\alpha^2 \tilde{G}^2}{c^4 r_{12}{}^4}\frac{2\zeta}{1 -  \zeta}  m_{2}{}^2 \bar{\delta}_2 \lambda_1^{(0)} y_{1}{}^{i} + [1\leftrightarrow 2]\,, \\
&G_{\rm NNLO}^i= \frac{\alpha^2 \tilde{G}^2}{c^6 r_{12}{}^3} \frac{\zeta}{1 -  \zeta} m_{2}{}^2 \bar{\delta}_2 \lambda_1^{(0)} \Biggl[2 n_{12}{}^{i} \Bigl(7 (n_{12} v_{1})^2 - 2 v_{1}^2 - 14 (n_{12} v_{1}) (n_{12} v_{2}) + 5 (n_{12} v_{2})^2 + 4 (v_{1} v_{2}) - 2 v_{2}\Bigr) \nn\\
& \hspace{1cm}+ \frac{ y_{1}{}^{i}}{r_{12}{}}\Bigl(-2 (n_{12} v_{1})^2 + v_{1}^2 + 4 (n_{12} v_{1}) (n_{12} v_{2}) + 2 (n_{12} v_{2})^2\Bigr)- 4(v_{1}{}^{i}-v_{2}{}^{i}) \Bigl((n_{12} v_{1}) -  (n_{12} v_{2})\Bigr) \Biggr]\nn\\
& \hspace{1.2cm}
+\frac{\alpha^3 \tilde{G}^3}{c^6 r_{12}{}^4} \frac{\zeta}{1 -  \zeta} \Biggl[n_{12}{}^{i} \Biggl(\bar{\delta}_2 \lambda_1^{(0)} \Bigl(\frac{m_{1}{} m_{2}{}^2 (13 \bar{\gamma} + 8 \bar{\gamma}^2 + 8 \bar{\beta}_2)}{\bar{\gamma}} + m_{2}{}^3 \bigl(\tfrac{1}{2} (-16 - 13 \bar{\gamma}) + (5 \zeta - 4 \lambda_1) \frac{(2 + \bar{\gamma}) (1 - 2 s_2)}{2 (-1 + \zeta)}\bigr)\Bigr) \nn\\
& \hspace{1cm}+  \bar{\delta}_2 \phi_{0}{}\lambda_1^{(1)}  m_{2}{}^3 \frac{\zeta (2 + \bar{\gamma})  (1 - 2 s_2)}{-1 + \zeta}\Biggr) \nn\\
& \hspace{1cm}+\frac{y_{1}{}^{i}}{r_{12}{}}\Biggl(\bar{\delta}_2 \lambda_1^{(0)} \Bigl(- \frac{4 m_{1}{} m_{2}{}^2 (\bar{\gamma} - 4 \bar{\beta}_2)}{\bar{\gamma}} + m_{2}{}^3 \bigl(-3 (2 + \bar{\gamma}) -  (5 \zeta - 4 \lambda_1)\frac{(2 + \bar{\gamma})  (1 - 2 s_2)}{-1 + \zeta}\bigr)\Bigr) \nn\\
& \hspace{1cm}-  \bar{\delta}_2\phi_{0}{}  \lambda_1^{(1)}  m_{2}{}^3\frac{2 \zeta (2 + \bar{\gamma}) (1 - 2 s_2)}{-1 + \zeta}\Biggr) \Biggr] + [1\leftrightarrow 2] \,.
\end{align}
\end{subequations}
We note that at leading order, the center of mass is zero and the first correction enters only at NLO. From this, we determine the center of mass frame by solving the equation $G^{i}=0$ iteratively at each PN order. We obtain the positions $y_{a,CM}^i$ in the center of mass frame as,
\begin{align}
&y_{1,CM}^i= \Bigl(\frac{m_2}{m}+ \nu \mathcal{P}\Bigr)n^ir + \nu \mathcal{Q}v^i\,, \\
&y_{2,CM}^i= \Bigl(-\frac{m_1}{m}+ \nu \mathcal{P}\Bigr)n^ir + \nu \mathcal{Q}v^i\,,
\end{align}
Dividing the quantities $\mathcal{P}$ and $\mathcal{Q}$ into a 2PN point-particle contribution and a tidal one, $\mathcal{P}=\mathcal{P}_{\rm 2PN}+\mathcal{P}_{\rm tidal}$ and $\mathcal{Q}=\mathcal{Q}_{\rm 2PN}+\mathcal{Q}_{\rm tidal}$. The point-particle results up to the 2PN order $\mathcal{P}_{\rm 2PN}$ and $\mathcal{Q}_{\rm 2PN}$ can be found in \cite{Bernard:2018ivi} and we display here the tidal corrections up the NNLO further dividing $\mathcal{P}_{\rm tidal}$ and $\mathcal{Q}_{\rm tidal}$ as
\begin{align}
&\mathcal{P}_{\rm tidal}=\mathcal{P}_{LO}+\mathcal{P}_{NLO}+\mathcal{P}_{NNLO}\,, \\
&\mathcal{Q}_{\rm tidal}=\mathcal{Q}_{LO}+\mathcal{Q}_{NLO}+\mathcal{Q}_{NNLO} \,.
\end{align}
We then get
\begin{subequations}
\begin{align}
&\mathcal{P}_{LO}=0 \,,\\
&\mathcal{P}_{NLO}= \frac{\alpha^2 \tilde{G}^2 }{c^4  r^4} \frac{- \zeta}{1 -\zeta} m (\lambda_-^{(0)} -  \delta \lambda_+^{(0)}) \,,\\
&\mathcal{P}_{NNLO}= \frac{\alpha^2 \tilde{G}^2}{c^6 r^4} \frac{\zeta }{2 (-1 + \zeta)}m \Bigl[ v^2\bigl((-7 - 6 \nu) \lambda_-^{(0)} -  \delta (1 - 6 \nu) \lambda_+^{(0)}\bigr) + (n v)^2 \bigl(2 (11 + 10 \nu) \lambda_-^{(0)} + 2 \delta (-1 + 6 \nu) \lambda_+^{(0)}\bigr) \Bigr]\nn\\
& \hspace{1.3cm}+\frac{\alpha^3 \tilde{G}^3}{c^6 r^5} \frac{\zeta}{4 (-1 + \zeta)} m^2 \Biggl\{ \frac{1}{\bar{\gamma}} \biggl[\biggl(16 \bar{\beta}^+{} - 2 \bar{\gamma} - 3 \bar{\gamma}^2 - 16 \bar{\beta}^-{} \delta + (64 \bar{\beta}^+{} + 12 \bar{\gamma} + 12 \bar{\gamma}^2) \nu \biggr) \lambda_-^{(0)} \nn\\
& \hspace{1cm}+ \biggl(-16 \bar{\beta}^-{} + (16 \bar{\beta}^+{} + 54 \bar{\gamma} + 35 \bar{\gamma}^2) \delta + (-64 \bar{\beta}^-{} - 8 \bar{\gamma} \delta) \nu \biggr) \lambda_+^{(0)} \biggr]\nn\\
& \hspace{1cm}+ \frac{(2 + \bar{\gamma}) \biggl(4 \lambda_1 (\Lambda_-^{(0)} - 4 \nu \Lambda_-^{(0)} -  \delta \Lambda_+^{(0)}) + \zeta \Bigl(5 (-1 + 4 \nu) \Lambda_-^{(0)} + 5 \delta \Lambda_+^{(0)} + 2 \phi_{0}{} \bigl((-1 + 4 \nu) \Lambda_-^{(1)} + \delta \Lambda_+^{(1)}\bigr)\Bigr)\biggr)}{-1 + \zeta} \Biggr\} \,,
\end{align}
\end{subequations}
and,
\begin{subequations}
\begin{align}
&\mathcal{Q}_{LO}=0 \,, \\
&\mathcal{Q}_{NLO}=0 \,, \\
&\mathcal{Q}_{NNLO}=\frac{ \alpha^2\tilde{G}^2}{c^6  r^3}\frac{4\zeta}{1 - \zeta} m (n v) \lambda_-^{(0)} \,.
\end{align}
\end{subequations}

We have introduced some \textit{plus} and \textit{minus} quantities for the ST and tidal parameters as defined in the notation section~\ref{notations}.

\subsection{Acceleration in the CM}
\label{Sec:EOMinCOM}

Once we have determined the coordinates in the center of mass frame, we can inject it in the relative acceleration $a^{i}\equiv a_1^i-a_2^i$. In the CM frame we get up to the NLO,
\begin{subequations}
\begin{align}
&a_{\rm CM, LO}^i= \frac{\alpha^2 \tilde{G}^2 m}{c^2 r^5} \frac{8  \zeta   }{ 1 - \zeta }\lambda_+^{(0)} n^{i} \,,\\
&a_{\rm CM, NLO}^i=\frac{\alpha^2 \tilde{G}^2}{c^4 r^5} \frac{\zeta }{1 -  \zeta}m\Biggl[ n^{i} \Bigl( 12 (n v)^2 (\delta \lambda_-^{(0)} - 2 \nu \lambda_+^{(0)}) -2 v^2 (\delta \lambda_-^{(0)} + 3 \lambda_+^{(0)} - 12 \nu \lambda_+^{(0)})\Bigr)  -4 (n v) v^{i}(\delta \lambda_-^{(0)} + \lambda_+^{(0)} - 4 \nu \lambda_+^{(0)}) \Biggr]\nn\\
& \hspace{1cm}+ \frac{\alpha^3 \tilde{G}^3}{c^4 r^6} \frac{\zeta }{2 (1 -  \zeta)} m^2 n^i \Biggl[- \frac{80 \bar{\beta}^-{} \lambda_-^{(0)}}{\bar{\gamma}} + \frac{(80 \
\bar{\beta}^+{} - 96 \bar{\gamma} - 47 \bar{\gamma}^2 - 52 \
\bar{\gamma} \nu) \lambda_+^{(0)}}{\bar{\gamma}} + \delta
\bigl(\frac{(80 \bar{\beta}^+{} + 16 \bar{\gamma} + 15 \
\bar{\gamma}^2) \lambda_-^{(0)}}{\bar{\gamma}} -  \frac{80 \
\bar{\beta}^-{} \lambda_+^{(0)}}{\bar{\gamma}}\bigr) \nn\\
& \hspace{4.5cm}+ \frac{5 (2 + \bar{\gamma}) \bigl(4 \lambda_1 (- \delta \Lambda_-^{(0)} + \Lambda_+^{(0)}) + \zeta (5 \delta \Lambda_-^{(0)} + 2 \phi_{0}{} \delta \Lambda_-^{(1)} - 5 \Lambda_+^{(0)} - 2 \phi_{0}{} \Lambda_+^{(1)})\bigr)}{-1 + \zeta}\Biggr]\,.
\end{align}
\end{subequations}
The expression for the NNLO tidal correction to the relative acceleration is displayed in App.~\ref{App:CMeomNNLO}.

\subsection{Conserved quantities in the CM}

\subsubsection{Energy}

In the center of mass frame, the conserved energy at NLO is given by 
\begin{subequations}
\begin{align}
&E_{LO}= \frac{\alpha^2 \tilde{G}^2 m_{}^2 \nu}{c^2 r^4} \frac{2  \zeta   }{ 1 - \zeta } \lambda_+^{(0)}\,,\\
&E_{NLO}= \frac{\alpha^2  \tilde{G}^2 }{c^4 r^4}\frac{-\zeta}{2 (1 - \zeta) }m^2 \nu \Biggl[(n v)^2 (-4 \delta \lambda_-^{(0)} + 8 \nu \lambda_+^{(0)}) + v^2 \bigl(\delta \lambda_-^{(0)} + (-1 + 2 \nu) \lambda_+^{(0)}\bigr)\Biggr]\nn\\
& \hspace{1cm} +\frac{\alpha^3 \tilde{G}^3}{c^4 r^5} \frac{\zeta}{1 -  \zeta} m^3 \nu \Biggl[ \biggl\{- \frac{8 \bar{\beta}^-}{\bar{\gamma}} + \frac{\bigl(16 \bar{\beta}^+ + \bar{\gamma} (2 + 3 \bar{\gamma})\bigr) \delta}{2 \bar{\gamma}}\biggr\} \lambda_-^{(0)} + \biggl\{\frac{16 \bar{\beta}^+ -  \bar{\gamma} (10 + 3 \bar{\gamma})}{2 \bar{\gamma}} -  \frac{8 \bar{\beta}^- \delta}{\bar{\gamma}}\biggr\} \lambda_+^{(0)} \nn\\
& \hspace{3.8cm}+\frac{(2 + \bar{\gamma}) \bigl(4 \lambda_1 (- \delta \Lambda_-^{(0)} + \Lambda_+^{(0)}) + \zeta (5 \delta \Lambda_-^{(0)} + 2 \phi_{0}{} \delta \Lambda_-^{(1)} - 5 \Lambda_+^{(0)} - 2 \phi_{0}{} \Lambda_+^{(1)})\bigr)}{2 (-1 + \zeta)}\Biggr]\,. 
\end{align}
  \end{subequations}
Again, to lighten the main text, we have relegated the NNLO tidal contribution to the CM energy in the App.~\ref{App:CMenergyNNLO}

\subsubsection{Angular momentum}

Finally, the tidal correction to conserved angular momentum in the center of mass frame up to the NNLO can be written as
\begin{align}
J_{\rm tidal}^i=(\mathcal{J}_{LO}+\mathcal{J}_{NLO}+\mathcal{J}_{NNLO})(n \times v)^i
\end{align}
where $(n \times v)^i$ denotes a vector product and with
\begin{subequations}
\begin{align}
&\mathcal{J}_{LO}= 0\,,\\
&\mathcal{J}_{NLO}=\frac{\alpha^2 \tilde{G}^2}{c^4 r^3} \frac{\zeta}{1 -  \zeta} m^2 \nu \Big( - \delta \lambda_-^{(0)} + (1 - 2 \nu) \lambda_+^{(0)} \Bigr)\,,\\
&\mathcal{J}_{NNLO}^i= \frac{\alpha^2 \tilde{G}^2}{c^6 r^3} \frac{-\zeta}{2 (1 - \zeta)} m^2 \nu \Biggl(2 (n v)^2 \bigl(\delta (21 + 8 \nu) \lambda_-^{(0)} + (23 + 6 \nu - 20 \nu^2) \lambda_+^{(0)}\bigr) + v^2 \bigl(- \delta (11 + 6 \nu) \lambda_-^{(0)} -  (13 - 8 \nu + 18 \nu^2) \lambda_+^{(0)}\bigr) \Biggr)  \nn\\
& \hspace{1.3cm} +\frac{\alpha^3 \tilde{G}^3}{c^6 r^4} \frac{-\zeta}{2 (1 - \zeta)} m^3 \nu  \Biggl[ \biggl\{(6 + \bar{\gamma}) \delta + \bigl(\frac{48 \bar{\beta}^-{}}{\bar{\gamma}} + \frac{(16 \bar{\beta}^+{} + 4 \bar{\gamma} + 3 \bar{\gamma}^2) \delta}{\bar{\gamma}}\bigr) \nu \biggr\} \lambda_-^{(0)} \nn\\
& \hspace{1cm}+ \biggl\{2 -  \bar{\gamma} + \bigl(- \frac{3 (16 \bar{\beta}^+{} + 52 \bar{\gamma} + 33 \bar{\gamma}^2)}{\bar{\gamma}} -  \frac{16 \bar{\beta}^-{} \delta}{\bar{\gamma}}\bigr) \nu + 8 \nu^2\biggr\} \lambda_+^{(0)} \nn\\
& \hspace{1cm}+\frac{(2 + \bar{\gamma}) (-1 + \nu) \bigl(4 \lambda_1 (- \delta \Lambda_-^{(0)} + \Lambda_+^{(0)}) + \zeta (5 \delta \Lambda_-^{(0)} + 2  \delta \phi_{0}{}\Lambda_-^{(1)} - 5 \Lambda_+^{(0)} - 2 \phi_{0}{} \Lambda_+^{(1)})\bigr)}{-1 + \zeta} \Biggr]\,.
\end{align}
\end{subequations}

\section{Tidal effects in Einstein scalar Gauss Bonnet theory}
\label{Sec:EsGB}

While our results have been computed within the simple framework of generalized Brans-Dicke theories, see Sec.~\eqref{sec:STtheory}, it can easily be generalized to more involved theories. An example that has been extensively studied in the past years is Einstein scalar-Gauss-Bonnet theory, which arises as a low-energy limit of several quantum gravity completion of gravity~\cite{Nojiri:2018ouv,Gross:1986mw,Boulware:1985wk}. An interesting feature of these theories is that black hole solutions are different from that of GR, having non-trivial scalar solutions~\cite{Kanti:1995vq}.

In these theories, the scalar field is non-minimally coupled to gravity through the Gauss-Bonnet topological invariant, ${\cal R}_{\rm GB}^2\equiv R^2-4R_{\mu\nu}R^{\mu\nu}+R_{\mu\nu\rho\sigma}R^{\mu\nu\rho\sigma}$. The action is
\be\label{EsGBaction}
S_{\mathrm{EsGB}} = \frac{c^{3}}{16\pi G} \int\ud^{4}x\,\sqrt{-g}\left[R - 2g^{\alpha\beta}\p_{\alpha}\hat{\psi}\p_{\beta}\hat{\psi} + \alpha\,f(\hat{\psi})\,{\cal R}_{\rm GB}^2\right] +S^{\rm EsGB}_{\mathrm{m}}\left[\mathfrak{m},{\cal A}^2(\hat{\psi})g_{\alpha\beta}\right]\,,
\ee
where $\alpha$ is the coupling constant and has a dimension of $[{\rm length}]^2$. The action is directly in the Einstein frame with a canonical kinetic term for the scalar field and a coupling to matter through the conformal metric ${\cal A}^2(\hat{\psi})g_{\alpha\beta}$. As displayed in App.~A of~\cite{Julie:2022qux}, the action~\eqref{EsGBaction} has been obtained from Eq.~\eqref{STactionJF} by using the following redefinitions:
\begin{align}
\hat{g}_{\mu \nu}&={\cal A}^2(\hat{\psi})g_{\mu \nu}\,,\\
3+2\omega(\phi)&=\left(\frac{\ud \ln {\cal A} }{\ud \hat{\psi}} \right)^{-2}
\end{align}
where ${\cal A}=1/\sqrt{\phi}$. Note that in this section, we call $\hat{g}_{\mu \nu}$ the metric associated to the ST action~\eqref{STactionJF} in the Jordan frame and $g_{\mu \nu}$ the metric associated to the EsGB action~\eqref{EsGBaction} in the Einstein frame. As for the scalar fields, $\phi$ is the same as the one displayed in Eq.~\eqref{STactionJF} and we call $\hat{\psi}$ the scalar field in the EsGB theory.

In recent years, waveform modelling in EsGB theories have been developed for all the stages of the coalescence. Advances in the well-posed formulation of initial conditions have allowed to perform numerical simulations and study the merger of these objects~\cite{Okounkova:2020rqw,Witek:2018dmd,East:2020hgw,Corman:2022xqg}. Results for the inspiral evolution of compact binaries have also been obtained, relying on the fact that the leading order correction due to the new GB term is at 3PN order~\cite{Julie:2017rpw,Julie:2018lfp,Julie:2019sab,Julie:2022qux,Shiralilou:2020gah,Shiralilou:2021mfl,vanGemeren:2023rhh}~\footnote{Formally, corrections from the Gauss Bonnet invariant scale as $\frac{\alpha}{c^2}$ but as $\alpha$ has a dimension of $[{\rm length}]^2$, it reduces to a 3PN correction when introducing a dimensionless coupling constant.}. Hence, using previous results in ``simple" ST theories and adding only the leading order correction allow to have the full dynamics at 3PN order and the waveform up to 2.5PN order, including the leading tidal effect.

Following the convention in Ref.~\cite{vanGemeren:2023rhh}, the matter action can be further decomposed in a point-particle part and a tidal one, $S^{\rm EsGB}_m = S^{\rm EsGB}_{\rm pp} + S^{\rm EsGB}_{\rm tid}$, with
\begin{align}\label{Sppbis}
&    S^{\rm EsGB}_{\rm pp} = -c\sum_{a=1,2}\,\int\ud s_a\,m_a^{\rm EsGB}\left(\hat{\psi}\right)\,,\\ \nn
&    S^{\rm EsGB}_{\rm fs} = -\frac{c}{2}\,\sum_{a=1,2}\,\int\ud s_a\,\biggl\{\lambda_a^{\rm EsGB}\left(\hat{\psi}\right) \left(\nabla_{\alpha}^{\perp}\hat{\psi}\right)_a\,\left(\nabla^{\alpha}_{\perp}\hat{\psi} \right)_a  + \frac{1}{2} \mu_a^{\rm EsGB}\left(\hat{\psi}\right) \left(\nabla_{\alpha\beta}^{\perp}\hat{\psi}\right)_a\,\left(\nabla^{\alpha\beta}_{\perp}\hat{\psi} \right)_a \\
& \hspace{6cm} + \nu_a^{\rm EsGB}\left(\hat{\psi}\right) \left(\nabla_{\alpha\beta}^{\perp}\hat{\psi}\right)_a\,G_a^{\alpha\beta} - \frac{1}{2c^2} c_a^{\rm EsGB}(\hat{\psi})  \,  G^a_{\alpha\beta} \, G_a^{\alpha\beta} \; \biggr\} \,,
\end{align}
where $\ud\tau_a\equiv{\cal A}(\hat{\psi})\,\ud s_a$ and we have added the superscript (EsGB) to the masses and tidal deformability parameters to indicate that they are not the same as in the rest of the paper. Notably, we use the definition $m^{\rm EsGB}_a(\hat{\psi}) \equiv {\cal A}(\hat{\psi})\,m_a^{\rm ST}(\phi)$~\cite{Julie:2022qux} which highlights the fact that in this section the matter is coupled to the Einstein frame metric.

Using the results obtained in the present article, one can directly get the NNLO order for tidal effects in EsGB gravity. To do so, one needs to translate the notation in the present work to the one used in Refs.~\cite{Julie:2019sab,Julie:2022qux}. Such a map has already been computed and can be found in App.~A of Ref.~\cite{Julie:2022qux} where the point-particle Lagrangian at 2PN order is presented for EsGB in App.~B. Also, the leading order tidal correction was derived in~\cite{vanGemeren:2023rhh} for EsGB and was found to be in perfect agreement with~\cite{Bernard:2019yfz}.
In order to derive such a result to the NNLO, we need to extend the map for the tidal coefficients. The result is presented in Table~\ref{table3}.  At this point, we stress that the parametrization used in~\cite{vanGemeren:2023rhh} is not exactly the same as in~\cite{Julie:2022qux}. In this section, our results are presented using the parametrization and the ST coefficients introduced in~\cite{Julie:2022qux}. However, as the latter work did not focus on tidal effect, the tidal deformation parameters are the one introduced in~\cite{vanGemeren:2023rhh} or in Eq.~(3.12) of~\cite{Creci:2023cfx} for the higher order ones.
\hspace{0.5cm}\begin{small}
\begin{center}
\begin{tabular}{|c||c|c|}
	\hline
	& \textbf{Simple ST -- Refs.~\cite{Bernard:2018hta,Bernard:2018ivi,Bernard:2019yfz}} & \textbf{EsGB -- Refs.~\cite{Julie:2019sab,Julie:2022qux,vanGemeren:2023rhh}} \\[2pt]
	\hline 
	\hline 
	LO & $\lambda_a^{(0)}$ & $\frac{{\cal A}_{0}}{4\alpha_0^2}\,{\lambda_a^{(0)}}_{|{\rm EsGB}}$ \\[12pt]
	\hline
	NLO & $\lambda_a^{(1)}$ & $\frac{{\cal A}_{0}^3}{8 \alpha_0^3}\left({\lambda_a^{(0)}}_{|{\rm EsGB}}(2 \frac{\beta_0}{\alpha_0}-5 \alpha_0) -  {\lambda_a^{(1)}}_{|{\rm EsGB}}\right) $ \\[8pt]
	\hline
	NNLO & $\lambda_a^{(2)}$ & $\frac{{\cal A}_{0}^5}{16 \alpha_0^4}\left({\lambda_a^{(0)}}_{|{\rm EsGB}}(35 \alpha_0^2 - 24\beta_0 + 8 \frac{\beta_0^2}{\alpha_0^2}-2 \frac{\beta_0'}{\alpha_0})+{\lambda_a^{(1)}}_{|{\rm EsGB}}(12 \alpha_0 - 5 \frac{\beta_0}{\alpha_0})+{\lambda_a^{(2)}}_{|{\rm EsGB}}\right)$  \\[5pt]
	& $\mu_a^{(0)}$ & $\frac{{\cal A}_{0}^3}{4\alpha_0^2} \,({\mu_a^{(0)}}_{|{\rm EsGB}}+2  \alpha_0 c^2{\nu_a^{(0)}}_{|{\rm EsGB}} -  \alpha_0^2 c^2{c_a^{(0)}}_{|{\rm EsGB}})$ \\[8pt]
    & $\nu_a^{(0)}$ &$-\frac{{\cal A}_{0}^3}{2\alpha_0} \,({\nu_a^{(0)}}_{|{\rm EsGB}}-\alpha_0{c_a^{(0)}}_{|{\rm EsGB}} )$  \\[8pt]
	& $c_a^{(0)}$ & ${\cal A}_{0}^3 \, {c_a^{(0)}}_{|{\rm EsGB}}$ \\[8pt]
\hline
\end{tabular}
\captionof{table}{Translation map for tidal coefficients between simple ST and EsGB theories. \label{table3}}
\end{center}
\end{small}
Note that there are some mixing between the different types of Love number. This is due to the fact that we are coupling the matter to the metric in different frames. 

To illustrate the use of this map, we present the LO and NLO tidal corrections to the acceleration in the CM frame in EsGB :
\begin{subequations}
\begin{align}
&{a_{\rm CM,LO}^{i}}_{|{\rm EsGB}}= \frac{G_{12}^2 M}{c^2 \nu r^5} \Biggl[- (-1 + m_{-} + 2 \nu) \delta_2 {\lambda_1^{(0)}}_{|{\rm EsGB}} + (1 + m_{-} - 2 \nu) \delta_1 {\lambda_2^{(0)}}_{|{\rm EsGB}}\Biggr] n^{i}\,, \label{EOM_EsGB_LO}\\ 
&{a_{\rm CM,NLO}^{i}}_{|{\rm EsGB}}= \frac{G_{12}^3 M^2}{c^4 r^6}  \Biggl[\biggl(- \frac{80 \bar{\beta}^-{} - 80 \bar{\beta}^+{} - 102 \bar{\gamma}_{12} - 77 \bar{\gamma}_{12}^2}{8 \bar{\gamma}_{12}} + \frac{5 (16 \bar{\beta}^-{} - 16 \bar{\beta}^+{} + 2 \bar{\gamma}_{12} - 3 \bar{\gamma}_{12}^2) m_{-}}{8 \bar{\gamma}_{12}} \nn\\
& \hspace{1cm}+ \frac{\tfrac{1}{8} (-56 - 31 \bar{\gamma}_{12}) + \tfrac{1}{8} (56 + 31 \bar{\gamma}_{12}) m_{-}}{\nu} + \tfrac{13}{2} \nu\biggr) \delta_2 {\lambda_1^{(0)}}_{|{\rm EsGB}} \nn\\
& \hspace{1cm}+ \biggl(\tfrac{15}{8} (2 + \bar{\gamma}_{12}) \alpha_2^0 -  \tfrac{5}{8} (2 + \bar{\gamma}_{12}) m_{-} \alpha_2^0 -  \frac{\tfrac{5}{8} (2 + \bar{\gamma}_{12}) \alpha_2^0 -  \tfrac{5}{8} (2 + \bar{\gamma}_{12}) m_{-} \alpha_2^0}{\nu}\biggr) \delta_2 {\lambda_1^{(1)}}_{|{\rm EsGB}} \nn\\
& \hspace{1cm}+ \biggl(\frac{80 \bar{\beta}^-{} + 80 \bar{\beta}^+{} + 102 \bar{\gamma}_{12} + 77 \bar{\gamma}_{12}^2}{8 \bar{\gamma}_{12}} + \frac{5 (16 \bar{\beta}^-{} + 16 \bar{\beta}^+{} - 2 \bar{\gamma}_{12} + 3 \bar{\gamma}_{12}^2) m_{-}}{8 \bar{\gamma}_{12}} \nn\\
& \hspace{1cm}+ \frac{\tfrac{1}{8} (-56 - 31 \bar{\gamma}_{12}) + \tfrac{1}{8} (-56 - 31 \bar{\gamma}_{12}) m_{-}}{\nu} + \tfrac{13}{2} \nu\biggr) \delta_1 {\lambda_2^{(0)}}_{|{\rm EsGB}} \nn\\
& \hspace{1cm}+ \biggl(\tfrac{15}{8} (2 + \bar{\gamma}_{12}) \alpha_1^0 + \tfrac{5}{8} (2 + \bar{\gamma}_{12}) m_{-} \alpha_1^0 -  \frac{\tfrac{5}{8} (2 + \bar{\gamma}_{12}) \alpha_1^0 + \tfrac{5}{8} (2 + \bar{\gamma}_{12}) m_{-} \alpha_1^0}{\nu}\biggr) \delta_1 {\lambda_2^{(1)}}_{|{\rm EsGB}}\Biggr] n^{i}\nn\\
& \hspace{1cm}+ \frac{G_{12}^2 M}{c^4 r^5} \Biggl[n^{i} \Biggl(\biggl((3 -  \frac{\tfrac{3}{2} -  \tfrac{3}{2} m_{-}}{\nu} + 6 \nu) \delta_2 {\lambda_1^{(0)}}_{|{\rm EsGB}} + (3 -  \frac{\tfrac{3}{2} + \tfrac{3}{2} m_{-}}{\nu} + 6 \nu) \delta_1 {\lambda_2^{(0)}}_{|{\rm EsGB}}\biggr) (n v)^2 \nn\\
& \hspace{1cm}+ \biggl((\tfrac{7}{2} -  \tfrac{5}{2} m_{-} -  \frac{\tfrac{1}{2} -  \tfrac{1}{2} m_{-}}{\nu} - 6 \nu) \delta_2 {\lambda_1^{(0)}}_{|{\rm EsGB}} + (\tfrac{7}{2} + \tfrac{5}{2} m_{-} -  \frac{\tfrac{1}{2} + \tfrac{1}{2} m_{-}}{\nu} - 6 \nu) \delta_1 {\lambda_2^{(0)}}_{|{\rm EsGB}}\bigr) v^2 \Biggr) \nn\\
& \hspace{1cm}+ \Biggl((1 -  m_{-} - 4 \nu) \delta_2 {\lambda_1^{(0)}}_{|{\rm EsGB}} + (1 + m_{-} - 4 \nu) \delta_1 {\lambda_2^{(0)}}_{|{\rm EsGB}}\Biggr) (n v) v^{i}\Biggr]\,.  
\end{align}
\end{subequations}

The full result for tidal effects in EsGB gravity at NNLO can be found in the supplementary material.  As expected, the leading order correction~\eqref{EOM_EsGB_LO} agrees with Ref.~\cite{vanGemeren:2023rhh} while the higher order results are new to this article.

An important point is that there are no specific corrections coming from the Gauss-Bonnet term as, mimicking the reasoning performed for tidal corrections in App.~\ref{App:Fokker}, such a correction would enter at the order ${\cal O}(\epsilon_{\rm tid}\alpha^2)$ which is at least a second order correction~\footnote{Note however that strictly speaking it is not really the case as in EsGB, the scalar charge is sourced by the Gauss-Bonnet term.}.

\section{Conclusions}

Tidal effects are one of the most promising tool to perform tests of gravity with the next generation of gravitational wave detectors. Indeed, when an additional scalar field is present in a theory, it will induce a varying dipole moment that will in turn induce a tidal deformation of the other body. Such an effect starts at 3PN order, a much lower effect than in GR, which makes it very important to include these new effects with a sufficient accuracy. Notably, by taking into account current constraints on the ST parameters, coming from the non-observation of dipolar emission, it was shown that such an effect could contribute up to ${\cal O}(1)$ cycle in the waveform for suitable binaries in the LISA band~\cite{Bernard:2019yfz}. In the case of EsGB, it is even more important to take those effects into account as one expect black holes to have a scalar hair and hence to have non-vanishing tidal Love numbers. 

In the present work we have tackled this program up to the NNLO in the dynamics both in generalized BD theory and in EsGB gravity. Such an accuracy allowed us to reach the order at which the usual gravitational tidal deformability enters. In a subsequent work, we will extend our result to the computation of the fluxes and waveform modes at the same NNLO in the tidal corrections~\cite{Dones:2023}. It will then permit to perform a more quantitative analysis of the impact of tidal effects based on the comparison between waveforms. 

An important other direction to be taken is to compute the value of the scalar and gravitational Love numbers for realistic models of compact objects in these theories. Such a program has been initiated in Ref.~\cite{Creci:2023cfx}. Notably, they found that the $l=1$ scalar Love number can be comparable to the gravitational one, showing the importance of introducing such terms in the waveforms.

\acknowledgments

L. B. and S. M. acknowledges financial support by the ANR PRoGRAM project, grant ANR-21-CE31-0003-001. L. B. and E.D. acknowledges financial support from the EU Horizon 2020 Research and Innovation Programme under the Marie Sklodowska-Curie Grant Agreement no. 101007855.
L. B. and E.D. are also grateful for the hospitality of Perimeter Institute where part of this work was carried out. Research at Perimeter Institute is supported in part by the Government of Canada through the Department of Innovation, Science and Economic Development Canada and by the Province of Ontario through the Ministry of Colleges and Universities. This research was also supported in part by the Simons Foundation through the Simons Foundation Emmy Noether Fellows Program at Perimeter Institute.

\appendix

\section{The Fokker approach}
\label{App:Fokker}

In this appendix, we present the Fokker reasoning in the case of a tidal perturbation w.r.t the point-particle contribution. Our goal is to show that it is sufficient to solve the point-particle field equations (\textit{i.e.} the Eqs. (\ref{rEFE}) in which we neglect the tidal contributions) in order to derive the NNLO tidal corrections to the Fokker Lagrangian. We start with the following action:
\begin{align}\label{Stotal}
& S [y_a,v_a,h,\psi] = S_{\mathrm{ST}}[y_a,v_a,h,\psi]+S_{\mathrm{pp}}[y_a,v_a,h,\psi] + S_{\mathrm{tidal}}[y_a,v_a,h,\psi]\,,
\end{align}
where $S_{\mathrm{tidal}}$ is the tidal action defined by Eq.~\eqref{Stidal}. By varying the action (\ref{Stotal}) w.r.t the metric and scalar perturbations, and solving iteratively the field equations 
\begin{subequations}\label{EFE_V2}
\begin{align}
\frac{ \delta S }{\delta h}[y_a,v_a,h,\psi]&=0 \,, \label{EFE_V2_1} \\
\frac{ \delta S }{\delta \psi}[y_a,v_a,h,\psi]&=0 \label{EFE_V2_2}\,,
\end{align}  
\end{subequations}
we obtain the solutions 
\begin{subequations}\label{EFE_V2_solutions}
\begin{align}
& h=h_{\mathrm{pp}}+h_{\mathrm{tidal}} \,, \label{metric_solution} \\
& \psi=\psi_{\mathrm{pp}}+\psi_{\mathrm{tidal}}\label{scalar_solution}\,.
\end{align}
\end{subequations}
Their tidal correction should be at least of order
\begin{align}
&\left(h^{00ii}_{\mathrm{tidal}}, h^{0i}_{\mathrm{tidal}}, h^{ij}_{\mathrm{tidal}};\,\psi_{\mathrm{tidal}}\right) = \mathcal{O}\Bigl( \frac{\epsilon_{\mathrm{tidal}}}{c^2},\frac{\epsilon_{\mathrm{tidal}}}{c^3},\frac{\epsilon_{\mathrm{tidal}}}{c^4};\frac{\epsilon_{\mathrm{tidal}}}{c^2}\Bigr)\,,
\end{align}
with $\mathcal{O}(\epsilon_{\mathrm{tidal}})=1/c^6 $, since the tidal effects in ST theories enter formally at 3PN order \cite{Bernard:2019yfz}. Because the Eqs.~\eqref{EFE_V2_solutions} are exact solutions to the field equations~\eqref{EFE_V2}, this implies that the functional derivative of the Fokker action w.r.t the scalar and metric perturbations, evaluated with the point-particle solutions, will have the orders
\begin{subequations}
\begin{align}
\frac{ \delta S }{\delta h}[y_a,v_a,h_{\rm pp},\psi_{\rm pp}]&= \mathcal{O}( c^2 \epsilon_{\mathrm{tidal}}, c \epsilon_{\mathrm{tidal}},\epsilon_{\mathrm{tidal}}; c^2 \epsilon_{\mathrm{tidal}}) \,, \\
\frac{ \delta S }{\delta \psi}[y_a,v_a,h_{\rm pp},\psi_{\rm pp}]&= \mathcal{O}( c^2 \epsilon_{\mathrm{tidal}}, c \epsilon_{\mathrm{tidal}},\epsilon_{\mathrm{tidal}}; c^2 \epsilon_{\mathrm{tidal}}) \,.
\end{align}  
\end{subequations}
By Taylor expanding the Fokker action around the point-particle solution, we get 
\begin{align}
S_{F}[y_a,v_a,h,\psi]&=S_{F}[y_a,v_a,h_{\mathrm{pp}},\psi_{\mathrm{pp}}] + \int \ud^{4}x \left(\frac{ \delta S }{\delta h}[y_a,v_a,h_{\mathrm{pp}},\psi_{\mathrm{pp}}] h_{\mathrm{tidal}} + \frac{ \delta S }{\delta \psi}[y_a,v_a,h_{\mathrm{pp}},\psi_{\mathrm{pp}}] \psi_{\mathrm{tidal}} \right) + \mathcal{O}(h^2_{\mathrm{tidal}},\psi^2_{\mathrm{tidal}}) \, \nn \\ 
& = S_{F}[y_a,v_a,h_{\mathrm{pp}},\psi_{\mathrm{pp}}] + \mathcal{O}(\epsilon_{\mathrm{tidal}}^2)
\end{align}
The contributions of the order of $\mathcal{O}(\epsilon^2_{\mathrm{tidal}})=1/c^{12} $ would be at least equivalent to a NNNLO (or 6PN) tidal effect. This shows that it is sufficient to inject in the action the point-particle solutions to the field equations to know the Fokker action up to the NNLO in the tidal effects.

\section{Feynman rules for the PN-EFT calculations}
\label{App:Feyn}

In this appendix, we give the Feynman rules that are relevant for the computation of tidal effects at NNLO. The starting point for the PNEFT machinery is the action~\eqref{Left} accompanied with the Kaluza-Klein decomposition of the metric~\eqref{KKdecomp}. Since we know that at this sector we don't encounter divergences, we can readily set $d=3$ for simplicity. Below are presented the Feynman rules with respect to the Kaluza-Klein modes and the canonically normalized scalar field $\psi$ relevant for the computation at this order.
\newline
The propagators for the potential modes are
$$
\vcenter{\hbox{

\begin{tikzpicture}[scale=0.6][baseline=(current bounding box.center)]

\begin{feynman}
\vertex  (a) at (0,0);
\vertex (b) at (-1,1);
\vertex (c) at (1,1);
\vertex (d) at (-1,-1);
\vertex  (e) at (1,-1);
\vertex (g) at (0,-1);
\vertex (f) at (0,1);
\diagram{
(d)--[horizontal](e);
};
\end{feynman}
\end{tikzpicture}
}}
  =\frac{-i}{8}\frac{1}{k^2-\partial_{t_a}\partial_{t_b}},\ \ \
\\
\vcenter{\hbox{

\begin{tikzpicture}[scale=0.6][baseline=(current bounding box.center)]

\begin{feynman}
\vertex  (a) at (0,0);
\vertex (b) at (-1,1);
\vertex (c) at (1,1);
\vertex (d) at (-1,0){\tiny${i}$};
\vertex  (e) at (1,0){\tiny${j}$};
\vertex (g) at (0,-1);
\vertex (f) at (0,1);
\diagram{
(d)--[horizontal,dashed](e);
};
\end{feynman}
\end{tikzpicture}
}}
 =\frac{i}{2}\frac{\delta_{ij}}{k^2-\partial_{t_a}\partial_{t_b}},\ \ \
\\
\vcenter{\hbox{

\begin{tikzpicture}[scale=0.6][baseline=(current bounding box.center)]

\begin{feynman}
\vertex  (a) at (0,0);
\vertex (b) at (-1,0);
\vertex (c) at (1,1);
\vertex (d) at (-1,0){\tiny${ij}$};
\vertex  (e) at (1,0){\tiny${kl}$};
\vertex (g) at (0,-1);
\vertex (f) at (0,1);
\diagram{
(d)--[horizontal,double](e);
};
\end{feynman}
\end{tikzpicture}
}} =\frac{i}{2}\frac{P_{ijkl}}{k^2-\partial_{t_a}\partial_{t_b}},\ \ \
\\
\vcenter{\hbox{

\begin{tikzpicture}[scale=0.6][baseline=(current bounding box.center)]

\begin{feynman}
\vertex  (a) at (0,0);
\vertex (b) at (-1,1);
\vertex (c) at (1,1);
\vertex (d) at (-1,-1);
\vertex  (e) at (1,-1);
\vertex (g) at (0,-1);
\vertex (f) at (0,1);
\diagram{
(d)--[horizontal,red](e);
};
\end{feynman}
\end{tikzpicture}
}}=-i\frac{1}{k^2-\partial_{t_a}\partial_{t_b}}$$
Then, the worldline vertices are
$$
  \vcenter{\hbox{\begin{tikzpicture}[scale=0.6][baseline=(current bounding box.center)]
\begin{feynman}
\vertex  (a) at (0,0);
\vertex (b) at (-1,1);
\vertex (c) at (1,1);
\vertex (d) at (-1,-1);
\vertex  (e) at (1,-1);
\vertex (g) at (0,-1);
\vertex (f) at (0,1);
\vertex (r) at (0,0);
\diagram{
(d)--[horizontal,very thick](e);
(g)--[horizontal](r);
};
\end{feynman}
\end{tikzpicture}}}
  =-2\frac{im_a}{\tilde{M}_{pl}}(1+\frac{3}{2}v_a^2),\ \ \ 
  \\
   \vcenter{\hbox{\begin{tikzpicture}[scale=0.6][baseline=(current bounding box.center)]
\begin{feynman}
\vertex  (a) at (0,0);
\vertex (b) at (-1,1);
\vertex (c) at (1,1);
\vertex (d) at (-1,-1);
\vertex  (e) at (1,-1);
\vertex (g) at (0,-1);
\vertex (f) at (0,1);
\vertex (r) at (0,0);
\diagram{
(d)--[horizontal,very thick](e);
(g)--[horizontal,dashed](r);
};
\end{feynman}
\end{tikzpicture}}}
  =2\frac{im_a}{\tilde{M}_{pl}}v_a^i(1+\frac{1}{2}v_a^2),\ \ \ 
  \\
 \vcenter{\hbox{\begin{tikzpicture}[scale=0.6][baseline=(current bounding box.center)]
\begin{feynman}
\vertex  (a) at (0,0);
\vertex (b) at (-1,1);
\vertex (c) at (1,1);
\vertex (d) at (-1,-1);
\vertex  (e) at (1,-1);
\vertex (g) at (0,-1);
\vertex (f) at (0,1);
\vertex (r) at (0,0);
\diagram{
(d)--[horizontal,very thick](e);
(g)--[horizontal,double](r);
};
\end{feynman}
\end{tikzpicture}}}
  =\frac{im_a}{\tilde{M}_{pl}}v_a^iv_a^j(1+\frac{1}{2}v_a^2),\ \ \ 
  \\
   \vcenter{\hbox{\begin{tikzpicture}[scale=0.6][baseline=(current bounding box.center)]
\begin{feynman}
\vertex  (a) at (0,0);
\vertex (b) at (-1,1);
\vertex (c) at (1,1);
\vertex (d) at (-1,-1);
\vertex  (e) at (1,-1);
\vertex (g) at (0,-1);
\vertex (f) at (0,1);
\vertex (r) at (0,0);
\diagram{
(d)--[horizontal,very thick](e);
(g)--[horizontal,red](r);
};
\end{feynman}
\end{tikzpicture}}}
  =-\tilde{d}_1^{(a)}\frac{im_a}{\tilde{M}_{pl}}(1-\frac{1}{2}v_a^2-\frac{1}{8}v_a^4),$$

$$
  \vcenter{\hbox{\begin{tikzpicture}[scale=0.6][baseline=(current bounding box.center)]
\begin{feynman}
\vertex  (a) at (0,0);
\vertex (b) at (-1,1);
\vertex (c) at (1,1);
\vertex (d) at (-1,-1);
\vertex  (e) at (1,-1);
\vertex (g) at (0,-1);
\vertex (f) at (0,1);
\vertex (r) at (-0.5,0);
\vertex (s) at (0.5,0);
\diagram{
(d)--[horizontal,very thick](e);
(g)--[horizontal](r);
(g)--[horizontal](s);
};
\end{feynman}
\end{tikzpicture}}}
  =-2\frac{im_a}{\tilde{M}_{pl}^2}\Big(1-\frac{c_a^{(0)}}{m_a}\frac{(k_1\cdot k_2)^2}{2}\Big),\ \ \ 
  \\
   \vcenter{\hbox{\begin{tikzpicture}[scale=0.6][baseline=(current bounding box.center)]
\begin{feynman}
\vertex  (a) at (0,0);
\vertex (b) at (-1,1);
\vertex (c) at (1,1);
\vertex (d) at (-1,-1);
\vertex  (e) at (1,-1);
\vertex (g) at (0,-1);
\vertex (f) at (0,1);
\vertex (r) at (-0.5,0);
\vertex (s) at (0.5,0);
\diagram{
(d)--[horizontal,very thick](e);
(g)--[horizontal,red](r);
(g)--[horizontal](s);
};
\end{feynman}
\end{tikzpicture}}}
  =-2\tilde{d}_1^{(a)}\frac{im_a}{\tilde{M}_{pl}^2}\Big(1+\frac{3}{2}v_a^2+\frac{(c_a^{(0)}-2\nu_a^{(0)})}{m_a d_1^{(a)}}\frac{(k_1\cdot k_2)^2}{2}\Big),
  $$
  $$
     \vcenter{\hbox{\begin{tikzpicture}[scale=0.6][baseline=(current bounding box.center)]
\begin{feynman}
\vertex  (a) at (0,0);
\vertex (b) at (-1,1);
\vertex (c) at (1,1);
\vertex (d) at (-1,-1);
\vertex  (e) at (1,-1);
\vertex (g) at (0,-1);
\vertex (f) at (0,1);
\vertex (r) at (-0.5,0);
\vertex (s) at (0.5,0);
\diagram{
(d)--[horizontal,very thick](e);
(g)--[horizontal,red](r);
(g)--[horizontal,dashed](s);
};
\end{feynman}
\end{tikzpicture}}}
  =2\tilde{d}_1^{(a)}\frac{im_a}{\tilde{M}_{pl}^2}v_a^i(1+\frac{1}{2}v_a^2),$$

  $$
 \vcenter{\hbox{\begin{tikzpicture}[scale=0.6][baseline=(current bounding box.center)]
\begin{feynman}
\vertex  (a) at (0,0);
\vertex (b) at (-1,1);
\vertex (c) at (1,1);
\vertex (d) at (-1,-1);
\vertex  (e) at (1,-1);
\vertex (g) at (0,-1);
\vertex (f) at (0,1);
\vertex (r) at (-0.5,0);
\vertex (s) at (0.5,0);
\diagram{
(d)--[horizontal,very thick](e);
(g)--[horizontal,red](r);
(g)--[horizontal,red](s);
};
\end{feynman}
\end{tikzpicture}}}
  =-\frac{1}{2}\tilde{d}_2^{(a)}\frac{im_a}{\tilde{M}_{pl}^2}(1-\frac{1}{2}v_a^2)+\frac{1}{2}\frac{i\tilde{f}_0^{(a)}\lambda_a^{(0)}}{\tilde{M}_{pl}^2}\Big(k_1\cdot k_2(1-\frac{v_a^2}{2}-\frac{v_a^4}{8})+k_1\cdot v_a\ k_2\cdot v_a(1+\frac{v_a^2}{2})-\partial_{t_1}\partial_{t_2}v_a^2-v_a^i\big(k_1(i\partial_{t_2})+k_2(i\partial_{t_1})\big)_i(1+\frac{v_a^2}{2})\Big)\ \ \
$$
$$
-\frac{i\tilde{f}_0^{(a)}}{\tilde{M}_{pl}^2}(4\mu_a^{(0)}+4\nu_a^{(0)}-c_a^{(0)})\frac{(k_1\cdot k_2)^2}{16},
$$

$$
   \vcenter{\hbox{\begin{tikzpicture}[scale=0.6][baseline=(current bounding box.center)]
\begin{feynman}
\vertex  (a) at (0,0){\tiny${k_2}$};
\vertex (b) at (-1,1);
\vertex (c) at (1,1);
\vertex (d) at (-1,-1);
\vertex  (e) at (1,-1);
\vertex (g) at (0,-1);
\vertex (f) at (0,1);
\vertex (r) at (-0.5,0){\tiny${k_1}$};
\vertex (s) at (0.5,0){\tiny${k_3}$};
\diagram{
(d)--[horizontal,very thick](e);
(g)--[horizontal,red](r);
(g)--[horizontal,red](s);
(g)--[horizontal,red](a);
};
\end{feynman}
\end{tikzpicture}}}
  =-\frac{\tilde{d}_3^{(a)}}{3!}\frac{im_a}{\tilde{M}_{pl}^3}+\frac{1}{3!}\frac{i\tilde{f}_1^{(a)}\lambda_a^{(0)}}{\tilde{M}_{pl}^3}\Big((k_{1}k_{2}+k_1k_3+k_2k_3)_{ij}(\delta^{ij}+v_a^iv_a^j)-iv_a^i\big(k_1 (\partial_{t_2}+\partial_{t_3})+k_2 (\partial_{t_1}+\partial_{t_3})+k_3 (\partial_{t_1}+\partial_{t_2})\big)_i\Big),
  $$

  $$
  \vcenter{\hbox{\begin{tikzpicture}[scale=0.6][baseline=(current bounding box.center)]
\begin{feynman}
\vertex  (a) at (0,0){\tiny${k_2}$};
\vertex (b) at (-1,1);
\vertex (c) at (1,1);
\vertex (d) at (-1,-1);
\vertex  (e) at (1,-1);
\vertex (g) at (0,-1);
\vertex (f) at (0,1);
\vertex (r) at (-0.5,0){\tiny${k_1}$};
\vertex (s) at (0.5,0){\tiny${k_3}$};
\diagram{
(d)--[horizontal,very thick](e);
(g)--[horizontal,red](r);
(g)--[horizontal](s);
(g)--[horizontal,red](a);
};
\end{feynman}
\end{tikzpicture}}}
  =-\tilde{d}_2^{(a)}\frac{im_a}{\tilde{M}_{pl}^3}+\frac{i\tilde{f}_0^{(a)}\lambda_a^{(0)}}{\tilde{M}_{pl}^3}\Big(k_1\cdot k_2(1+v_a^2/2)-k_1\cdot v_1\ k_2\cdot v_2+2v_a^i\big(k_1(i\partial_{t_2})+k_2(i\partial_{t_1})\big)_i\Big),
  $$

  $$
     \vcenter{\hbox{\begin{tikzpicture}[scale=0.6][baseline=(current bounding box.center)]
\begin{feynman}
\vertex  (a) at (0,0){\tiny${k_2}$};
\vertex (b) at (-1,1);
\vertex (c) at (1,1);
\vertex (d) at (-1,-1);
\vertex  (e) at (1,-1);
\vertex (g) at (0,-1);
\vertex (f) at (0,1);
\vertex (r) at (-0.5,0){\tiny${k_1}$};
\vertex (s) at (0.5,0){\tiny${k_3}$};
\diagram{
(d)--[horizontal,very thick](e);
(g)--[horizontal,red](r);
(g)--[horizontal,dashed](s);
(g)--[horizontal,red](a);
};
\end{feynman}
\end{tikzpicture}}}
  =-\frac{i\tilde{f}_0^{(a)}\lambda_a^{(0)}}{\tilde{M}_{pl}^3}\Big(k_1^i(i\partial_{t_2})+k_2^i(i\partial_{t_1})+v_a^ik_1\cdot k_2\Big),\ \ \
  \\
     \vcenter{\hbox{\begin{tikzpicture}[scale=0.6][baseline=(current bounding box.center)]
\begin{feynman}
\vertex  (a) at (0,0){\tiny${k_2}$};
\vertex (b) at (-1,1);
\vertex (c) at (1,1);
\vertex (d) at (-1,-1);
\vertex  (e) at (1,-1);
\vertex (g) at (0,-1);
\vertex (f) at (0,1);
\vertex (r) at (-0.5,0){\tiny${k_1}$};
\vertex (s) at (0.5,0){\tiny${k_3}$};
\diagram{
(d)--[horizontal,very thick](e);
(g)--[horizontal,red](r);
(g)--[horizontal,double](s);
(g)--[horizontal,red](a);
};
\end{feynman}
\end{tikzpicture}}}
  =-\frac{i\tilde{f}_0^{(a)}\lambda_a^{(0)}}{\tilde{M}_{pl}^3}k_1^ik_2^j,\ \ \
\\
     \vcenter{\hbox{\begin{tikzpicture}[scale=0.6][baseline=(current bounding box.center)]
\begin{feynman}
\vertex  (a) at (0,0);
\vertex (b) at (-1,1);
\vertex (c) at (1,1);
\vertex (d) at (-1,-1);
\vertex  (e) at (1,-1);
\vertex (g) at (0,-1);
\vertex (f) at (0,1);
\vertex (r) at (-0.5,0);
\vertex (s) at (0.5,0);
\vertex (u) at (0.2,0);
\vertex (w) at (0.6,0);
\diagram{
(d)--[horizontal,very thick](e);
(g)--[horizontal,red](r);
(g)--[horizontal](s);
(g)--[horizontal](a);
};
\end{feynman}
\end{tikzpicture}}}
  =-2\tilde{d}_1^{(a)}\frac{im_a}{\tilde{M}_{pl}^3}
  $$

  $$
    \vcenter{\hbox{\begin{tikzpicture}[scale=0.6][baseline=(current bounding box.center)]
\begin{feynman}
\vertex  (a) at (0,0);
\vertex (b) at (-1,1);
\vertex (c) at (1,1);
\vertex (d) at (-1,-1);
\vertex  (e) at (1,-1);
\vertex (g) at (0,-1);
\vertex (f) at (0,1);
\vertex (r) at (-0.6,0){\tiny${k_1}$};
\vertex (s) at (-0.2,0){\tiny${k_2}$};
\vertex (u) at (0.2,0){\tiny${k_3}$};
\vertex (w) at (0.6,0){\tiny${k_4}$};
\diagram{
(d)--[horizontal,very thick](e);
(g)--[horizontal,red](r);
(g)--[horizontal,red](s);
(g)--[horizontal](u);
(g)--[horizontal](w);
};
\end{feynman}
\end{tikzpicture}}}
  =9\frac{i\tilde{f}_0^{(a)}\lambda_a^{(0)}}{\tilde{M}_{pl}^4}(k_1\cdot k_2),\ \ \ 
  \\
   \vcenter{\hbox{\begin{tikzpicture}[scale=0.6][baseline=(current bounding box.center)]
\begin{feynman}
\vertex  (a) at (0,0);
\vertex (b) at (-1,1);
\vertex (c) at (1,1);
\vertex (d) at (-1,-1);
\vertex  (e) at (1,-1);
\vertex (g) at (0,-1);
\vertex (f) at (0,1);
\vertex (r) at (-0.6,0){\tiny${k_1}$};
\vertex (s) at (-0.2,0){\tiny${k_2}$};
\vertex (u) at (0.2,0){\tiny${k_3}$};
\vertex (w) at (0.6,0){\tiny${k_4}$};
\diagram{
(d)--[horizontal,very thick](e);
(g)--[horizontal,red](r);
(g)--[horizontal,red](s);
(g)--[horizontal,red](u);
(g)--[horizontal](w);
};
\end{feynman}
\end{tikzpicture}}}
  =\frac{i\tilde{f}_1^{(a)}\lambda_a^{(0)}}{\tilde{M}_{pl}^4}\frac{(k_1+k_2+k_3)^2-k_1^2-k_2^2-k_3^2}{2},
$$

$$
   \vcenter{\hbox{\begin{tikzpicture}[scale=0.6][baseline=(current bounding box.center)]
\begin{feynman}
\vertex  (a) at (0,0);
\vertex (b) at (-1,1);
\vertex (c) at (1,1);
\vertex (d) at (-1,-1);
\vertex  (e) at (1,-1);
\vertex (g) at (0,-1);
\vertex (f) at (0,1);
\vertex (r) at (-0.6,0){\tiny${k_1}$};
\vertex (s) at (-0.2,0){\tiny${k_2}$};
\vertex (u) at (0.2,0){\tiny${k_3}$};
\vertex (w) at (0.6,0){\tiny${k_4}$};
\diagram{
(d)--[horizontal,very thick](e);
(g)--[horizontal,red](r);
(g)--[horizontal,red](s);
(g)--[horizontal,red](u);
(g)--[horizontal,red](w);
};
\end{feynman}
\end{tikzpicture}}}
  =\frac{i\tilde{f}_2^{(a)}\lambda_a^{(0)}}{\tilde{M}_{pl}^4}\frac{(k_1+k_2+k_3+k_4)^2-k_1^2-k_2^2-k_3^2-k_4^2}{4!2},
  $$
where all the momenta are outgoing. Finally, the bulk interaction vertices are

  $$
 \vcenter{\hbox{\begin{tikzpicture}[scale=0.6][baseline=(current bounding box.center)]
\begin{feynman}
\vertex  (a) at (0,0);
\vertex (b) at (-1,1){\tiny$k_{1}$};
\vertex (c) at (1,1);
\vertex (d) at (-1,-1){\tiny$k_{2}$};
\vertex  (e) at (1,-1);
\vertex (g) at (0,-1);
\vertex (f) at (0,1);
\vertex (r) at (-0.5,0);
\vertex (s) at (0.5,0);
\vertex (u) at (1,0){\tiny$k_{3}$};
\diagram{
(b)--[horizontal,red](a);
(d)--[horizontal,red](a);
(u)--[horizontal,red](a);
};
\end{feynman}
\end{tikzpicture}}}
  =-\frac{i}{\tilde{M}_{pl}}\frac{x_1}{2}\Big((k_1^2+k_2^2+k_3^2)-2(\partial_{t_1}\partial_{t_2}+\partial_{t_1}\partial_{t_3}+\partial_{t_2}\partial_{t_3})\Big),
$$

$$
   \vcenter{\hbox{\begin{tikzpicture}[scale=0.6][baseline=(current bounding box.center)]
\begin{feynman}
\vertex  (a) at (0,0);
\vertex (b) at (-1,1){\tiny$k_{1}$};
\vertex (c) at (1,1);
\vertex (d) at (-1,-1){\tiny$k_{2}$};
\vertex  (e) at (1,-1);
\vertex (g) at (0,-1);
\vertex (f) at (0,1);
\vertex (r) at (-0.5,0);
\vertex (s) at (0.5,0);
\vertex (u) at (1,0){\tiny$k_{3}$};
\diagram{
(b)--[horizontal,red](a);
(d)--[horizontal,red](a);
(u)--[horizontal](a);
};
\end{feynman}
\end{tikzpicture}}}
  =-8\frac{i}{\tilde{M}_{pl}}\partial_{t_1}\partial_{t_2},\ \ \ 
  \\
   \vcenter{\hbox{\begin{tikzpicture}[scale=0.6][baseline=(current bounding box.center)]
\begin{feynman}
\vertex  (a) at (0,0);
\vertex (b) at (-1,1){\tiny$k_{1}$};
\vertex (c) at (1,1);
\vertex (d) at (-1,-1){\tiny$k_{2}$};
\vertex  (e) at (1,-1);
\vertex (g) at (0,-1);
\vertex (f) at (0,1);
\vertex (r) at (-0.5,0);
\vertex (s) at (0.5,0);
\vertex (u) at (1,0){\tiny$k_{3i}$};
\diagram{
(b)--[horizontal,red](a);
(d)--[horizontal,red](a);
(u)--[horizontal,dashed](a);
};
\end{feynman}
\end{tikzpicture}}}
  =-2\frac{i}{\tilde{M}_{pl}}\Big(k_1(i\partial_{t_2})+k_2(i\partial_{t_1})\Big)_i,\ \ \ 
  \\
  $$

    $$
 \vcenter{\hbox{\begin{tikzpicture}[scale=0.6][baseline=(current bounding box.center)]
\begin{feynman}
\vertex  (a) at (0,0);
\vertex (b) at (-1,1){\tiny$k_{1}$};
\vertex (c) at (1,1);
\vertex (d) at (-1,-1){\tiny$k_{2}$};
\vertex  (e) at (1,-1);
\vertex (g) at (0,-1);
\vertex (f) at (0,1);
\vertex (r) at (-0.5,0);
\vertex (s) at (0.5,0);
\vertex (u) at (1,0){\tiny$k_{3ij}$};
\diagram{
(b)--[horizontal,red](a);
(d)--[horizontal,red](a);
(u)--[horizontal,double](a);
};
\end{feynman}
\end{tikzpicture}}}
  =\frac{1}{2c_d}\times\ \ \ 
  \\
   \vcenter{\hbox{\begin{tikzpicture}[scale=0.6][baseline=(current bounding box.center)]
\begin{feynman}
\vertex  (a) at (0,0);
\vertex (b) at (-1,1){\tiny$k_{1}$};
\vertex (c) at (1,1);
\vertex (d) at (-1,-1){\tiny$k_{2}$};
\vertex  (e) at (1,-1);
\vertex (g) at (0,-1);
\vertex (f) at (0,1);
\vertex (r) at (-0.5,0);
\vertex (s) at (0.5,0);
\vertex (u) at (1,0){\tiny$k_{3ij}$};
\diagram{
(b)--[horizontal](a);
(d)--[horizontal](a);
(u)--[horizontal,double](a);
};
\end{feynman}
\end{tikzpicture}}}
  =-\frac{i}{\tilde{M}_{pl}}\Big(k_1^kk_2^lQ_{ijkl}+\partial_{t_1}\partial_{t_2}\delta_{ij}\Big),\ \ \ 
  \\
   \vcenter{\hbox{\begin{tikzpicture}[scale=0.6][baseline=(current bounding box.center)]
\begin{feynman}
\vertex  (a) at (0,0);
\vertex (b) at (-1,1){\tiny$k_{1}$};
\vertex (c) at (1,1){\tiny$k_{2}$};
\vertex (d) at (-1,-1){\tiny$k_{3}$};
\vertex  (e) at (1,-1){\tiny$k_{4}$};
\vertex (g) at (0,-1);
\vertex (f) at (0,1);
\vertex (r) at (-0.5,0);
\vertex (s) at (0.5,0);
\vertex (u) at (1,0);
\diagram{
(b)--[horizontal,red](a);
(d)--[horizontal,red](a);
(c)--[horizontal,red](a);
(e)--[horizontal,red](a);
};
\end{feynman}
\end{tikzpicture}}}
  =-\frac{i}{\tilde{M}_{pl}^2}\frac{x_2}{2}(k_1^2+k_2^2+k_3^2+k_4^2)
  $$
where $P_{ijkl}=-(\delta_{ik}\delta_{jl}+\delta_{il}\delta_{jk}+(2-c_d)\delta_{ij}\delta_{kl})$, $Q_{ijkl}=I_{ijkl}-\delta_{ij}\delta_{kl}$ and $I_{ijkl}=\delta_{il}\delta_{kj}+\delta_{ik}\delta_{jl}$.

\section{NNLO results in the CM frame}
\label{App:CMeom}
 
\subsection{Acceleration in the CM frame}
\label{App:CMeomNNLO}

The NNLO tidal correction to the relative acceleration in the center of mass frame is given, after separating it in power of $\tilde{G}$, by 
\begin{subequations}
\begin{align}
&a_{\rm CM, NNLO}^{i,(2)}=\alpha^2 \tilde{G}^2 \frac{\zeta}{1 -  \zeta} m \Biggl[- \frac{9 (2 + \bar{\gamma})^2 }{4 (1 - \zeta)^2 r^7}(c_+^{(0)}-4{\nu}_+^{(0)}) n^{i} + \frac{ 36 }{c^2 r^7} \mu_+^{(0)} n^{i}\nn\\
& \hspace{1cm}+ \frac{1}{c^6 r^5}\Biggl(v^{i} \Bigl((n v)v^2 \bigl(4 \delta (18 + \nu) \
\lambda_-^{(0)} - 4 (-2 + \nu) (9 + 4 \nu) \lambda_+^{(0)}\bigr) + (n v)^3 \bigl(-6 \delta (23 + 2 \nu) \lambda_-^{(0)} + 6 (-23 - 2 \nu \
+ 8 \nu^2) \lambda_+^{(0)}\bigr) \nn\\
& \hspace{1cm}+ n^{i} \Bigl((n v)^2 v^2 \bigl(-3 \delta (87 + 10 \nu) \lambda_-^{(0)} + 3 (3 + 2 \nu) (-29 + 16 \nu) \lambda_+^{(0)}\bigr) + v^4 \bigl(\tfrac{1}{2} \delta (45 + 4 \nu) \lambda_-^{(0)} + \tfrac{1}{2} (45 + 20 \nu - 64 \nu^2) \lambda_+^{(0)}\bigr) \nn\\
& \hspace{1cm}+ (n v)^4 \bigl(48 \delta (7 + \nu) \lambda_-^{(0)} - 48 (-7 -  \nu + 3 \nu^2) \lambda_+^{(0)}\bigr)\Bigr)  \Biggr]   \,,\\
&a_{\rm CM, NNLO}^{i,(3)}=\frac{\alpha^3 \tilde{G}^3}{c^6 r^6} \frac{\zeta}{1 -  \zeta} m^2 \Biggl[ (n v) v^{i} \biggl(\biggl\{\frac{40 \bar{\beta}^{-}}{\bar{\gamma}} -  \frac{2 (20 \bar{\beta}^+ + 19 \bar{\gamma}) \delta}{\bar{\gamma}} + \bigl(- \frac{240 \bar{\beta}^{-}}{\bar{\gamma}} + \frac{(80 \bar{\beta}^+ + 6 \bar{\gamma} + 15 \bar{\gamma}^2) \delta}{\bar{\gamma}}\bigr) \nu \biggr\}\lambda_-^{(0)}  \nn\\
& \hspace{1cm}  + \biggl\{- \frac{2 (20 \bar{\beta}^+ + 19 \bar{\gamma})}{\bar{\gamma}} + \frac{40 \bar{\beta}^{-} \delta}{\bar{\gamma}} + (\frac{240 \bar{\beta}^+ + 314 \bar{\gamma} + 229 \bar{\gamma}^2}{\bar{\gamma}} -  \frac{80 \bar{\beta}^{-} \delta}{\bar{\gamma}}) \nu + 52 \nu^2\biggr\} \lambda_+^{(0)}   \nn\\
& \hspace{1cm} \frac{5 (2 + \bar{\gamma}) \nu \bigl(4 \lambda_1 (- \delta \Lambda_-^{(0)} + \Lambda_+^{(0)}) + \zeta (5 \delta \Lambda_-^{(0)} + 2  \delta \phi_{0}{} \Lambda_-^{(1)} - 5 \Lambda_+^{(0)} - 2 \phi_{0}{} \Lambda_+^{(1)})\bigr)}{-1 + \zeta} \biggr)\nn\\
& \hspace{1cm}+ n^{i} \Biggl(v^2 \biggl(\biggl\{\frac{20 \bar{\beta}^{-}}{\bar{\gamma}} + \frac{(-40 \bar{\beta}^+ - 36 \bar{\gamma} + 15 \bar{\gamma}^2) \delta}{2 \bar{\gamma}} + \bigl(- \frac{200 \bar{\beta}^{-}}{\bar{\gamma}} + \frac{(240 \bar{\beta}^+ + 46 \bar{\gamma} + 45 \bar{\gamma}^2) \delta}{2 \bar{\gamma}}\bigr) \nu \biggr\} \lambda_-^{(0)} \nn\\
& \hspace{1cm}+ \biggl\{\frac{5 (-8 \bar{\beta}^+ - 4 \bar{\gamma} + 5 \bar{\gamma}^2)}{2 \bar{\gamma}} + \frac{20 \bar{\beta}^{-} \delta}{\bar{\gamma}} + (\frac{400 \bar{\beta}^+ - 54 \bar{\gamma} + 31 \bar{\gamma}^2}{2 \bar{\gamma}} -  \frac{120 \bar{\beta}^{-} \delta}{\bar{\gamma}}) \nu + 14 \nu^2\biggr\} \lambda_+^{(0)} \nn\\
& \hspace{1cm}- \frac{3 (2 + \bar{\gamma}) (1 - 5 \nu) \bigl(4 \lambda_1 (- \delta \Lambda_-^{(0)} + \Lambda_+^{(0)}) + \zeta (5 \delta \Lambda_-^{(0)} + 2 \phi_{0}{} \delta \Lambda_-^{(1)} - 5 \Lambda_+^{(0)} - 2 \phi_{0}{} \Lambda_+^{(1)})\bigr)}{2 (-1 + \zeta)}  \biggr)  \nn\\
& \hspace{1cm}+ (n v)^2 \biggl(\biggl\{- \tfrac{5}{2} (-32 + 17 \bar{\gamma}) \delta + \bigl(\frac{700 \bar{\beta}^{-}}{\bar{\gamma}} -  \frac{(560 \bar{\beta}^+ + 38 \bar{\gamma} + 105 \bar{\gamma}^2) \delta}{4 \bar{\gamma}}\bigr) \nu \biggr\} \lambda_-^{(0)}  \nn\\
& \hspace{1cm}+ \biggl\{- \tfrac{7}{2} (-32 + 5 \bar{\gamma}) + (- \frac{2800 \bar{\beta}^+ + 2906 \bar{\gamma} + 2295 \bar{\gamma}^2}{4 \bar{\gamma}} + \frac{140 \bar{\beta}^{-} \delta}{\bar{\gamma}}) \nu - 83 \nu^2\biggr\} \lambda_+^{(0)}   \nn\\
& \hspace{1cm} - \frac{7 (2 + \bar{\gamma}) (2 + 5 \nu) \bigl(4 \lambda_1 (- \delta \Lambda_-^{(0)} + \Lambda_+^{(0)}) + \zeta (5 \delta \Lambda_-^{(0)} + 2 \delta \phi_{0}{} \Lambda_-^{(1)} - 5 \Lambda_+^{(0)} - 2 \phi_{0}{} \Lambda_+^{(1)})\bigr)}{4 (-1 + \zeta)}\biggr)\Biggr) \Biggr]\,,\\
&a_{\rm CM, NNLO}^{i,(4)}= \frac{\alpha^4 \tilde{G}^4}{c^6 r^7} \frac{\zeta}{(1 -  \zeta)^2} m^3 n^{i}\Biggl[ (2 + \bar{\gamma}) \biggl(\biggl\{\zeta (34 + 19 \bar{\gamma}) \delta + (- \frac{144 \bar{\beta}^{-} \zeta }{\bar{\gamma}} + 16 \zeta \delta) \nu \biggr\} \phi_{0}{}\Lambda_-^{(1)} \nn\\
& \hspace{1cm}+ (5 \zeta - 4 \lambda_1) \biggl(\biggl\{\tfrac{1}{2} (34 + 19 \bar{\gamma}) \delta + (- \frac{72 \bar{\beta}^{-}}{\bar{\gamma}} + 8 \delta) \nu \biggr\} \Lambda_-^{(0)} + \biggl\{\tfrac{1}{2} (-34 - 19 \bar{\gamma}) + \frac{(72 \bar{\beta}^+ - 8 \bar{\gamma} + 9 \bar{\gamma}^2) \nu}{\bar{\gamma}}\biggr\} \Lambda_+^{(0)}\biggr) \nn\\
& \hspace{1cm}+ \biggl\{- \zeta (34 + 19 \bar{\gamma})  + \frac{2 \zeta (72 \bar{\beta}^+ - 8 \bar{\gamma} + 9 \bar{\gamma}^2)  \nu}{\bar{\gamma}}\biggr\} \phi_{0}{}\Lambda_+^{(1)}\biggr)\nn\\
& \hspace{1cm}+\frac{1}{\bar{\gamma}^2} \Biggl(\biggl\{\tfrac{1}{5} \Bigl(-960 \bar{\beta}^{-} \bar{\beta}^+ + 960 \bar{\beta}^{-} \bar{\beta}^+ \zeta + 1080 \bar{\beta}^{-} \bar{\gamma} - 1080 \bar{\beta}^{-} \zeta \bar{\gamma} + 370 \bar{\beta}^{-} \bar{\gamma}^2 - 370 \bar{\beta}^{-} \zeta \bar{\gamma}^2 - 240 \bar{\gamma} \bar{\chi}_{-}{} + 240 \zeta \bar{\gamma} \bar{\chi}_{-}{}\nn\\
& \hspace{1cm} - 96 \zeta \bar{\gamma} \mathcal{S}_{-}{} \mathcal{S}_{+}{} + 96 \zeta^2 \bar{\gamma} \mathcal{S}_{-}{} \mathcal{S}_{+}{} - 96 \zeta \bar{\gamma}^2 \mathcal{S}_{-}{} \mathcal{S}_{+}{} - 654 \zeta^2 \bar{\gamma}^2 \mathcal{S}_{-}{} \mathcal{S}_{+}{} - 24 \zeta \bar{\gamma}^3 \mathcal{S}_{-}{} \mathcal{S}_{+}{} - 351 \zeta^2 \bar{\gamma}^3 \mathcal{S}_{-}{} \mathcal{S}_{+}{} + 1740 \zeta \bar{\gamma}^2 \mathcal{S}_{-}{} \mathcal{S}_{+}{} \lambda_1 \nn\\
& \hspace{1cm}+ 870 \zeta \bar{\gamma}^3 \mathcal{S}_{-}{} \mathcal{S}_{+}{} \lambda_1 - 1200 \bar{\gamma}^2 \mathcal{S}_{-}{} \mathcal{S}_{+}{} (\lambda_1)^2 - 600 \bar{\gamma}^3 \mathcal{S}_{-}{} \mathcal{S}_{+}{} (\lambda_1)^2 + 240 \bar{\gamma}^2 \mathcal{S}_{-}{} \mathcal{S}_{+}{} \lambda_2 + 120 \bar{\gamma}^3 \mathcal{S}_{-}{} \mathcal{S}_{+}{} \lambda_2\Bigr) \nn\\
& \hspace{1cm}+ \tfrac{1}{20} \delta \Bigl(1920 \bar{\beta}^{-}{}^2 + 1920 \bar{\beta}^+{}^2 - 1920 \bar{\beta}^{-}{}^2 \zeta - 1920 \bar{\beta}^+{}^2 \zeta - 4320 \bar{\beta}^+ \bar{\gamma} + 4320 \bar{\beta}^+ \zeta \bar{\gamma} - 1056 \bar{\gamma}^2 \nn\\
& \hspace{1cm}- 1480 \bar{\beta}^+ \bar{\gamma}^2 + 1056 \zeta \bar{\gamma}^2 + 1480 \bar{\beta}^+ \zeta \bar{\gamma}^2 - 1596 \bar{\gamma}^3 + 1596 \zeta \bar{\gamma}^3 - 459 \bar{\gamma}^4 + 459 \zeta \bar{\gamma}^4 + 960 \bar{\gamma} \bar{\chi}_{+}{} - 960 \zeta \bar{\gamma} \bar{\chi}_{+}{} - 192 \zeta \bar{\gamma} \mathcal{S}_{-}{}^2 \nn\\
& \hspace{1cm}+ 192 \zeta^2 \bar{\gamma} \mathcal{S}_{-}{}^2 - 192 \zeta \bar{\gamma}^2 \mathcal{S}_{-}{}^2 - 1308 \zeta^2 \bar{\gamma}^2 \mathcal{S}_{-}{}^2 - 48 \zeta \bar{\gamma}^3 \mathcal{S}_{-}{}^2 - 702 \zeta^2 \bar{\gamma}^3 \mathcal{S}_{-}{}^2 - 192 \zeta \bar{\gamma} \mathcal{S}_{+}{}^2 + 192 \zeta^2 \bar{\gamma} \mathcal{S}_{+}{}^2 - 192 \zeta \bar{\gamma}^2 \mathcal{S}_{+}{}^2 \nn\\
& \hspace{1cm}- 1308 \zeta^2 \bar{\gamma}^2 \mathcal{S}_{+}{}^2 - 48 \zeta \bar{\gamma}^3 \mathcal{S}_{+}{}^2 - 702 \zeta^2 \bar{\gamma}^3 \mathcal{S}_{+}{}^2 + 3480 \zeta \bar{\gamma}^2 \mathcal{S}_{-}{}^2 \lambda_1 + 1740 \zeta \bar{\gamma}^3 \mathcal{S}_{-}{}^2 \lambda_1 + 3480 \zeta \bar{\gamma}^2 \mathcal{S}_{+}{}^2 \lambda_1 + 1740 \zeta \bar{\gamma}^3 \mathcal{S}_{+}{}^2 \lambda_1 \nn\\
& \hspace{1cm}- 2400 \bar{\gamma}^2 \mathcal{S}_{-}{}^2 (\lambda_1)^2 - 1200 \bar{\gamma}^3 \mathcal{S}_{-}{}^2 (\lambda_1)^2 - 2400 \bar{\gamma}^2 \mathcal{S}_{+}{}^2 (\lambda_1)^2 - 1200 \bar{\gamma}^3 \mathcal{S}_{+}{}^2 (\lambda_1)^2 \nn\\
& \hspace{1cm}+ 480 \bar{\gamma}^2 \mathcal{S}_{-}{}^2 \lambda_2 + 240 \bar{\gamma}^3 \mathcal{S}_{-}{}^2 \lambda_2 + 480 \bar{\gamma}^2 \mathcal{S}_{+}{}^2 \lambda_2 + 240 \bar{\gamma}^3 \mathcal{S}_{+}{}^2 \lambda_2\Bigr) \nn\\
& \hspace{1cm}+ \nu \Bigl(2 (-1 + \zeta) \bar{\gamma} (64 \bar{\beta}^+ + 11 \bar{\gamma} + 12 \bar{\gamma}^2) \delta -  \tfrac{2}{5} \bigl(-960 \bar{\beta}^{-} \bar{\beta}^+ + 960 \bar{\beta}^{-} \bar{\beta}^+ \zeta - 1120 \bar{\beta}^{-} \bar{\gamma} + 1120 \bar{\beta}^{-} \zeta \bar{\gamma} - 480 \bar{\beta}^{-} \bar{\gamma}^2 \nn\\
& \hspace{1cm}+ 480 \bar{\beta}^{-} \zeta \bar{\gamma}^2 - 240 \bar{\gamma} \bar{\chi}_{-}{} + 240 \zeta \bar{\gamma} \bar{\chi}_{-}{} - 96 \zeta \bar{\gamma} \mathcal{S}_{-}{} \mathcal{S}_{+}{} + 96 \zeta^2 \bar{\gamma} \mathcal{S}_{-}{} \mathcal{S}_{+}{} - 96 \zeta \bar{\gamma}^2 \mathcal{S}_{-}{} \mathcal{S}_{+}{} - 654 \zeta^2 \bar{\gamma}^2 \mathcal{S}_{-}{} \mathcal{S}_{+}{} - 24 \zeta \bar{\gamma}^3 \mathcal{S}_{-}{} \mathcal{S}_{+}{} \nn\\
& \hspace{1cm}- 351 \zeta^2 \bar{\gamma}^3 \mathcal{S}_{-}{} \mathcal{S}_{+}{} + 1740 \zeta \bar{\gamma}^2 \mathcal{S}_{-}{} \mathcal{S}_{+}{} \lambda_1 + 870 \zeta \bar{\gamma}^3 \mathcal{S}_{-}{} \mathcal{S}_{+}{} \lambda_1 - 1200 \bar{\gamma}^2 \mathcal{S}_{-}{} \mathcal{S}_{+}{} (\lambda_1)^2 - 600 \bar{\gamma}^3 \mathcal{S}_{-}{} \mathcal{S}_{+}{} (\lambda_1)^2 + 240 \bar{\gamma}^2 \mathcal{S}_{-}{} \mathcal{S}_{+}{} \lambda_2 \nn\\
& \hspace{1cm}+ 120 \bar{\gamma}^3 \mathcal{S}_{-}{} \mathcal{S}_{+}{} \lambda_2\bigr)\Bigr)\biggr\} \lambda_-^{(0)} \nn\\
& \hspace{1cm}+ \biggl\{-12 \zeta \bar{\gamma}^2 (2 + \bar{\gamma}) \mathcal{S}_{-}{} \mathcal{S}_{+}{} (6 \zeta - 5 \lambda_1) - 6 \zeta \bar{\gamma}^2 (2 + \bar{\gamma}) (\mathcal{S}_{-}{}^2 + \mathcal{S}_{+}{}^2) \delta (6 \zeta - 5 \lambda_1) + 24 \zeta \bar{\gamma}^2 (2 + \bar{\gamma}) \mathcal{S}_{-}{} \mathcal{S}_{+}{} \nu (6 \zeta - 5 \lambda_1)\biggr\} \phi_{0}{}\lambda_-^{(1)} \nn\\
& \hspace{1cm}+ \biggl\{-12 \zeta^2 \bar{\gamma}^2 (2 + \bar{\gamma})  \mathcal{S}_{-}{} \mathcal{S}_{+}{} - 6 \zeta^2 \bar{\gamma}^2 (2 + \bar{\gamma}) (\mathcal{S}_{-}{}^2 + \mathcal{S}_{+}{}^2) \delta + 24 \zeta^2 \bar{\gamma}^2 (2 + \bar{\gamma}) \mathcal{S}_{-}{} \mathcal{S}_{+}{} \nu \biggr\} \phi_{0}{}^2\lambda_-^{(2)} \nn\\
& \hspace{1cm}+ \biggl\{\tfrac{3}{5} \delta \Bigl(-320 \bar{\beta}^{-} \bar{\beta}^+ + 320 \bar{\beta}^{-} \bar{\beta}^+ \zeta + 360 \bar{\beta}^{-} \bar{\gamma} - 360 \bar{\beta}^{-} \zeta \bar{\gamma} + 130 \bar{\beta}^{-} \bar{\gamma}^2 - 130 \bar{\beta}^{-} \zeta \bar{\gamma}^2 - 80 \bar{\gamma} \bar{\chi}_{-}{} + 80 \zeta \bar{\gamma} \bar{\chi}_{-}{} - 32 \zeta \bar{\gamma} \mathcal{S}_{-}{} \mathcal{S}_{+}{} \nn\\
& \hspace{1cm}+ 32 \zeta^2 \bar{\gamma} \mathcal{S}_{-}{} \mathcal{S}_{+}{} + 48 \zeta \bar{\gamma}^2 \mathcal{S}_{-}{} \mathcal{S}_{+}{} + 202 \zeta^2 \bar{\gamma}^2 \mathcal{S}_{-}{} \mathcal{S}_{+}{} + 32 \zeta \bar{\gamma}^3 \mathcal{S}_{-}{} \mathcal{S}_{+}{} + 93 \zeta^2 \bar{\gamma}^3 \mathcal{S}_{-}{} \mathcal{S}_{+}{} - 580 \zeta \bar{\gamma}^2 \mathcal{S}_{-}{} \mathcal{S}_{+}{} \lambda_1 - 290 \zeta \bar{\gamma}^3 \mathcal{S}_{-}{} \mathcal{S}_{+}{} \lambda_1 \nn\\
& \hspace{1cm}+ 400 \bar{\gamma}^2 \mathcal{S}_{-}{} \mathcal{S}_{+}{} (\lambda_1)^2 + 200 \bar{\gamma}^3 \mathcal{S}_{-}{} \mathcal{S}_{+}{} (\lambda_1)^2 - 80 \bar{\gamma}^2 \mathcal{S}_{-}{} \mathcal{S}_{+}{} \lambda_2 - 40 \bar{\gamma}^3 \mathcal{S}_{-}{} \mathcal{S}_{+}{} \lambda_2\Bigr) \nn\\
& \hspace{1cm}+ \tfrac{1}{20} \Bigl(1920 \bar{\beta}^{-}{}^2 + 1920 \bar{\beta}^+{}^2 - 1920 \bar{\beta}^{-}{}^2 \zeta - 1920 \bar{\beta}^+{}^2 \zeta - 4320 \bar{\beta}^+ \bar{\gamma} + 4320 \bar{\beta}^+ \zeta \bar{\gamma} + 3264 \bar{\gamma}^2 - 1560 \bar{\beta}^+ \bar{\gamma}^2 - 3264 \zeta \bar{\gamma}^2 \nn\\
& \hspace{1cm}+ 1560 \bar{\beta}^+ \zeta \bar{\gamma}^2 + 3164 \bar{\gamma}^3 - 3164 \zeta \bar{\gamma}^3 + 851 \bar{\gamma}^4 - 851 \zeta \bar{\gamma}^4 + 960 \bar{\gamma} \bar{\chi}_{+}{} - 960 \zeta \bar{\gamma} \bar{\chi}_{+}{} - 192 \zeta \bar{\gamma} \mathcal{S}_{-}{}^2 + 192 \zeta^2 \bar{\gamma} \mathcal{S}_{-}{}^2 + 288 \zeta \bar{\gamma}^2 \mathcal{S}_{-}{}^2 \nn\\
& \hspace{1cm}+ 1212 \zeta^2 \bar{\gamma}^2 \mathcal{S}_{-}{}^2 + 192 \zeta \bar{\gamma}^3 \mathcal{S}_{-}{}^2 + 558 \zeta^2 \bar{\gamma}^3 \mathcal{S}_{-}{}^2 - 192 \zeta \bar{\gamma} \mathcal{S}_{+}{}^2 + 192 \zeta^2 \bar{\gamma} \mathcal{S}_{+}{}^2 + 288 \zeta \bar{\gamma}^2 \mathcal{S}_{+}{}^2 + 1212 \zeta^2 \bar{\gamma}^2 \mathcal{S}_{+}{}^2 + 192 \zeta \bar{\gamma}^3 \mathcal{S}_{+}{}^2 \nn\\
& \hspace{1cm}+ 558 \zeta^2 \bar{\gamma}^3 \mathcal{S}_{+}{}^2 - 3480 \zeta \bar{\gamma}^2 \mathcal{S}_{-}{}^2 \lambda_1 - 1740 \zeta \bar{\gamma}^3 \mathcal{S}_{-}{}^2 \lambda_1 - 3480 \zeta \bar{\gamma}^2 \mathcal{S}_{+}{}^2 \lambda_1 - 1740 \zeta \bar{\gamma}^3 \mathcal{S}_{+}{}^2 \lambda_1 + 2400 \bar{\gamma}^2 \mathcal{S}_{-}{}^2 (\lambda_1)^2 \nn\\
& \hspace{1cm}+ 1200 \bar{\gamma}^3 \mathcal{S}_{-}{}^2 (\lambda_1)^2 + 2400 \bar{\gamma}^2 \mathcal{S}_{+}{}^2 (\lambda_1)^2 + 1200 \bar{\gamma}^3 \mathcal{S}_{+}{}^2 (\lambda_1)^2 - 480 \bar{\gamma}^2 \mathcal{S}_{-}{}^2 \lambda_2 - 240 \bar{\gamma}^3 \mathcal{S}_{-}{}^2 \lambda_2 - 480 \bar{\gamma}^2 \mathcal{S}_{+}{}^2 \lambda_2 - 240 \bar{\gamma}^3 \mathcal{S}_{+}{}^2 \lambda_2\Bigr) \nn\\
& \hspace{1cm}+ \nu \Bigl(4 \bar{\beta}^{-} (-1 + \zeta) \bar{\gamma} (-32 + 11 \bar{\gamma}) \delta + \tfrac{1}{10} \bigl(-5760 \bar{\beta}^{-}{}^2 + 1920 \bar{\beta}^+{}^2 + 5760 \bar{\beta}^{-}{}^2 \zeta - 1920 \bar{\beta}^+{}^2 \zeta - 4480 \bar{\beta}^+ \bar{\gamma} + 4480 \bar{\beta}^+ \zeta \bar{\gamma} \nn\\
& \hspace{1cm}+ 4176 \bar{\gamma}^2 - 600 \bar{\beta}^+ \bar{\gamma}^2 - 4176 \zeta \bar{\gamma}^2 + 600 \bar{\beta}^+ \zeta \bar{\gamma}^2 + 1216 \bar{\gamma}^3 - 1216 \zeta \bar{\gamma}^3 - 231 \bar{\gamma}^4 + 231 \zeta \bar{\gamma}^4 - 960 \bar{\gamma} \bar{\chi}_{+}{} + 960 \zeta \bar{\gamma} \bar{\chi}_{+}{} \nn\\
& \hspace{1cm}+ 192 \zeta \bar{\gamma} \mathcal{S}_{-}{}^2 - 192 \zeta^2 \bar{\gamma} \mathcal{S}_{-}{}^2 - 288 \zeta \bar{\gamma}^2 \mathcal{S}_{-}{}^2 - 1212 \zeta^2 \bar{\gamma}^2 \mathcal{S}_{-}{}^2 - 192 \zeta \bar{\gamma}^3 \mathcal{S}_{-}{}^2 - 558 \zeta^2 \bar{\gamma}^3 \mathcal{S}_{-}{}^2 + 192 \zeta \bar{\gamma} \mathcal{S}_{+}{}^2 - 192 \zeta^2 \bar{\gamma} \mathcal{S}_{+}{}^2 \nn\\
& \hspace{1cm}- 288 \zeta \bar{\gamma}^2 \mathcal{S}_{+}{}^2 - 1212 \zeta^2 \bar{\gamma}^2 \mathcal{S}_{+}{}^2 - 192 \zeta \bar{\gamma}^3 \mathcal{S}_{+}{}^2 - 558 \zeta^2 \bar{\gamma}^3 \mathcal{S}_{+}{}^2 + 3480 \zeta \bar{\gamma}^2 \mathcal{S}_{-}{}^2 \lambda_1 \nn\\
& \hspace{1cm}+ 1740 \zeta \bar{\gamma}^3 \mathcal{S}_{-}{}^2 \lambda_1 + 3480 \zeta \bar{\gamma}^2 \mathcal{S}_{+}{}^2 \lambda_1 + 1740 \zeta \bar{\gamma}^3 \mathcal{S}_{+}{}^2 \lambda_1 - 2400 \bar{\gamma}^2 \mathcal{S}_{-}{}^2 (\lambda_1)^2 - 1200 \bar{\gamma}^3 \mathcal{S}_{-}{}^2 (\lambda_1)^2 - 2400 \bar{\gamma}^2 \mathcal{S}_{+}{}^2 (\lambda_1)^2\nn\\
& \hspace{1cm} - 1200 \bar{\gamma}^3 \mathcal{S}_{+}{}^2 (\lambda_1)^2 + 480 \bar{\gamma}^2 \mathcal{S}_{-}{}^2 \lambda_2 + 240 \bar{\gamma}^3 \mathcal{S}_{-}{}^2 \lambda_2 + 480 \bar{\gamma}^2 \mathcal{S}_{+}{}^2 \lambda_2 + 240 \bar{\gamma}^3 \mathcal{S}_{+}{}^2 \lambda_2\bigr)\Bigr)\biggr\} \lambda_+^{(0)} \nn\\
& \hspace{1cm}+ \biggl\{6 \zeta \bar{\gamma}^2 (2 + \bar{\gamma})  (\mathcal{S}_{-}{}^2 + \mathcal{S}_{+}{}^2) (6 \zeta - 5 \lambda_1) + \delta \Bigl(12 \zeta \bar{\gamma}^2 (2 + \bar{\gamma})  \mathcal{S}_{-}{} \mathcal{S}_{+}{}  (6 \zeta - 5 \lambda_1)\Bigr) \nn\\
& \hspace{1cm}- \nu \Bigl( 12 \zeta \bar{\gamma}^2 (2 + \bar{\gamma}) \phi_{0}{} (\mathcal{S}_{-}{}^2 + \mathcal{S}_{+}{}^2)  (6 \zeta - 5 \lambda_1) \Bigr)\biggr\} \phi_{0}{} \lambda_+^{(1)} \nn\\
& \hspace{1cm}+ \biggl\{6 \zeta^2 \bar{\gamma}^2 (2 + \bar{\gamma})  (\mathcal{S}_{-}{}^2 + \mathcal{S}_{+}{}^2) + \delta \Bigl(12 \zeta^2 \bar{\gamma}^2 (2 + \bar{\gamma})  \mathcal{S}_{-}{} \mathcal{S}_{+}{} \Bigr) - \nu \Bigl( 12 \zeta^2 \bar{\gamma}^2 (2 + \bar{\gamma})  (\mathcal{S}_{-}{}^2 + \mathcal{S}_{+}{}^2) \Bigr) \biggr\} \phi_{0}{}^2\lambda_+^{(2)}\Biggr)\Biggr]\,.
\end{align}
\end{subequations}

\subsection{Energy in the CM frame}
\label{App:CMenergyNNLO}

The NNLO tidal correction to the conserved energy in the center of mass frame is given, after separating it in power of $\tilde{G}$, by 
\begin{subequations}
\begin{align}
&E_{NNLO}^{(2)}= \alpha^2 \tilde{G}^2  \frac{\zeta}{1 -  \zeta} m^2 \nu \Biggl[- \frac{3 (2 + \bar{\gamma})^2 (c_+^{(0)}-4{\nu}_+^{(0)})}{8 (-1 + \zeta)^2 r^6} + \frac{ 6 \mu_+^{(0)}}{c^2 r^6} \nn\\
& \hspace{1cm}+ \frac{1}{c^6 r^4}\Bigl((n v)^4 \bigl((42 \delta + 6 \delta \nu) \lambda_-^{(0)} + (42 + 6 \nu - 18 \nu^2) \lambda_+^{(0)}\bigr) + v^4 \bigl((\tfrac{29}{8} \delta + \tfrac{9}{4} \delta \nu) \lambda_-^{(0)} + (\tfrac{35}{8} - 3 \nu + \tfrac{27}{4} \nu^2) \lambda_+^{(0)}\bigr) \nn\\
& \hspace{1cm}+ (n v)^2 v^2 \bigl((- \tfrac{71}{2} \delta - 12 \delta \nu) \lambda_-^{(0)} + (- \tfrac{77}{2} - 9 \nu + 30 \nu^2) \lambda_+^{(0)}\bigr)\Bigr)   \Biggr]\,,\\
&E_{NNLO}^{(3)}= \frac{\alpha^3 \tilde{G}^3}{c^6 r^5} \frac{\zeta}{1 -  \zeta} m^3 \nu \Biggl[ v^2 \biggl(\Bigl(\tfrac{1}{4} (-2 -  \bar{\gamma}) \delta + \bigl(- \frac{12 \bar{\beta}^-}{\bar{\gamma}} -  \frac{(16 \bar{\beta}^+ + 4 \bar{\gamma} + 3 \bar{\gamma}^2) \delta}{4 \bar{\gamma}}\bigr) \nu \Bigr) \lambda_-^{(0)} \nn\\
& \hspace{1cm}+ \bigl(\tfrac{1}{4} (2 + \bar{\gamma}) + (\frac{48 \bar{\beta}^+ + 212 \bar{\gamma} + 131 \bar{\gamma}^2}{4 \bar{\gamma}} + \frac{4 \bar{\beta}^- \delta}{\bar{\gamma}}) \nu - 2 \nu^2\bigr) \lambda_+^{(0)} \nn \\ 
& \hspace{1cm} - \frac{(2 + \bar{\gamma}) (-1 + \nu)\bigl(4 \lambda_1 (- \delta \Lambda_-^{(0)} + \Lambda_+^{(0)}) + \zeta (5 \delta \Lambda_-^{(0)} + 2 \phi_{0}{} \delta \Lambda_-^{(1)} - 5 \Lambda_+^{(0)} - 2 \phi_{0}{} \Lambda_+^{(1)})\bigr)}{4 (-1 + \zeta)} \biggr) \nn\\
& \hspace{1cm}+ (n v)^2 \biggl(\Bigl(\tfrac{1}{2} (10 - 3 \bar{\gamma}) \delta + \bigl(\frac{100 \bar{\beta}^-}{\bar{\gamma}} -  \frac{(80 \bar{\beta}^+ - 4 \bar{\gamma} + 15 \bar{\gamma}^2) \delta}{4 \bar{\gamma}}\bigr) \nu \Bigr) \lambda_-^{(0)}  \nn\\
& \hspace{1cm}+ \bigl(\tfrac{1}{2} (6 - 5 \bar{\gamma}) + (- \frac{400 \bar{\beta}^+ + 484 \bar{\gamma} + 369 \bar{\gamma}^2}{4 \bar{\gamma}} + \frac{20 \bar{\beta}^- \delta}{\bar{\gamma}}) \nu - 18 \nu^2\bigr) \lambda_+^{(0)} \nn\\
& \hspace{1cm} - \frac{(2 + \bar{\gamma}) (2 + 5 \nu) \bigl(4 \lambda_1 (- \delta \Lambda_-^{(0)} + \Lambda_+^{(0)}) + \zeta (5 \delta \Lambda_-^{(0)} + 2 \phi_{0}{} \delta \Lambda_-^{(1)} - 5 \Lambda_+^{(0)} - 2 \phi_{0}{} \Lambda_+^{(1)})\bigr)}{4 (-1 + \zeta)}\biggr)\Biggr]\,,\\
&E_{NNLO}^{(4)}= \frac{\alpha^4 \tilde{G}^4}{c^6 r^6} \frac{\zeta}{(1 -  \zeta)^2} m^4 \nu \Biggl[ (2 + \bar{\gamma})^2 \bigl(\tfrac{3}{2} \zeta \phi_{0}{} \delta \Lambda_-^{(1)} + (5 \zeta - 4 \lambda_1) (\tfrac{3}{4} \delta \Lambda_-^{(0)} -  \tfrac{3}{4} \Lambda_+^{(0)}) -  \tfrac{3}{2} \zeta \phi_{0}{} \Lambda_+^{(1)}\bigr) \nn\\
& \hspace{1cm}+ (2 + \bar{\gamma}) \bigl(- \frac{12 \bar{\beta}^-{} \nu (5 \zeta - 4 \lambda_1) \Lambda_-^{(0)}}{\bar{\gamma}} -  \frac{24 \bar{\beta}^-{} \zeta \phi_{0}{} \nu \Lambda_-^{(1)}}{\bar{\gamma}} + \frac{3 (8 \bar{\beta}^+{} + \bar{\gamma}^2) \nu (5 \zeta - 4 \lambda_1) \Lambda_+^{(0)}}{2 \bar{\gamma}} + \frac{3 \zeta (8 \bar{\beta}^+{} + \bar{\gamma}^2) \phi_{0}{} \nu \Lambda_+^{(1)}}{\bar{\gamma}}\bigr)\nn\\
& \hspace{1cm}\frac{1}{\bar{\gamma}^2}\Biggl(\biggl\{\tfrac{1}{5} \nu \bigl(320 \bar{\beta}^-{} \bar{\beta}^+{} - 320 \bar{\beta}^-{} \bar{\beta}^+{} \zeta + 160 \bar{\beta}^-{} \bar{\gamma} - 160 \bar{\beta}^-{} \zeta \bar{\gamma} + 180 \bar{\beta}^-{} \bar{\gamma}^2 - 180 \bar{\beta}^-{} \zeta \bar{\gamma}^2 + 80 \bar{\gamma} \bar{\chi}_{-}{} \nn\\
& \hspace{1cm}- 80 \zeta \bar{\gamma} \bar{\chi}_{-}{} + 32 \zeta \bar{\gamma} \mathcal{S}_{-}{} \mathcal{S}_{+}{} - 32 \zeta^2 \bar{\gamma} \mathcal{S}_{-}{} \mathcal{S}_{+}{} + 32 \zeta \bar{\gamma}^2 \mathcal{S}_{-}{} \mathcal{S}_{+}{} + 218 \zeta^2 \bar{\gamma}^2 \mathcal{S}_{-}{} \mathcal{S}_{+}{} + 8 \zeta \bar{\gamma}^3 \mathcal{S}_{-}{} \mathcal{S}_{+}{} + 117 \zeta^2 \bar{\gamma}^3 \mathcal{S}_{-}{} \mathcal{S}_{+}{} - 580 \zeta \bar{\gamma}^2 \mathcal{S}_{-}{} \mathcal{S}_{+}{} \lambda_1 \nn\\
& \hspace{1cm}- 290 \zeta \bar{\gamma}^3 \mathcal{S}_{-}{} \mathcal{S}_{+}{} \lambda_1 + 400 \bar{\gamma}^2 \mathcal{S}_{-}{} \mathcal{S}_{+}{} (\lambda_1)^2 + 200 \bar{\gamma}^3 \mathcal{S}_{-}{} \mathcal{S}_{+}{} (\lambda_1)^2 - 80 \bar{\gamma}^2 \mathcal{S}_{-}{} \mathcal{S}_{+}{} \lambda_2 - 40 \bar{\gamma}^3 \mathcal{S}_{-}{} \mathcal{S}_{+}{} \lambda_2\bigr) \nn\\
& \hspace{1cm}+ \tfrac{1}{10} \bigl(-320 \bar{\beta}^-{} \bar{\beta}^+{} + 320 \bar{\beta}^-{} \bar{\beta}^+{} \zeta + 160 \bar{\beta}^-{} \bar{\gamma} - 160 \bar{\beta}^-{} \zeta \bar{\gamma} - 20 \bar{\beta}^-{} \bar{\gamma}^2 + 20 \bar{\beta}^-{} \zeta \bar{\gamma}^2 - 80 \bar{\gamma} \bar{\chi}_{-}{} + 80 \zeta \bar{\gamma} \bar{\chi}_{-}{} - 32 \zeta \bar{\gamma} \mathcal{S}_{-}{} \mathcal{S}_{+}{} \nn\\
& \hspace{1cm}+ 32 \zeta^2 \bar{\gamma} \mathcal{S}_{-}{} \mathcal{S}_{+}{} - 32 \zeta \bar{\gamma}^2 \mathcal{S}_{-}{} \mathcal{S}_{+}{} - 218 \zeta^2 \bar{\gamma}^2 \mathcal{S}_{-}{} \mathcal{S}_{+}{} - 8 \zeta \bar{\gamma}^3 \mathcal{S}_{-}{} \mathcal{S}_{+}{} - 117 \zeta^2 \bar{\gamma}^3 \mathcal{S}_{-}{} \mathcal{S}_{+}{} + 580 \zeta \bar{\gamma}^2 \mathcal{S}_{-}{} \mathcal{S}_{+}{} \lambda_1 + 290 \zeta \bar{\gamma}^3 \mathcal{S}_{-}{} \mathcal{S}_{+}{} \lambda_1\nn\\
& \hspace{1cm} - 400 \bar{\gamma}^2 \mathcal{S}_{-}{} \mathcal{S}_{+}{} (\lambda_1)^2 - 200 \bar{\gamma}^3 \mathcal{S}_{-}{} \mathcal{S}_{+}{} (\lambda_1)^2 + 80 \bar{\gamma}^2 \mathcal{S}_{-}{} \mathcal{S}_{+}{} \lambda_2 + 40 \bar{\gamma}^3 \mathcal{S}_{-}{} \mathcal{S}_{+}{} \lambda_2\bigr) \nn\\
& \hspace{1cm}+ \tfrac{1}{40} \delta \bigl(640 \bar{\beta}^-{}^2 + 640 \bar{\beta}^+{}^2 - 640 \bar{\beta}^-{}^2 \zeta - 640 \bar{\beta}^+{}^2 \zeta - 640 \bar{\beta}^+{} \bar{\gamma} + 640 \bar{\beta}^+{} \zeta \bar{\gamma} - 172 \bar{\gamma}^2 + 80 \bar{\beta}^+{} \bar{\gamma}^2 + 172 \zeta \bar{\gamma}^2 - 80 \bar{\beta}^+{} \zeta \bar{\gamma}^2 \nn\\
& \hspace{1cm}- 212 \bar{\gamma}^3 + 212 \zeta \bar{\gamma}^3 - 53 \bar{\gamma}^4 + 53 \zeta \bar{\gamma}^4 + 320 \bar{\gamma} \bar{\chi}_{+}{} - 320 \zeta \bar{\gamma} \bar{\chi}_{+}{} - 64 \zeta \bar{\gamma} \mathcal{S}_{-}{}^2 + 64 \zeta^2 \bar{\gamma} \mathcal{S}_{-}{}^2 - 64 \zeta \bar{\gamma}^2 \mathcal{S}_{-}{}^2 - 436 \zeta^2 \bar{\gamma}^2 \mathcal{S}_{-}{}^2 \nn\\
& \hspace{1cm}- 16 \zeta \bar{\gamma}^3 \mathcal{S}_{-}{}^2 - 234 \zeta^2 \bar{\gamma}^3 \mathcal{S}_{-}{}^2 - 64 \zeta \bar{\gamma} \mathcal{S}_{+}{}^2 + 64 \zeta^2 \bar{\gamma} \mathcal{S}_{+}{}^2 - 64 \zeta \bar{\gamma}^2 \mathcal{S}_{+}{}^2 - 436 \zeta^2 \bar{\gamma}^2 \mathcal{S}_{+}{}^2 - 16 \zeta \bar{\gamma}^3 \mathcal{S}_{+}{}^2 - 234 \zeta^2 \bar{\gamma}^3 \mathcal{S}_{+}{}^2 \nn\\
& \hspace{1cm}+ 1160 \zeta \bar{\gamma}^2 \mathcal{S}_{-}{}^2 \lambda_1 + 580 \zeta \bar{\gamma}^3 \mathcal{S}_{-}{}^2 \lambda_1 + 1160 \zeta \bar{\gamma}^2 \mathcal{S}_{+}{}^2 \lambda_1 + 580 \zeta \bar{\gamma}^3 \mathcal{S}_{+}{}^2 \lambda_1 - 800 \bar{\gamma}^2 \mathcal{S}_{-}{}^2 (\lambda_1)^2 - 400 \bar{\gamma}^3 \mathcal{S}_{-}{}^2 (\lambda_1)^2\nn\\
& \hspace{1cm} - 800 \bar{\gamma}^2 \mathcal{S}_{+}{}^2 (\lambda_1)^2 - 400 \bar{\gamma}^3 \mathcal{S}_{+}{}^2 (\lambda_1)^2 + 160 \bar{\gamma}^2 \mathcal{S}_{-}{}^2 \lambda_2 + 80 \bar{\gamma}^3 \mathcal{S}_{-}{}^2 \lambda_2 + 160 \bar{\gamma}^2 \mathcal{S}_{+}{}^2 \lambda_2 + 80 \bar{\gamma}^3 \mathcal{S}_{+}{}^2 \lambda_2\bigr)\biggr\} \lambda_-^{(0)} \nn\\
& \hspace{1cm}+ \biggl\{-2 \zeta \bar{\gamma}^2 (2 + \bar{\gamma})  \mathcal{S}_{-}{} \mathcal{S}_{+}{} (6 \zeta - 5 \lambda_1) -  \zeta \bar{\gamma}^2 (2 + \bar{\gamma})  (\mathcal{S}_{-}{}^2 + \mathcal{S}_{+}{}^2) \delta (6 \zeta - 5 \lambda_1) + 4 \zeta \bar{\gamma}^2 (2 + \bar{\gamma}) \mathcal{S}_{-}{} \mathcal{S}_{+}{} \nu (6 \zeta - 5 \lambda_1)\biggr\} \phi_{0}{}\lambda_-^{(1)} \nn\\
& \hspace{1cm}+ \biggl\{-2 \zeta^2 \bar{\gamma}^2 (2 + \bar{\gamma})  \mathcal{S}_{-}{} \mathcal{S}_{+}{} -  \zeta^2 \bar{\gamma}^2 (2 + \bar{\gamma}) (\mathcal{S}_{-}{}^2 + \mathcal{S}_{+}{}^2) \delta + 4 \zeta^2 \bar{\gamma}^2 (2 + \bar{\gamma}) \mathcal{S}_{-}{} \mathcal{S}_{+}{} \nu \biggr\} \phi_{0}{}^2 \lambda_-^{(2)} \nn\\
& \hspace{1cm}+ \biggl\{\tfrac{1}{10} \delta \bigl(-320 \bar{\beta}^-{} \bar{\beta}^+{} + 320 \bar{\beta}^-{} \bar{\beta}^+{} \zeta + 160 \bar{\beta}^-{} \bar{\gamma} - 160 \bar{\beta}^-{} \zeta \bar{\gamma} - 20 \bar{\beta}^-{} \bar{\gamma}^2 + 20 \bar{\beta}^-{} \zeta \bar{\gamma}^2 - 80 \bar{\gamma} \bar{\chi}_{-}{} + 80 \zeta \bar{\gamma} \bar{\chi}_{-}{} - 32 \zeta \bar{\gamma} \mathcal{S}_{-}{} \mathcal{S}_{+}{} \nn\\
& \hspace{1cm}+ 32 \zeta^2 \bar{\gamma} \mathcal{S}_{-}{} \mathcal{S}_{+}{} + 48 \zeta \bar{\gamma}^2 \mathcal{S}_{-}{} \mathcal{S}_{+}{} + 202 \zeta^2 \bar{\gamma}^2 \mathcal{S}_{-}{} \mathcal{S}_{+}{} + 32 \zeta \bar{\gamma}^3 \mathcal{S}_{-}{} \mathcal{S}_{+}{} + 93 \zeta^2 \bar{\gamma}^3 \mathcal{S}_{-}{} \mathcal{S}_{+}{} - 580 \zeta \bar{\gamma}^2 \mathcal{S}_{-}{} \mathcal{S}_{+}{} \lambda_1 - 290 \zeta \bar{\gamma}^3 \mathcal{S}_{-}{} \mathcal{S}_{+}{} \lambda_1 \nn\\
& \hspace{1cm}+ 400 \bar{\gamma}^2 \mathcal{S}_{-}{} \mathcal{S}_{+}{} (\lambda_1)^2 + 200 \bar{\gamma}^3 \mathcal{S}_{-}{} \mathcal{S}_{+}{} (\lambda_1)^2 - 80 \bar{\gamma}^2 \mathcal{S}_{-}{} \mathcal{S}_{+}{} \lambda_2 - 40 \bar{\gamma}^3 \mathcal{S}_{-}{} \mathcal{S}_{+}{} \lambda_2\bigr) \nn\\
& \hspace{1cm}+ \tfrac{1}{40} \bigl(640 \bar{\beta}^-{}^2 + 640 \bar{\beta}^+{}^2 - 640 \bar{\beta}^-{}^2 \zeta - 640 \bar{\beta}^+{}^2 \zeta - 640 \bar{\beta}^+{} \bar{\gamma} + 640 \bar{\beta}^+{} \zeta \bar{\gamma} + 348 \bar{\gamma}^2 + 80 \bar{\beta}^+{} \bar{\gamma}^2 - 348 \zeta \bar{\gamma}^2 - 80 \bar{\beta}^+{} \zeta \bar{\gamma}^2 + 308 \bar{\gamma}^3 \nn\\
& \hspace{1cm}- 308 \zeta \bar{\gamma}^3 + 77 \bar{\gamma}^4 - 77 \zeta \bar{\gamma}^4 + 320 \bar{\gamma} \bar{\chi}_{+}{} - 320 \zeta \bar{\gamma} \bar{\chi}_{+}{} - 64 \zeta \bar{\gamma} \mathcal{S}_{-}{}^2 + 64 \zeta^2 \bar{\gamma} \mathcal{S}_{-}{}^2 + 96 \zeta \bar{\gamma}^2 \mathcal{S}_{-}{}^2 + 404 \zeta^2 \bar{\gamma}^2 \mathcal{S}_{-}{}^2 + 64 \zeta \bar{\gamma}^3 \mathcal{S}_{-}{}^2 \nn\\
& \hspace{1cm}+ 186 \zeta^2 \bar{\gamma}^3 \mathcal{S}_{-}{}^2 - 64 \zeta \bar{\gamma} \mathcal{S}_{+}{}^2 + 64 \zeta^2 \bar{\gamma} \mathcal{S}_{+}{}^2 + 96 \zeta \bar{\gamma}^2 \mathcal{S}_{+}{}^2 + 404 \zeta^2 \bar{\gamma}^2 \mathcal{S}_{+}{}^2 + 64 \zeta \bar{\gamma}^3 \mathcal{S}_{+}{}^2 + 186 \zeta^2 \bar{\gamma}^3 \mathcal{S}_{+}{}^2 - 1160 \zeta \bar{\gamma}^2 \mathcal{S}_{-}{}^2 \lambda_1 \nn\\
& \hspace{1cm}- 580 \zeta \bar{\gamma}^3 \mathcal{S}_{-}{}^2 \lambda_1 - 1160 \zeta \bar{\gamma}^2 \mathcal{S}_{+}{}^2 \lambda_1 - 580 \zeta \bar{\gamma}^3 \mathcal{S}_{+}{}^2 \lambda_1 + 800 \bar{\gamma}^2 \mathcal{S}_{-}{}^2 (\lambda_1)^2 \nn\\
& \hspace{1cm}+ 400 \bar{\gamma}^3 \mathcal{S}_{-}{}^2 (\lambda_1)^2 + 800 \bar{\gamma}^2 \mathcal{S}_{+}{}^2 (\lambda_1)^2 + 400 \bar{\gamma}^3 \mathcal{S}_{+}{}^2 (\lambda_1)^2 - 160 \bar{\gamma}^2 \mathcal{S}_{-}{}^2 \lambda_2 - 80 \bar{\gamma}^3 \mathcal{S}_{-}{}^2 \lambda_2 - 160 \bar{\gamma}^2 \mathcal{S}_{+}{}^2 \lambda_2 - 80 \bar{\gamma}^3 \mathcal{S}_{+}{}^2 \lambda_2\bigr) \nn\\
& \hspace{1cm}+ \tfrac{1}{20} \nu \bigl(-1920 \bar{\beta}^-{}^2 + 640 \bar{\beta}^+{}^2 + 1920 \bar{\beta}^-{}^2 \zeta - 640 \bar{\beta}^+{}^2 \zeta - 640 \bar{\beta}^+{} \bar{\gamma} + 640 \bar{\beta}^+{} \zeta \bar{\gamma} + 1232 \bar{\gamma}^2 - 400 \bar{\beta}^+{} \bar{\gamma}^2 - 1232 \zeta \bar{\gamma}^2 \nn\\
& \hspace{1cm}+ 400 \bar{\beta}^+{} \zeta \bar{\gamma}^2 + 512 \bar{\gamma}^3 - 512 \zeta \bar{\gamma}^3 - 77 \bar{\gamma}^4 + 77 \zeta \bar{\gamma}^4 - 320 \bar{\gamma} \bar{\chi}_{+}{} + 320 \zeta \bar{\gamma} \bar{\chi}_{+}{} + 64 \zeta \bar{\gamma} \mathcal{S}_{-}{}^2 - 64 \zeta^2 \bar{\gamma} \mathcal{S}_{-}{}^2 - 96 \zeta \bar{\gamma}^2 \mathcal{S}_{-}{}^2 \nn\\
& \hspace{1cm}- 404 \zeta^2 \bar{\gamma}^2 \mathcal{S}_{-}{}^2 - 64 \zeta \bar{\gamma}^3 \mathcal{S}_{-}{}^2 - 186 \zeta^2 \bar{\gamma}^3 \mathcal{S}_{-}{}^2 + 64 \zeta \bar{\gamma} \mathcal{S}_{+}{}^2 - 64 \zeta^2 \bar{\gamma} \mathcal{S}_{+}{}^2 - 96 \zeta \bar{\gamma}^2 \mathcal{S}_{+}{}^2 - 404 \zeta^2 \bar{\gamma}^2 \mathcal{S}_{+}{}^2 - 64 \zeta \bar{\gamma}^3 \mathcal{S}_{+}{}^2 \nn\\
& \hspace{1cm}- 186 \zeta^2 \bar{\gamma}^3 \mathcal{S}_{+}{}^2 + 1160 \zeta \bar{\gamma}^2 \mathcal{S}_{-}{}^2 \lambda_1 + 580 \zeta \bar{\gamma}^3 \mathcal{S}_{-}{}^2 \lambda_1 + 1160 \zeta \bar{\gamma}^2 \mathcal{S}_{+}{}^2 \lambda_1 + 580 \zeta \bar{\gamma}^3 \mathcal{S}_{+}{}^2 \lambda_1 - 800 \bar{\gamma}^2 \mathcal{S}_{-}{}^2 (\lambda_1)^2 - 400 \bar{\gamma}^3 \mathcal{S}_{-}{}^2 (\lambda_1)^2 \nn\\
& \hspace{1cm}- 800 \bar{\gamma}^2 \mathcal{S}_{+}{}^2 (\lambda_1)^2 - 400 \bar{\gamma}^3 \mathcal{S}_{+}{}^2 (\lambda_1)^2 + 160 \bar{\gamma}^2 \mathcal{S}_{-}{}^2 \lambda_2 + 80 \bar{\gamma}^3 \mathcal{S}_{-}{}^2 \lambda_2 + 160 \bar{\gamma}^2 \mathcal{S}_{+}{}^2 \lambda_2 + 80 \bar{\gamma}^3 \mathcal{S}_{+}{}^2 \lambda_2\bigr)\biggr\} \lambda_+^{(0)} \nn\\
& \hspace{1cm}+ \biggl\{\zeta \bar{\gamma}^2 (2 + \bar{\gamma})  (\mathcal{S}_{-}{}^2 + \mathcal{S}_{+}{}^2) (6 \zeta - 5 \lambda_1) + 2 \zeta \bar{\gamma}^2 (2 + \bar{\gamma})  \mathcal{S}_{-}{} \mathcal{S}_{+}{} \delta (6 \zeta - 5 \lambda_1) - 2 \zeta \bar{\gamma}^2 (2 + \bar{\gamma}) (\mathcal{S}_{-}{}^2 + \mathcal{S}_{+}{}^2) \nu (6 \zeta - 5 \lambda_1)\biggr\} \phi_{0}{}\lambda_+^{(1)} \nn\\
& \hspace{1cm}+ \biggl\{\zeta^2 \bar{\gamma}^2 (2 + \bar{\gamma})  (\mathcal{S}_{-}{}^2 + \mathcal{S}_{+}{}^2) + 2 \zeta^2 \bar{\gamma}^2 (2 + \bar{\gamma})  \mathcal{S}_{-}{} \mathcal{S}_{+}{} \delta - 2 \zeta^2 \bar{\gamma}^2 (2 + \bar{\gamma}) (\mathcal{S}_{-}{}^2 + \mathcal{S}_{+}{}^2) \nu \biggr\} \phi_{0}{}^2\lambda_+^{(2)} \Biggr)\Biggr]\,. 
\end{align}
  \end{subequations}

\bibliography{BDM23}

@article{Kuntz:2019zef,
    author = "Kuntz, Adrien and Piazza, Federico and Vernizzi, Filippo",
    title = "{Effective field theory for gravitational radiation in scalar-tensor gravity}",
    eprint = "1902.04941",
    archivePrefix = "arXiv",
    primaryClass = "gr-qc",
    doi = "10.1088/1475-7516/2019/05/052",
    journal = "JCAP",
    volume = "05",
    pages = "052",
    year = "2019"
}

@article{Brax:2021qqo,
    author = "Brax, Philippe and Davis, Anne-Christine and Melville, Scott and Wong, Leong Khim",
    title = "{Spin-orbit effects for compact binaries in scalar-tensor gravity}",
    eprint = "2107.10841",
    archivePrefix = "arXiv",
    primaryClass = "gr-qc",
    doi = "10.1088/1475-7516/2021/10/075",
    journal = "JCAP",
    volume = "10",
    pages = "075",
    year = "2021"
}

@article{Barausse:2012da,
    author = "Barausse, Enrico and Palenzuela, Carlos and Ponce, Marcelo and Lehner, Luis",
    title = "{Neutron-star mergers in scalar-tensor theories of gravity}",
    eprint = "1212.5053",
    archivePrefix = "arXiv",
    primaryClass = "gr-qc",
    doi = "10.1103/PhysRevD.87.081506",
    journal = "Phys. Rev. D",
    volume = "87",
    pages = "081506",
    year = "2013"
}

@article{Bernard:2015njp,
    author = "Bernard, Laura and Blanchet, Luc and Boh\'e, Alejandro and Faye, Guillaume and Marsat, Sylvain",
    title = "{Fokker action of nonspinning compact binaries at the fourth post-Newtonian approximation}",
    eprint = "1512.02876",
    archivePrefix = "arXiv",
    primaryClass = "gr-qc",
    doi = "10.1103/PhysRevD.93.084037",
    journal = "Phys. Rev. D",
    volume = "93",
    number = "8",
    pages = "084037",
    year = "2016"
}

@article{Bernard:2018hta,
    author = "Bernard, Laura",
    title = "{Dynamics of compact binary systems in scalar-tensor theories: Equations of motion to the third post-Newtonian order}",
    eprint = "1802.10201",
    archivePrefix = "arXiv",
    primaryClass = "gr-qc",
    doi = "10.1103/PhysRevD.98.044004",
    journal = "Phys. Rev. D",
    volume = "98",
    number = "4",
    pages = "044004",
    year = "2018"
}

@article{Bernard:2018ivi,
    author = "Bernard, Laura",
    title = "{Dynamics of compact binary systems in scalar-tensor theories: II. Center-of-mass and conserved quantities to 3PN order}",
    eprint = "1812.04169",
    archivePrefix = "arXiv",
    primaryClass = "gr-qc",
    doi = "10.1103/PhysRevD.99.044047",
    journal = "Phys. Rev. D",
    volume = "99",
    number = "4",
    pages = "044047",
    year = "2019"
}

@article{Bernard:2019yfz,
    author = "Bernard, Laura",
    title = "{Dipolar tidal effects in scalar-tensor theories}",
    eprint = "1906.10735",
    archivePrefix = "arXiv",
    primaryClass = "gr-qc",
    doi = "10.1103/PhysRevD.101.021501",
    journal = "Phys. Rev. D",
    volume = "101",
    number = "2",
    pages = "021501",
    year = "2020",
    note = "[Erratum: Phys.Rev.D 107, 069901 (2023)]"
}

@article{Bernard:2022noq,
    author = "Bernard, Laura and Blanchet, Luc and Trestini, David",
    title = "{Gravitational waves in scalar-tensor theory to one-and-a-half post-Newtonian order}",
    eprint = "2201.10924",
    archivePrefix = "arXiv",
    primaryClass = "gr-qc",
    doi = "10.1088/1475-7516/2022/08/008",
    journal = "JCAP",
    volume = "08",
    number = "08",
    pages = "008",
    year = "2022"
}

@article{Blanchet:2013haa,
    author = "Blanchet, Luc",
    title = "{Gravitational Radiation from Post-Newtonian Sources and Inspiralling Compact Binaries}",
    eprint = "1310.1528",
    archivePrefix = "arXiv",
    primaryClass = "gr-qc",
    doi = "10.12942/lrr-2014-2",
    journal = "Living Rev. Rel.",
    volume = "17",
    pages = "2",
    year = "2014"
}

@article{Boulware:1985wk,
    author = "Boulware, David G. and Deser, Stanley",
    title = "{String Generated Gravity Models}",
    reportNumber = "DOE-ER-40048-27 P5",
    doi = "10.1103/PhysRevLett.55.2656",
    journal = "Phys. Rev. Lett.",
    volume = "55",
    pages = "2656",
    year = "1985"
}

@article{Corman:2022xqg,
    author = "Corman, Maxence and Ripley, Justin L. and East, William E.",
    title = "{Nonlinear studies of binary black hole mergers in Einstein-scalar-Gauss-Bonnet gravity}",
    eprint = "2210.09235",
    archivePrefix = "arXiv",
    primaryClass = "gr-qc",
    doi = "10.1103/PhysRevD.107.024014",
    journal = "Phys. Rev. D",
    volume = "107",
    number = "2",
    pages = "024014",
    year = "2023"
}

@article{Creci:2023cfx,
    author = "Creci, Gast\'on and Hinderer, Tanja and Steinhoff, Jan",
    title = "{Tidal properties of neutron stars in scalar-tensor theories of gravity}",
    journal = "",
    eprint = "2308.11323",
    archivePrefix = "arXiv",
    primaryClass = "gr-qc",
    year = "2023"
}

@article{Damour:1985mt,
    author = {Damour, Thibault and Sch\"afer, Gerhard},
    title = "{Lagrangians for $n$ point masses at the second post-Newtonian approximation of general relativity}",
    doi = "10.1007/BF00773685",
    journal = "Gen. Rel. Grav.",
    volume = "17",
    pages = "879--905",
    year = "1985"
}

@article{Damour:1992we,
    author = "Damour, Thibault and Esposito-Farese, Gilles",
    title = "{Tensor multiscalar theories of gravitation}",
    reportNumber = "IHES-P-91-93, CPT-91-PE-2542",
    doi = "10.1088/0264-9381/9/9/015",
    journal = "Class. Quant. Grav.",
    volume = "9",
    pages = "2093--2176",
    year = "1992"
}

@article{Damour:2009wj,
    author = "Damour, Thibault and Nagar, Alessandro",
    title = "{Effective One Body description of tidal effects in inspiralling compact binaries}",
    eprint = "0911.5041",
    archivePrefix = "arXiv",
    primaryClass = "gr-qc",
    doi = "10.1103/PhysRevD.81.084016",
    journal = "Phys. Rev. D",
    volume = "81",
    pages = "084016",
    year = "2010"
}

@article{deAndrade:2000gf,
    author = "de Andrade, Vanessa C. and Blanchet, Luc and Faye, Guillaume",
    title = "{Third postNewtonian dynamics of compact binaries: Noetherian conserved quantities and equivalence between the harmonic coordinate and ADM Hamiltonian formalisms}",
    eprint = "gr-qc/0011063",
    archivePrefix = "arXiv",
    doi = "10.1088/0264-9381/18/5/301",
    journal = "Class. Quant. Grav.",
    volume = "18",
    pages = "753--778",
    year = "2001"
}

@article{DeFelice:2010aj,
    author = "De Felice, Antonio and Tsujikawa, Shinji",
    title = "{f(R) theories}",
    eprint = "1002.4928",
    archivePrefix = "arXiv",
    primaryClass = "gr-qc",
    doi = "10.12942/lrr-2010-3",
    journal = "Living Rev. Rel.",
    volume = "13",
    pages = "3",
    year = "2010"
}

@article{Dones:2023,
    author = "Dones, Eve and Bernard, Laura",
    title = "{Tidal effects in gravitational waveforms and fluxes up to next-to-next-to leading post-Newtonian order in massless scalar-tensor theories}",
    journal = "to be published",
    year = "2023"
}

@ARTICLE{1975ApJ...196L..59E,
       author = {{Eardley}, D.~M.},
        title = "{Observable effects of a scalar gravitational field in a binary pulsar.}",
      journal = {"Astrophys. J. Lett."},
         year = 1975,
        month = mar,
       volume = {196},
        pages = {L59-L62},
          doi = {10.1086/181744},
       adsurl = {https://ui.adsabs.harvard.edu/abs/1975ApJ...196L..59E}
}

@article{East:2020hgw,
    author = "East, William E. and Ripley, Justin L.",
    title = "{Evolution of Einstein-scalar-Gauss-Bonnet gravity using a modified harmonic formulation}",
    eprint = "2011.03547",
    archivePrefix = "arXiv",
    primaryClass = "gr-qc",
    doi = "10.1103/PhysRevD.103.044040",
    journal = "Phys. Rev. D",
    volume = "103",
    number = "4",
    pages = "044040",
    year = "2021"
}

@article{Figueras:2021abd,
    author = "Figueras, Pau and Fran\c{c}a, Tiago",
    title = "{Black hole binaries in cubic Horndeski theories}",
    eprint = "2112.15529",
    archivePrefix = "arXiv",
    primaryClass = "gr-qc",
    doi = "10.1103/PhysRevD.105.124004",
    journal = "Phys. Rev. D",
    volume = "105",
    number = "12",
    pages = "124004",
    year = "2022"
}

@article{Flanagan:2007ix,
    author = "Flanagan, Eanna E. and Hinderer, Tanja",
    title = "{Constraining neutron star tidal Love numbers with gravitational wave detectors}",
    eprint = "0709.1915",
    archivePrefix = "arXiv",
    primaryClass = "astro-ph",
    doi = "10.1103/PhysRevD.77.021502",
    journal = "Phys. Rev. D",
    volume = "77",
    pages = "021502",
    year = "2008"
}

@article{Goldberger:2004jt,
    author = "Goldberger, Walter D. and Rothstein, Ira Z.",
    title = "{An Effective field theory of gravity for extended objects}",
    eprint = "hep-th/0409156",
    archivePrefix = "arXiv",
    reportNumber = "UCSD-PTH-04-17, CMU-HEP-04-06",
    doi = "10.1103/PhysRevD.73.104029",
    journal = "Phys. Rev. D",
    volume = "73",
    pages = "104029",
    year = "2006"
}

@article{Gross:1986mw,
    author = "Gross, David J. and Sloan, John H.",
    title = "{The Quartic Effective Action for the Heterotic String}",
    reportNumber = "NSF-ITP-87-02",
    doi = "10.1016/0550-3213(87)90465-2",
    journal = "Nucl. Phys. B",
    volume = "291",
    pages = "41--89",
    year = "1987"
}

@article{Henry:2019xhg,
    author = "Henry, Quentin and Faye, Guillaume and Blanchet, Luc",
    title = "{Tidal effects in the equations of motion of compact binary systems to next-to-next-to-leading post-Newtonian order}",
    eprint = "1912.01920",
    archivePrefix = "arXiv",
    primaryClass = "gr-qc",
    doi = "10.1103/PhysRevD.101.064047",
    journal = "Phys. Rev. D",
    volume = "101",
    number = "6",
    pages = "064047",
    year = "2020"
}

@article{Julie:2017rpw,
    author = "Juli\'e, F\'elix-Louis",
    title = "{On the motion of hairy black holes in Einstein-Maxwell-dilaton theories}",
    eprint = "1711.10769",
    archivePrefix = "arXiv",
    primaryClass = "gr-qc",
    doi = "10.1088/1475-7516/2018/01/026",
    journal = "JCAP",
    volume = "01",
    pages = "026",
    year = "2018"
}

@article{Julie:2018lfp,
    author = "Juli\'e, F\'elix-Louis",
    title = "{Gravitational radiation from compact binary systems in Einstein-Maxwell-dilaton theories}",
    eprint = "1809.05041",
    archivePrefix = "arXiv",
    primaryClass = "gr-qc",
    doi = "10.1088/1475-7516/2018/10/033",
    journal = "JCAP",
    volume = "10",
    pages = "033",
    year = "2018"
}

@article{Julie:2019sab,
    author = "Juli\'e, F\'elix-Louis and Berti, Emanuele",
    title = "{Post-Newtonian dynamics and black hole thermodynamics in Einstein-scalar-Gauss-Bonnet gravity}",
    eprint = "1909.05258",
    archivePrefix = "arXiv",
    primaryClass = "gr-qc",
    doi = "10.1103/PhysRevD.100.104061",
    journal = "Phys. Rev. D",
    volume = "100",
    number = "10",
    pages = "104061",
    year = "2019"
}

@article{Julie:2022qux,
    author = "Juli\'e, F\'elix-Louis and Baibhav, Vishal and Berti, Emanuele and Buonanno, Alessandra",
    title = "{Third post-Newtonian effective-one-body Hamiltonian in scalar-tensor and Einstein-scalar-Gauss-Bonnet gravity}",
    eprint = "2212.13802",
    archivePrefix = "arXiv",
    primaryClass = "gr-qc",
    doi = "10.1103/PhysRevD.107.104044",
    journal = "Phys. Rev. D",
    volume = "107",
    number = "10",
    pages = "104044",
    year = "2023"
}

@article{Kanti:1995vq,
    author = "Kanti, P. and Mavromatos, N. E. and Rizos, J. and Tamvakis, K. and Winstanley, E.",
    title = "{Dilatonic black holes in higher curvature string gravity}",
    eprint = "hep-th/9511071",
    archivePrefix = "arXiv",
    reportNumber = "CERN-TH-95-297, OUTP-95-43-P, CERN-TH/95-297, OUTP-95-43P",
    doi = "10.1103/PhysRevD.54.5049",
    journal = "Phys. Rev. D",
    volume = "54",
    pages = "5049--5058",
    year = "1996"
}

@article{Langlois:2018dxi,
    author = "Langlois, David",
    title = "{Dark energy and modified gravity in degenerate higher-order scalar\textendash{}tensor (DHOST) theories: A review}",
    eprint = "1811.06271",
    archivePrefix = "arXiv",
    primaryClass = "gr-qc",
    doi = "10.1142/S0218271819420069",
    journal = "Int. J. Mod. Phys. D",
    volume = "28",
    number = "05",
    pages = "1942006",
    year = "2019"
}

@article{Mirshekari:2013vb,
    author = "Mirshekari, Saeed and Will, Clifford M.",
    title = "{Compact binary systems in scalar-tensor gravity: Equations of motion to 2.5 post-Newtonian order}",
    eprint = "1301.4680",
    archivePrefix = "arXiv",
    primaryClass = "gr-qc",
    doi = "10.1103/PhysRevD.87.084070",
    journal = "Phys. Rev. D",
    volume = "87",
    number = "8",
    pages = "084070",
    year = "2013"
}

@article{Mougiakakos:2023,
    author = "Mougiakakos, Stavros and Bernard, Laura",
    journal = "{to be published}",
    year = "2023"
}

@article{Moura:2006pz,
    author = "Moura, Filipe and Schiappa, Ricardo",
    title = "{Higher-derivative corrected black holes: Perturbative stability and absorption cross-section in heterotic string theory}",
    eprint = "hep-th/0605001",
    archivePrefix = "arXiv",
    reportNumber = "CPHT-RR047-0805, SPHT-T05-51, ITFA-2006-19, CERN-PH-TH-2006-076",
    doi = "10.1088/0264-9381/24/2/006",
    journal = "Class. Quant. Grav.",
    volume = "24",
    pages = "361--386",
    year = "2007"
}

@article{Nojiri:2018ouv,
    author = "Nojiri, S. and Odintsov, S. D. and Oikonomou, V. K.",
    title = "{Ghost-free Gauss-Bonnet Theories of Gravity}",
    eprint = "1811.07790",
    archivePrefix = "arXiv",
    primaryClass = "gr-qc",
    doi = "10.1103/PhysRevD.99.044050",
    journal = "Phys. Rev. D",
    volume = "99",
    number = "4",
    pages = "044050",
    year = "2019"
}

@article{Okounkova:2020rqw,
    author = "Okounkova, Maria",
    title = "{Numerical relativity simulation of GW150914 in Einstein dilaton Gauss-Bonnet gravity}",
    eprint = "2001.03571",
    archivePrefix = "arXiv",
    primaryClass = "gr-qc",
    doi = "10.1103/PhysRevD.102.084046",
    journal = "Phys. Rev. D",
    volume = "102",
    number = "8",
    pages = "084046",
    year = "2020"
}

@article{Palenzuela:2013hsa,
    author = "Palenzuela, Carlos and Barausse, Enrico and Ponce, Marcelo and Lehner, Luis",
    title = "{Dynamical scalarization of neutron stars in scalar-tensor gravity theories}",
    eprint = "1310.4481",
    archivePrefix = "arXiv",
    primaryClass = "gr-qc",
    doi = "10.1103/PhysRevD.89.044024",
    journal = "Phys. Rev. D",
    volume = "89",
    number = "4",
    pages = "044024",
    year = "2014"
}

@article{Porto:2016pyg,
      author         = "Porto, Rafael A.",
      title          = "{The effective field theorist's approach to
                        gravitational dynamics}",
      journal        = "Phys. Rept.",
      volume         = "633",
      year           = "2016",
      pages          = "1-104",
      doi            = "10.1016/j.physrep.2016.04.003",
      eprint         = "1601.04914",
      archivePrefix  = "arXiv",
      primaryClass   = "hep-th",
      SLACcitation   = "%%CITATION = ARXIV:1601.04914;%%"
}

@book{Schwartz:1978,
     author        = {Schwartz, L.},
     title         = {Th\'eorie des distributions},
     publisher     = {Hermann},
     address       = {Paris},
     year          = {1978}
}

@article{Shibata:2013pra,
    author = "Shibata, Masaru and Taniguchi, Keisuke and Okawa, Hirotada and Buonanno, Alessandra",
    title = "{Coalescence of binary neutron stars in a scalar-tensor theory of gravity}",
    eprint = "1310.0627",
    archivePrefix = "arXiv",
    primaryClass = "gr-qc",
    doi = "10.1103/PhysRevD.89.084005",
    journal = "Phys. Rev. D",
    volume = "89",
    number = "8",
    pages = "084005",
    year = "2014"
}

@article{Shiralilou:2020gah,
    author = "Shiralilou, Banafsheh and Hinderer, Tanja and Nissanke, Samaya and Ortiz, N\'estor and Witek, Helvi",
    title = "{Nonlinear curvature effects in gravitational waves from inspiralling black hole binaries}",
    eprint = "2012.09162",
    archivePrefix = "arXiv",
    primaryClass = "gr-qc",
    doi = "10.1103/PhysRevD.103.L121503",
    journal = "Phys. Rev. D",
    volume = "103",
    number = "12",
    pages = "L121503",
    year = "2021"
}

@article{Shiralilou:2021mfl,
    author = "Shiralilou, Banafsheh and Hinderer, Tanja and Nissanke, Samaya M. and Ortiz, N\'estor and Witek, Helvi",
    title = "{Post-Newtonian gravitational and scalar waves in scalar-Gauss\textendash{}Bonnet gravity}",
    eprint = "2105.13972",
    archivePrefix = "arXiv",
    primaryClass = "gr-qc",
    doi = "10.1088/1361-6382/ac4196",
    journal = "Class. Quant. Grav.",
    volume = "39",
    number = "3",
    pages = "035002",
    year = "2022"
}

@article{vanGemeren:2023rhh,
    author = "van Gemeren, Iris and Shiralilou, Banafsheh and Hinderer, Tanja",
    title = "{Dipolar tidal effects in gravitational waves from scalarized black hole binary inspirals in quadratic gravity}",
    eprint = "2302.08480",
    archivePrefix = "arXiv",
    primaryClass = "gr-qc",
    doi = "10.1103/PhysRevD.108.024026",
    journal = "Phys. Rev. D",
    volume = "108",
    number = "2",
    pages = "024026",
    year = "2023"
}

@article{Will:1972zz,
    author = "Will, Clifford M. and Nordtvedt, Jr., Kenneth",
    title = "{Conservation Laws and Preferred Frames in Relativistic Gravity. I. Preferred-Frame Theories and an Extended PPN Formalism}",
    doi = "10.1086/151754",
    journal = "Astrophys. J.",
    volume = "177",
    pages = "757",
    year = "1972"
}

@book{Will:2018bme,
    author = "Will, Clifford M.",
    title = "{Theory and Experiment in Gravitational Physics}",
    isbn = "978-1-108-67982-4, 978-1-107-11744-0",
    publisher = "Cambridge University Press",
    month = "9",
    year = "2018"
}

@article{Witek:2018dmd,
    author = "Witek, Helvi and Gualtieri, Leonardo and Pani, Paolo and Sotiriou, Thomas P.",
    title = "{Black holes and binary mergers in scalar Gauss-Bonnet gravity: scalar field dynamics}",
    eprint = "1810.05177",
    archivePrefix = "arXiv",
    primaryClass = "gr-qc",
    doi = "10.1103/PhysRevD.99.064035",
    journal = "Phys. Rev. D",
    volume = "99",
    number = "6",
    pages = "064035",
    year = "2019"
}

@article{Witek:2020uzz,
    author = "Witek, Helvi and Gualtieri, Leonardo and Pani, Paolo",
    title = "{Towards numerical relativity in scalar Gauss-Bonnet gravity: $3+1$ decomposition beyond the small-coupling limit}",
    eprint = "2004.00009",
    archivePrefix = "arXiv",
    primaryClass = "gr-qc",
    doi = "10.1103/PhysRevD.101.124055",
    journal = "Phys. Rev. D",
    volume = "101",
    number = "12",
    pages = "124055",
    year = "2020"
}

\end{document}


\title{Supplementary material: Tidal effects up to next-to-next-to leading post-Newtonian order in massless scalar-tensor theories}

\author{Laura Bernard}
 \email{laura.bernard@obspm.fr}
 \affiliation{%
 Laboratoire Univers et Théories, Observatoire de Paris, Université PSL, Université Paris Cité, CNRS, F-92190 Meudon, France
}%

\author{Eve Dones}
 \email{eve.dones@obspm.fr}
 \affiliation{%
 Laboratoire Univers et Théories, Observatoire de Paris, Université PSL, Université Paris Cité, CNRS, F-92190 Meudon, France
}%

\author{Stavros Mougiakakos}
 \email{stavros.mougiakakos@obspm.fr}
 \affiliation{%
 Laboratoire Univers et Théories, Observatoire de Paris, Université PSL, Université Paris Cité, CNRS, F-92190 Meudon, France
}%

\date{\today}

\begin{abstract}
In this file, we have put all the lengthy expressions for the results, as well as the intermediate steps such as the calculations of Feynman diagrams. In addition, most of the quantities computed during this work will be available in a Mathematica file directly available in the
Supplemental Material.
\end{abstract}



\maketitle

\newpage

\section{Results in ST theories up to the NNLO in harmonic coordinates} 

\subsection{Acceleration}
\label{EOM_harmonic_coordinates}


In this section is displayed the tidal correction to the acceleration in harmonic coordinates up to the NNLO. The NNLO correction term is further split in increasing powers of $\tilde{G}$ as 
\be\nn
a_1^{(NNLO),i}=\tilde{G}^2 a_1^{(NNLO),i,(2)} + \tilde{G}^3 a_1^{(NNLO),i,(3)} + \tilde{G}^4 a_1^{(NNLO),i,(4)}
\ee
\begin{subequations}
\begin{align}
&a_1^{(LO),i} =- \frac{\alpha^2 \tilde{G}^2 m_2}{ c^2 r_{12}{}^5} \frac{-8\zeta}{1-\zeta} n_{12}{}^{i}\Bigl( \tfrac{m_{2}{} }{ m_{1}{}}\bar{\delta}_2
\lambda_1^{(0)}  + \tfrac{m_{1}{} }{m_2} \bar{\delta}_1 \lambda_2^{(0)}\Bigr) \,, \\ 
&a_1^{(NLO),i} = \frac{\alpha^2 \tilde{G}^2}{c^4 r_{12}{}^5} \frac{\zeta}{1 -  \zeta} \Biggl[m_{1}{} \bar{\delta}_1 \lambda_2^{(0)} \Bigl(n_{12}{}^{i} \bigl(12 (n_{12} v_{1})^2 - 8 v_{1}^2 - 24 (n_{12} v_{1}) (n_{12} v_{2}) - 12 (n_{12} v_{2})^2\bigr) - 8 (n_{12} v_{1}) v_{1}{}^{i} + 8 (n_{12} v_{1}) v_{2}{}^{i}\Bigr) \nn \\
& \hspace{1cm}+ \frac{m_{2}{}^2}{m_{1}{}} \bar{\delta}_2 \lambda_1^{(0)} \Bigl(n_{12}{}^{i} \bigl(-12 (n_{12} v_{1})^2 - 4 v_{1}^2 + 24 (n_{12} v_{1}) (n_{12} v_{2}) - 36 (n_{12} v_{2})^2 - 8 (v_{1} v_{2}) + 4 v_{2}^2)\bigr) - 8 (n_{12} v_{2}) v_{1}{}^{i} \nn \\
& \hspace{1cm}+ 8 (n_{12} v_{2}) v_{2}{}^{i}\Bigr) \Biggr]\nn \\
& \hspace{1cm}+\frac{\alpha^3 \tilde{G}^3}{c^4 r_{12}{}^6} \frac{\zeta}{1 -  \zeta} n_{12}{}^{i} \biggl(\bar{\delta}_1 \lambda_2^{(0)} \Bigl(- \frac{8 m_{1}{} m_{2}{} (5 \bar{\gamma} + 2 \bar{\gamma}^2 - 10 \bar{\beta}_1)}{\bar{\gamma}} + m_{1}{}^2 \bigl(-66 - 31 \bar{\gamma} -  \frac{5 (2 + \bar{\gamma}) (5 \zeta - 4 \lambda_1) (1 - 2 s_1)}{-1 + \zeta}\bigr)\Bigr) \nn \\
& \hspace{1cm}-  \frac{10 \zeta (2 + \bar{\gamma}) m_{1}{}^2 \phi_{0}{} \bar{\delta}_1 \lambda_2^{(1)} (1 - 2 s_1)}{-1 + \zeta} \nn \\
& \hspace{1cm}+ \bar{\delta}_2 \lambda_1^{(0)} \Bigl(- \frac{8 m_{2}{}^2 (7 \bar{\gamma} + 2 \bar{\gamma}^2 - 10 \bar{\beta}_2)}{\bar{\gamma}} + \frac{m_{2}{}^3}{m_{1}{}} \bigl(-56 - 31 \bar{\gamma} -  \frac{5 (2 + \bar{\gamma}) (5 \zeta - 4 \lambda_1) (1 - 2 s_2)}{-1 + \zeta}\bigr)\Bigr) \nn \\
& \hspace{1cm}-  \frac{10 \zeta (2 + \bar{\gamma}) m_{2}{}^3 \phi_{0}{} \bar{\delta}_2 \lambda_1^{(1)} (1 - 2 s_2)}{(-1 + \zeta) m_{1}{}}\biggr)\,, \\
&a_1^{(NNLO),i,(2)} =\alpha^2 \Biggl[\frac{n_{12}{}^{i}}{c^2 r_{12}{}^7}\bigl(- \frac{ 36\zeta m_{2}{}^2 \bar{\delta}_2 \mu_1^{(0)}}{(-1 + \zeta) m_{1}{}} -  \frac{36 \zeta m_{1}{} \bar{\delta}_1 \mu_2^{(0)}}{-1 + \zeta}\bigr)  \nn \\
& \hspace{1cm}+ \frac{n_{12}{}^{i}}{r_{12}{}^7} \Bigl(m_{1}{} (c_2^{(0)} -  \frac{4 \zeta \nu_2^0 (1 - 2 s_1)}{1 - 2 \zeta  s_1}) \bigl(\frac{9 \zeta \bar{\delta}_1}{-1 + \zeta} -  \frac{9 (2 + \bar{\gamma})^2 (-1 -  \zeta + 4 \zeta  s_1)}{4 (-1 + \zeta)}\bigr) \nn\\
& \hspace{1cm}+ \frac{m_{2}{}^2}{m_{1}{}} (c_1^{(0)} -  \frac{4 \zeta \nu_1^0 (1 - 2 s_2)}{1 - 2 \zeta  s_2}) \bigl(\frac{9 \zeta \bar{\delta}_2}{-1 + \zeta} -  \frac{9 (2 + \bar{\gamma})^2 (-1 -  \zeta + 4 \zeta  s_2)}{4 (-1 + \zeta)}\bigr)\Bigr) \nn \\
& \hspace{1cm}+ \frac{n_{12}{}^{i}}{c^6 r_{12}{}^5} \Biggl(m_{1}{} \bar{\delta}_1 \lambda_2^{(0)} \biggl(- \frac{672 \zeta (n_{12} v_{1})^4}{-1 + \zeta} -  \frac{45 \zeta v_{1}^4}{-1 + \zeta} + \frac{2688 \zeta (n_{12} v_{1})^3 (n_{12} v_{2})}{-1 + \zeta} -  \frac{672 \zeta (n_{12} v_{2})^4}{-1 + \zeta} -  \frac{180 \zeta (v_{1} v_{2})^2}{-1 + \zeta} \nn \\
& \hspace{1cm}+ \frac{180 \zeta (v_{1} v_{2}) v_{2}^2}{-1 + \zeta} -  \frac{45 \zeta v_{2}^4}{-1 + \zeta} + v_{1}^2 \bigl(\frac{498 \zeta (n_{12} v_{2})^2}{-1 + \zeta} + \frac{180 \zeta (v_{1} v_{2})}{-1 + \zeta} -  \frac{90 \zeta v_{2}^2}{-1 + \zeta}\bigr) \nn \\
& \hspace{1cm}+ (n_{12} v_{1})^2 \bigl(\frac{522 \zeta v_{1}^2}{-1 + \zeta} -  \frac{3984 \zeta (n_{12} v_{2})^2}{-1 + \zeta} -  \frac{1044 \zeta (v_{1} v_{2})}{-1 + \zeta} + \frac{522 \zeta v_{2}^2}{-1 + \zeta}\bigr) + (n_{12} v_{2})^2 \bigl(- \frac{1020 \zeta (v_{1} v_{2})}{-1 + \zeta} + \frac{522 \zeta v_{2}^2}{-1 + \zeta}\bigr) \nn \\
& \hspace{1cm}+ (n_{12} v_{1}) \Bigl(- \frac{1044 \zeta v_{1}^2 (n_{12} v_{2})}{-1 + \zeta} + \frac{2592 \zeta (n_{12} v_{2})^3}{-1 + \zeta} + (n_{12} v_{2}) \bigl(\frac{2064 \zeta (v_{1} v_{2})}{-1 + \zeta} -  \frac{1020 \zeta v_{2}^2}{-1 + \zeta}\bigr)\Bigr)\biggr)\nn  \\
& \hspace{1cm}+ \frac{m_{2}{}^2}{m_{1}{}} \bar{\delta}_2 \lambda_1^{(0)} \biggl(- \frac{48 \zeta (n_{12} v_{1})^2 (n_{12} v_{2})^2}{-1 + \zeta} -  \frac{12 \zeta v_{1}^2 (n_{12} v_{2})^2}{-1 + \zeta} -  \frac{96 \zeta (n_{12} v_{2})^4}{-1 + \zeta} -  \frac{4 \zeta (v_{1} v_{2})^2}{-1 + \zeta} + \frac{8 \zeta (v_{1} v_{2}) v_{2}^2}{-1 + \zeta} -  \frac{4 \zeta v_{2}^4}{-1 + \zeta}\nn\\
& \hspace{1cm} + (n_{12} v_{2})^2 \bigl(- \frac{48 \zeta (v_{1} v_{2})}{-1 + \zeta} + \frac{60 \zeta v_{2}^2}{-1 + \zeta}\bigr) + (n_{12} v_{1}) \Bigl(\frac{96 \zeta (n_{12} v_{2})^3}{-1 + \zeta} + (n_{12} v_{2}) \bigl(\frac{24 \zeta (v_{1} v_{2})}{-1 + \zeta} -  \frac{24 \zeta v_{2}^2}{-1 + \zeta}\bigr)\Bigr)\biggr)\Biggr)  \nn \\
& \hspace{1cm}+ \biggl(\frac{m_{2}{}^2}{m_{1}{}} \bar{\delta}_2 \lambda_1^{(0)} \Bigl(- \frac{12 \zeta (n_{12} v_{1})^2 (n_{12} v_{2})}{-1 + \zeta} -  \frac{4 \zeta v_{1}^2 (n_{12} v_{2})}{-1 + \zeta} + \frac{24 \zeta (n_{12} v_{1}) (n_{12} v_{2})^2}{-1 + \zeta} -  \frac{36 \zeta (n_{12} v_{2})^3}{-1 + \zeta}\nn \\
& \hspace{1cm}+ (n_{12} v_{2}) \bigl(- \frac{8 \zeta (v_{1} v_{2})}{-1 + \zeta} + \frac{12 \zeta v_{2}^2}{-1 + \zeta}\bigr)\Bigr) \nn \\
& \hspace{1cm}+ m_{1}{} \bar{\delta}_1 \lambda_2^{(0)} \Bigl(\frac{276 \zeta (n_{12} v_{1})^3}{-1 + \zeta} -  \frac{816 \zeta (n_{12} v_{1})^2 (n_{12} v_{2})}{-1 + \zeta} + \frac{136 \zeta v_{1}^2 (n_{12} v_{2})}{-1 + \zeta}-  \frac{264 \zeta (n_{12} v_{2})^3}{-1 + \zeta} \nn \\
& \hspace{1cm}+ (n_{12} v_{1}) \bigl(- \frac{144 \zeta v_{1}^2}{-1 + \zeta} + \frac{780 \zeta (n_{12} v_{2})^2}{-1 + \zeta} + \frac{280 \zeta (v_{1} v_{2})}{-1 + \zeta} -  \frac{136 \zeta v_{2}^2}{-1 + \zeta}\bigr) + (n_{12} v_{2}) \bigl(- \frac{272 \zeta (v_{1} v_{2})}{-1 + \zeta} + \frac{136 \zeta v_{2}^2}{-1 + \zeta}\bigr)\Bigr)\biggr) v_{1}{}^{i}\nn \\
& \hspace{1cm}+ \biggl(\frac{m_{2}{}^2}{m_{1}{}} \bar{\delta}_2 \lambda_1^{(0)} \Bigl(\frac{12 \zeta (n_{12} v_{1})^2 (n_{12} v_{2})}{-1 + \zeta} + \frac{4 \zeta v_{1}^2 (n_{12} v_{2})}{-1 + \zeta} -  \frac{24 \zeta (n_{12} v_{1}) (n_{12} v_{2})^2}{-1 + \zeta} + \frac{36 \zeta (n_{12} v_{2})^3}{-1 + \zeta}\nn \\
& \hspace{1cm}+ (n_{12} v_{2}) \bigl(\frac{8 \zeta (v_{1} v_{2})}{-1 + \zeta} -  \frac{12 \zeta v_{2}^2}{-1 + \zeta}\bigr)\Bigr) \nn \\
& \hspace{1cm}+ m_{1}{} \bar{\delta}_1 \lambda_2^{(0)} \Bigl(- \frac{276 \zeta (n_{12} v_{1})^3}{-1 + \zeta} + \frac{816 \zeta (n_{12} v_{1})^2 (n_{12} v_{2})}{-1 + \zeta} -  \frac{136 \zeta v_{1}^2 (n_{12} v_{2})}{-1 + \zeta} + \frac{264 \zeta (n_{12} v_{2})^3}{-1 + \zeta} \nn \\
& \hspace{1cm}+ (n_{12} v_{2}) \bigl(\frac{272 \zeta (v_{1} v_{2})}{-1 + \zeta} -  \frac{136 \zeta v_{2}^2}{-1 + \zeta}\bigr) + (n_{12} v_{1}) \bigl(\frac{144 \zeta v_{1}^2}{-1 + \zeta} -  \frac{780 \zeta (n_{12} v_{2})^2}{-1 + \zeta} -  \frac{280 \zeta (v_{1} v_{2})}{-1 + \zeta} + \frac{136 \zeta v_{2}^2}{-1 + \zeta}\bigr)\Bigr)\biggr) v_{2}{}^{i}\Biggr]\,, \\
&a_1^{(NNLO),i,(3)} =\frac{\alpha^3}{c^6 r_{12}{}^6} \Biggl[n_{12}{}^{i} \Biggl(m_{1}{} m_{2}{} \bar{\delta}_1 \lambda_2^{(0)} \bigl(\frac{12 \zeta (-16 + 5 \bar{\gamma}) (n_{12} v_{1})^2}{-1 + \zeta} -  \frac{4 \zeta (-7 \bar{\gamma} + 5 \bar{\gamma}^2 - 10 \bar{\beta}_1) v_{1}^2}{(-1 + \zeta) \bar{\gamma}} \nn \\
& \hspace{1cm}-  \frac{24 \zeta (-16 + 5 \bar{\gamma}) (n_{12} v_{1}) (n_{12} v_{2})}{-1 + \zeta} + \frac{4 \zeta (-83 \bar{\gamma} + \bar{\gamma}^2 + 70 \bar{\beta}_1) (n_{12} v_{2})^2}{(-1 + \zeta) \bar{\gamma}} + \frac{8 \zeta (-17 \bar{\gamma} + \bar{\gamma}^2 + 10 \bar{\beta}_1) (v_{1} v_{2})}{(-1 + \zeta) \bar{\gamma}} \nn\\
& \hspace{1cm} -  \frac{4 \zeta (-17 \bar{\gamma} + \bar{\gamma}^2 + 10 \bar{\beta}_1) v_{2}^2}{(-1 + \zeta) \bar{\gamma}}\bigr) \nn \\
& \hspace{1cm}+ m_{2}{}^2 \bar{\delta}_2 \lambda_1^{(0)} \bigl(\frac{5 \zeta (103 \bar{\gamma} + 76 \bar{\gamma}^2 + 112 \bar{\beta}_2) (n_{12} v_{1})^2}{(-1 + \zeta) \bar{\gamma}} -  \frac{\zeta (91 \bar{\gamma} + 60 \bar{\gamma}^2 + 40 \bar{\beta}_2) v_{1}^2}{(-1 + \zeta) \bar{\gamma}} \nn \\
& \hspace{1cm}-  \frac{2 \zeta (507 \bar{\gamma} + 380 \bar{\gamma}^2 + 560 \bar{\beta}_2) (n_{12} v_{1}) (n_{12} v_{2})}{(-1 + \zeta) \bar{\gamma}} + \frac{\zeta (275 \bar{\gamma} + 324 \bar{\gamma}^2 + 840 \bar{\beta}_2) (n_{12} v_{2})^2}{(-1 + \zeta) \bar{\gamma}} + \frac{2 \zeta (35 \bar{\gamma} + 44 \bar{\gamma}^2 + 120 \bar{\beta}_2) (v_{1} v_{2})}{(-1 + \zeta) \bar{\gamma}}  \nn \\
& \hspace{1cm}-  \frac{\zeta (35 \bar{\gamma} + 44 \bar{\gamma}^2 + 120 \bar{\beta}_2) v_{2}^2}{(-1 + \zeta) \bar{\gamma}}\bigr) \nn \\
& \hspace{1cm}+ m_{1}{}^2 \biggl(\bar{\delta}_1 \lambda_2^{(0)} \Bigl(\bigl(\frac{\zeta (213 + 311 \bar{\gamma})}{-1 + \zeta} -  \frac{35 \zeta^2 (2 + \bar{\gamma}) (1 - 2 s_1)}{(-1 + \zeta)^2} + \frac{28 \zeta (2 + \bar{\gamma}) \lambda_1 (1 - 2 s_1)}{(-1 + \zeta)^2}\bigr) (n_{12} v_{1})^2 \nn \\
& \hspace{1cm}+ \bigl(- \frac{\zeta (29 + 51 \bar{\gamma})}{-1 + \zeta} -  \frac{15 \zeta^2 (2 + \bar{\gamma}) (1 - 2 s_1)}{(-1 + \zeta)^2} + \frac{12 \zeta (2 + \bar{\gamma}) \lambda_1 (1 - 2 s_1)}{(-1 + \zeta)^2}\bigr) v_{1}^2 + \bigl(- \frac{2 \zeta (221 + 311 \bar{\gamma})}{-1 + \zeta} + \frac{70 \zeta^2 (2 + \bar{\gamma}) (1 - 2 s_1)}{(-1 + \zeta)^2} \nn \\
& \hspace{1cm}-  \frac{56 \zeta (2 + \bar{\gamma}) \lambda_1 (1 - 2 s_1)}{(-1 + \zeta)^2}\bigr) (n_{12} v_{1}) (n_{12} v_{2}) + \bigl(\frac{15 \zeta (-4 + 27 \bar{\gamma})}{2 (-1 + \zeta)} -  \frac{245 \zeta^2 (2 + \bar{\gamma}) (1 - 2 s_1)}{2 (-1 + \zeta)^2} + \frac{98 \zeta (2 + \bar{\gamma}) \lambda_1 (1 - 2 s_1)}{(-1 + \zeta)^2}\bigr) (n_{12} v_{2})^2 \nn \\
& \hspace{1cm}+ \bigl(\frac{2 \zeta (-37 + 20 \bar{\gamma})}{-1 + \zeta} -  \frac{20 \zeta^2 (2 + \bar{\gamma}) (1 - 2 s_1)}{(-1 + \zeta)^2} + \frac{16 \zeta (2 + \bar{\gamma}) \lambda_1 (1 - 2 s_1)}{(-1 + \zeta)^2}\bigr) (v_{1} v_{2}) + \bigl(- \frac{\zeta (-37 + 20 \bar{\gamma})}{-1 + \zeta} \nn \\
& \hspace{1cm}+ \frac{10 \zeta^2 (2 + \bar{\gamma}) (1 - 2 s_1)}{(-1 + \zeta)^2} -  \frac{8 \zeta (2 + \bar{\gamma}) \lambda_1 (1 - 2 s_1)}{(-1 + \zeta)^2}\bigr) v_{2}^2\Bigr)\nn  \\
& \hspace{1cm}+ \phi_{0}{} \bar{\delta}_1 \lambda_2^{(1)} \bigl(- \frac{14 \zeta^2 (2 + \bar{\gamma}) (1 - 2 s_1) (n_{12} v_{1})^2}{(-1 + \zeta)^2} -  \frac{6 \zeta^2 (2 + \bar{\gamma}) (1 - 2 s_1) v_{1}^2}{(-1 + \zeta)^2} + \frac{28 \zeta^2 (2 + \bar{\gamma}) (1 - 2 s_1) (n_{12} v_{1}) (n_{12} v_{2})}{(-1 + \zeta)^2} \nn \\
& \hspace{1cm}-  \frac{49 \zeta^2 (2 + \bar{\gamma}) (1 - 2 s_1) (n_{12} v_{2})^2}{(-1 + \zeta)^2} -  \frac{8 \zeta^2 (2 + \bar{\gamma}) (1 - 2 s_1) (v_{1} v_{2})}{(-1 + \zeta)^2} + \frac{4 \zeta^2 (2 + \bar{\gamma}) (1 - 2 s_1) v_{2}^2}{(-1 + \zeta)^2}\bigr)\biggr) \nn \\
& \hspace{1cm}+ \frac{m_{2}{}^3}{m_{1}{}} \biggl(\bar{\delta}_2 \lambda_1^{(0)} \Bigl(\bigl(- \frac{\zeta (32 + 25 \bar{\gamma})}{-1 + \zeta} -  \frac{35 \zeta^2 (2 + \bar{\gamma}) (1 - 2 s_2)}{(-1 + \zeta)^2} + \frac{28 \zeta (2 + \bar{\gamma}) \lambda_1 (1 - 2 s_2)}{(-1 + \zeta)^2}\bigr) (n_{12} v_{1})^2 + \bigl(- \frac{\zeta (8 + 5 \bar{\gamma})}{-1 + \zeta} \nn \\
& \hspace{1cm}-  \frac{15 \zeta^2 (2 + \bar{\gamma}) (1 - 2 s_2)}{(-1 + \zeta)^2} + \frac{12 \zeta (2 + \bar{\gamma}) \lambda_1 (1 - 2 s_2)}{(-1 + \zeta)^2}\bigr) v_{1}^2 + \bigl(\frac{2 \zeta (32 + 25 \bar{\gamma})}{-1 + \zeta} + \frac{70 \zeta^2 (2 + \bar{\gamma}) (1 - 2 s_2)}{(-1 + \zeta)^2} \nn \\
& \hspace{1cm}-  \frac{56 \zeta (2 + \bar{\gamma}) \lambda_1 (1 - 2 s_2)}{(-1 + \zeta)^2}\bigr) (n_{12} v_{1}) (n_{12} v_{2}) + \bigl(- \frac{3 \zeta (152 + 89 \bar{\gamma})}{2 (-1 + \zeta)} -  \frac{245 \zeta^2 (2 + \bar{\gamma}) (1 - 2 s_2)}{2 (-1 + \zeta)^2} + \frac{98 \zeta (2 + \bar{\gamma}) \lambda_1 (1 - 2 s_2)}{(-1 + \zeta)^2}\bigr) (n_{12} v_{2})^2 \nn \\
& \hspace{1cm} + \bigl(- \frac{4 \zeta (24 + 13 \bar{\gamma})}{-1 + \zeta} -  \frac{20 \zeta^2 (2 + \bar{\gamma}) (1 - 2 s_2)}{(-1 + \zeta)^2} + \frac{16 \zeta (2 + \bar{\gamma}) \lambda_1 (1 - 2 s_2)}{(-1 + \zeta)^2}\bigr) (v_{1} v_{2}) + \bigl(\frac{2 \zeta (24 + 13 \bar{\gamma})}{-1 + \zeta} + \frac{10 \zeta^2 (2 + \bar{\gamma}) (1 - 2 s_2)}{(-1 + \zeta)^2} \nn \\
& \hspace{1cm}-  \frac{8 \zeta (2 + \bar{\gamma}) \lambda_1 (1 - 2 s_2)}{(-1 + \zeta)^2}\bigr) v_{2}^2\Bigr) \nn \\
& \hspace{1cm}+ \phi_{0}{} \bar{\delta}_2 \lambda_1^{(1)} \bigl(- \frac{14 \zeta^2 (2 + \bar{\gamma}) (1 - 2 s_2) (n_{12} v_{1})^2}{(-1 + \zeta)^2} -  \frac{6 \zeta^2 (2 + \bar{\gamma}) (1 - 2 s_2) v_{1}^2}{(-1 + \zeta)^2} \nn \\
& \hspace{1cm}+ \frac{28 \zeta^2 (2 + \bar{\gamma}) (1 - 2 s_2) (n_{12} v_{1}) (n_{12} v_{2})}{(-1 + \zeta)^2} -  \frac{49 \zeta^2 (2 + \bar{\gamma}) (1 - 2 s_2) (n_{12} v_{2})^2}{(-1 + \zeta)^2} -  \frac{8 \zeta^2 (2 + \bar{\gamma}) (1 - 2 s_2) (v_{1} v_{2})}{(-1 + \zeta)^2} \nn \\
& \hspace{1cm}+ \frac{4 \zeta^2 (2 + \bar{\gamma}) (1 - 2 s_2) v_{2}^2}{(-1 + \zeta)^2}\bigr)\biggr)\Biggr)\nn  \\
& \hspace{1cm}+ \Biggl(m_{1}{} m_{2}{} \bar{\delta}_1 \lambda_2^{(0)} \bigl(\frac{4 \zeta (19 \bar{\gamma} + 20 \bar{\beta}_1) (n_{12} v_{1})}{(-1 + \zeta) \bar{\gamma}} -  \frac{4 \zeta (29 + 4 \bar{\gamma}) (n_{12} v_{2})}{-1 + \zeta}\bigr) \nn \\
& \hspace{1cm}+ m_{2}{}^2 \bar{\delta}_2 \lambda_1^{(0)} \bigl(- \frac{2 \zeta (105 \bar{\gamma} + 64 \bar{\gamma}^2 + 40 \bar{\beta}_2) (n_{12} v_{1})}{(-1 + \zeta) \bar{\gamma}} + \frac{2 \zeta (85 \bar{\gamma} + 64 \bar{\gamma}^2 + 80 \bar{\beta}_2) (n_{12} v_{2})}{(-1 + \zeta) \bar{\gamma}}\bigr) \nn \\
& \hspace{1cm}+ \frac{m_{2}{}^3}{m_{1}{}} \Bigl(\bar{\delta}_2 \lambda_1^{(0)} \bigl(- \frac{\zeta (56 + 31 \bar{\gamma})}{-1 + \zeta} -  \frac{25 \zeta^2 (2 + \bar{\gamma}) (1 - 2 s_2)}{(-1 + \zeta)^2} + \frac{20 \zeta (2 + \bar{\gamma}) \lambda_1 (1 - 2 s_2)}{(-1 + \zeta)^2}\bigr) (n_{12} v_{2}) \nn \\
& \hspace{1cm}-  \frac{10 \zeta^2 (2 + \bar{\gamma}) \phi_{0}{} \bar{\delta}_2 \lambda_1^{(1)} (1 - 2 s_2) (n_{12} v_{2})}{(-1 + \zeta)^2}\Bigr) \nn \\
& \hspace{1cm}+ m_{1}{}^2 \biggl(- \frac{10 \zeta^2 (2 + \bar{\gamma}) \phi_{0}{} \bar{\delta}_1 \lambda_2^{(1)} (1 - 2 s_1) (n_{12} v_{2})}{(-1 + \zeta)^2} \nn \\
& \hspace{1cm}+ \bar{\delta}_1 \lambda_2^{(0)} \Bigl(- \frac{2 \zeta (65 + 66 \bar{\gamma}) (n_{12} v_{1})}{-1 + \zeta} + \bigl(\frac{\zeta (88 + 117 \bar{\gamma})}{-1 + \zeta} -  \frac{25 \zeta^2 (2 + \bar{\gamma}) (1 - 2 s_1)}{(-1 + \zeta)^2} + \frac{20 \zeta (2 + \bar{\gamma}) \lambda_1 (1 - 2 s_1)}{(-1 + \zeta)^2}\bigr) (n_{12} v_{2})\Bigr)\biggr)\Biggr) v_{1}{}^{i} \nn \\
& \hspace{1cm}+ \Biggl(m_{1}{} m_{2}{} \bar{\delta}_1 \lambda_2^{(0)} \bigl(- \frac{4 \zeta (19 \bar{\gamma} + 20 \bar{\beta}_1) (n_{12} v_{1})}{(-1 + \zeta) \bar{\gamma}} + \frac{4 \zeta (29 + 4 \bar{\gamma}) (n_{12} v_{2})}{-1 + \zeta}\bigr) \nn \\
& \hspace{1cm}+ m_{2}{}^2 \bar{\delta}_2 \lambda_1^{(0)} \bigl(\frac{2 \zeta (105 \bar{\gamma} + 64 \bar{\gamma}^2 + 40 \bar{\beta}_2) (n_{12} v_{1})}{(-1 + \zeta) \bar{\gamma}} -  \frac{2 \zeta (85 \bar{\gamma} + 64 \bar{\gamma}^2 + 80 \bar{\beta}_2) (n_{12} v_{2})}{(-1 + \zeta) \bar{\gamma}}\bigr) \nn \\
& \hspace{1cm}+ \frac{m_{2}{}^3}{m_{1}{}} \Bigl(\bar{\delta}_2 \lambda_1^{(0)} \bigl(\frac{\zeta (56 + 31 \bar{\gamma})}{-1 + \zeta} + \frac{25 \zeta^2 (2 + \bar{\gamma}) (1 - 2 s_2)}{(-1 + \zeta)^2} -  \frac{20 \zeta (2 + \bar{\gamma}) \lambda_1 (1 - 2 s_2)}{(-1 + \zeta)^2}\bigr) (n_{12} v_{2}) \nn \\
& \hspace{1cm}+ \frac{10 \zeta^2 (2 + \bar{\gamma}) \phi_{0}{} \bar{\delta}_2 \lambda_1^{(1)} (1 - 2 s_2) (n_{12} v_{2})}{(-1 + \zeta)^2}\Bigr) \nn \\
& \hspace{1cm}+ m_{1}{}^2 \biggl(\frac{10 \zeta^2 (2 + \bar{\gamma}) \phi_{0}{} \bar{\delta}_1 \lambda_2^{(1)} (1 - 2 s_1) (n_{12} v_{2})}{(-1 + \zeta)^2} \nn \\
& \hspace{1cm}+ \bar{\delta}_1 \lambda_2^{(0)} \Bigl(\frac{2 \zeta (65 + 66 \bar{\gamma}) (n_{12} v_{1})}{-1 + \zeta} + \bigl(- \frac{\zeta (88 + 117 \bar{\gamma})}{-1 + \zeta} + \frac{25 \zeta^2 (2 + \bar{\gamma}) (1 - 2 s_1)}{(-1 + \zeta)^2} -  \frac{20 \zeta (2 + \bar{\gamma}) \lambda_1 (1 - 2 s_1)}{(-1 + \zeta)^2}\bigr) (n_{12} v_{2})\Bigr)\biggr)\Biggr) v_{2}{}^{i}\Biggr] \,, \\
&a_1^{(NNLO),i,(4)} =\frac{\alpha^4}{c^6 r_{12}{}^7} n_{12}{}^{i} \Biggl[m_{1}{} m_{2}{}^2 \Biggl(\lambda_1^{(0)}\Bigl(- \frac{3 \zeta \bar{\gamma} (2 + \bar{\gamma})^2 (-4 + 3 \bar{\gamma})}{10 (-1 + \zeta)} \nn \\
& \hspace{1cm}-  \frac{2 \zeta \bar{\delta}_2 \bigl(421 \bar{\gamma}^2 + 216 \bar{\gamma}^3 + 49 \bar{\gamma}^4 - 1560 \bar{\gamma} \bar{\beta}_2 - 300 \bar{\gamma}^2 \bar{\beta}_2 + 480 (\bar{\beta}_2)^2 + 240 \bar{\gamma} \bar{\chi}_2\bigr)}{5 (-1 + \zeta) \bar{\gamma}^2}\Bigr)  \nn \\
& \hspace{1cm}+ \lambda_2^{(0)}\Bigl(- \frac{3 \zeta \bar{\gamma} (2 + \bar{\gamma})^2 (-4 + 3 \bar{\gamma})}{10 (-1 + \zeta)} -  \frac{2 \zeta \bar{\delta}_1 \bigl(276 \bar{\gamma}^2 + 196 \bar{\gamma}^3 + 49 \bar{\gamma}^4 - 1080 \bar{\gamma} \bar{\beta}_1 - 380 \bar{\gamma}^2 \bar{\beta}_1 + 480 (\bar{\beta}_1)^2 + 240 \bar{\gamma} \bar{\chi}_1\bigr)}{5 (-1 + \zeta) \bar{\gamma}^2}\Bigr) \Biggr) \nn \\
& \hspace{1cm}+ m_{1}{}^3 \Biggl(\frac{24 \zeta^2 \phi_{0}{}^2 (\bar{\delta}_1)^2 \lambda_2^{(2)}}{(-1 + \zeta)^2} \nn \\
& \hspace{1cm}+ \lambda_2^{(0)} \biggl(\bar{\delta}_1 \bigl(- \frac{\zeta (584 + 516 \bar{\gamma} + 131 \bar{\gamma}^2 + 32 \bar{\beta}_2)}{2 (-1 + \zeta)} -  \frac{105 \zeta^2 (2 + \bar{\gamma})^2 (1 - 2 s_1)}{(-1 + \zeta)^2} \nn \\
& \hspace{1cm}+ \frac{84 \zeta (2 + \bar{\gamma})^2 \lambda_1 (1 - 2 s_1)}{(-1 + \zeta)^2}\bigr) + (\bar{\delta}_1)^2 \Bigl(\frac{6 \zeta (4 + 21 \zeta)}{(-1 + \zeta)^2} + \frac{240 (\lambda_1)^2}{(-1 + \zeta)^2} -  \frac{48 \lambda_2}{(-1 + \zeta)^2} + \lambda_1 \bigl(- \frac{348 \zeta}{(-1 + \zeta)^2} + \frac{32 \zeta (1 - 2 s_2)}{(-1 + \zeta)^2}\bigr) \nn \\
& \hspace{1cm}-  \frac{40 \zeta^2 (1 - 2 s_2)}{(-1 + \zeta)^2}\Bigr)\biggr) \nn \\
& \hspace{1cm}+ \phi_{0}{} \lambda_2^{(1)} \Bigl(- \frac{42 \zeta^2 (2 + \bar{\gamma})^2 \bar{\delta}_1 (1 - 2 s_1)}{(-1 + \zeta)^2} + (\bar{\delta}_1)^2 \bigl(\frac{144 \zeta^2}{(-1 + \zeta)^2} -  \frac{120 \zeta \lambda_1}{(-1 + \zeta)^2} -  \frac{16 \zeta^2 (1 - 2 s_2)}{(-1 + \zeta)^2}\bigr)\Bigr)\Biggr) \nn \\
& \hspace{1cm}+ m_{1}{}^2 m_{2}{} \Biggl(\lambda_2^{(0)} \Bigl(\bar{\delta}_1 \bigl(- \frac{2 \zeta (332 \bar{\gamma}^2 + 203 \bar{\gamma}^3 + 31 \bar{\gamma}^4 - 408 \bar{\gamma} \bar{\beta}_1 - 144 \bar{\gamma}^2 \bar{\beta}_1 + 20 \bar{\gamma}^2 \bar{\beta}_2 + 192 \bar{\beta}_1 \bar{\beta}_2)}{(-1 + \zeta) \bar{\gamma}^2}\nn \\
& \hspace{1cm} -  \frac{5 \zeta^2 (2 + \bar{\gamma}) (34 \bar{\gamma} + 17 \bar{\gamma}^2 - 72 \bar{\beta}_1) (1 - 2 s_1)}{(-1 + \zeta)^2 \bar{\gamma}} + \frac{4 \zeta (2 + \bar{\gamma}) (34 \bar{\gamma} + 17 \bar{\gamma}^2 - 72 \bar{\beta}_1) \lambda_1 (1 - 2 s_1)}{(-1 + \zeta)^2 \bar{\gamma}}\bigr) + (\bar{\delta}_1)^2 \bigl(- \frac{140 \zeta^2 (1 - 2 s_2)}{(-1 + \zeta)^2}\nn \\
& \hspace{1cm} + \frac{112 \zeta \lambda_1 (1 - 2 s_2)}{(-1 + \zeta)^2}\bigr)\Bigr) \nn \\
& \hspace{1cm}+ \phi_{0}{} \lambda_2^{(1)} \bigl(- \frac{2 \zeta^2 (2 + \bar{\gamma}) (34 \bar{\gamma} + 17 \bar{\gamma}^2 - 72 \bar{\beta}_1) \bar{\delta}_1 (1 - 2 s_1)}{(-1 + \zeta)^2 \bar{\gamma}} -  \frac{56 \zeta^2 (\bar{\delta}_1)^2 (1 - 2 s_2)}{(-1 + \zeta)^2}\bigr)\Biggr) \nn \\
& \hspace{1cm}+ \frac{m_{2}{}^4}{m_{1}{}} \Biggl(\frac{24 \zeta^2 \phi_{0}{}^2 (\bar{\delta}_2)^2 \lambda_1^{(2)}}{(-1 + \zeta)^2} \nn \\
& \hspace{1cm}+ \lambda_1^{(0)} \biggl((\bar{\delta}_2)^2 \Bigl(\frac{6 \zeta (4 + 21 \zeta)}{(-1 + \zeta)^2} + \frac{240 (\lambda_1)^2}{(-1 + \zeta)^2} -  \frac{48 \lambda_2}{(-1 + \zeta)^2} + \lambda_1 \bigl(- \frac{348 \zeta}{(-1 + \zeta)^2} -  \frac{32 \zeta (1 - 2 s_1)}{(-1 + \zeta)^2}\bigr) + \frac{40 \zeta^2 (1 - 2 s_1)}{(-1 + \zeta)^2}\Bigr) \nn \\
& \hspace{1cm}+ \bar{\delta}_2 \bigl(- \frac{\zeta (432 + 476 \bar{\gamma} + 131 \bar{\gamma}^2 - 8 \bar{\beta}_1)}{2 (-1 + \zeta)} -  \frac{85 \zeta^2 (2 + \bar{\gamma})^2 (1 - 2 s_2)}{(-1 + \zeta)^2} + \frac{68 \zeta (2 + \bar{\gamma})^2 \lambda_1 (1 - 2 s_2)}{(-1 + \zeta)^2}\bigr)\biggr) \nn \\
& \hspace{1cm}+ \phi_{0}{} \lambda_1^{(1)} \Bigl((\bar{\delta}_2)^2 \bigl(\frac{144 \zeta^2}{(-1 + \zeta)^2} -  \frac{120 \zeta \lambda_1}{(-1 + \zeta)^2} + \frac{16 \zeta^2 (1 - 2 s_1)}{(-1 + \zeta)^2}\bigr) -  \frac{34 \zeta^2 (2 + \bar{\gamma})^2 \bar{\delta}_2 (1 - 2 s_2)}{(-1 + \zeta)^2}\Bigr)\Biggr) \nn \\
& \hspace{1cm}+ m_{2}{}^3 \Biggl(\lambda_1^{(0)} \Bigl((\bar{\delta}_2)^2 \bigl(- \frac{220 \zeta^2 (1 - 2 s_1)}{(-1 + \zeta)^2} + \frac{176 \zeta \lambda_1 (1 - 2 s_1)}{(-1 + \zeta)^2}\bigr) \nn \\
& \hspace{1cm}+ \bar{\delta}_2 \bigl(- \frac{2 \zeta (345 \bar{\gamma}^2 + 221 \bar{\gamma}^3 + 31 \bar{\gamma}^4 + 32 \bar{\gamma}^2 \bar{\beta}_1 - 376 \bar{\gamma} \bar{\beta}_2 - 150 \bar{\gamma}^2 \bar{\beta}_2 + 192 \bar{\beta}_1 \bar{\beta}_2)}{(-1 + \zeta) \bar{\gamma}^2} -  \frac{15 \zeta^2 (2 + \bar{\gamma}) (14 \bar{\gamma} + 7 \bar{\gamma}^2 - 24 \bar{\beta}_2) (1 - 2 s_2)}{(-1 + \zeta)^2 \bar{\gamma}} \nn \\
& \hspace{1cm}+ \frac{12 \zeta (2 + \bar{\gamma}) (14 \bar{\gamma} + 7 \bar{\gamma}^2 - 24 \bar{\beta}_2) \lambda_1 (1 - 2 s_2)}{(-1 + \zeta)^2 \bar{\gamma}}\bigr)\Bigr) \nn \\
& \hspace{1cm}+ \phi_{0}{} \lambda_1^{(1)} \bigl(- \frac{88 \zeta^2 (\bar{\delta}_2)^2 (1 - 2 s_1)}{(-1 + \zeta)^2} -  \frac{6 \zeta^2 (2 + \bar{\gamma}) (14 \bar{\gamma} + 7 \bar{\gamma}^2 - 24 \bar{\beta}_2) \bar{\delta}_2 (1 - 2 s_2)}{(-1 + \zeta)^2 \bar{\gamma}}\bigr)\Biggr)\Biggr]
\end{align}
\end{subequations}



\subsection{Energy}
\label{Energy}


We display here the tidal correction to the conserved energy in harmonic coordinates up to the NNLO. The NNLO correction term is further split in increasing powers of $\tilde{G}$ as 
\be\nn
E_{\rm NNLO}=\tilde{G}^2 E_{\rm NNLO}^{(2)} + \tilde{G}^3 E_{\rm NNLO}^{(3)} + \tilde{G}^4 E_{\rm NNLO}^{(4)}
\ee

\begin{subequations}
\begin{align}
&E_{\rm LO} =- \frac{2 \alpha^2 \zeta \tilde{G}^2 (m_{2}{}^2 \bar{\delta}_2 \lambda_1^{(0)})}{c^2 (-1 + \zeta) r_{12}{}^4} + [1 \leftrightarrow 2]\,, \\
&E_{\rm NLO}=\frac{\alpha^3 \tilde{G}^3}{c^4 r_{12}{}^5} \Biggl[\frac{4 \zeta m_{1}{} m_{2}{}^2 (\bar{\gamma} - 4 \bar{\beta}_2) \bar{\delta}_2 \lambda_1^{(0)}}{(-1 + \zeta) \bar{\gamma}}  \nn \\
& \hspace{1cm}+ m_{2}{}^3 \Bigl(\bar{\delta}_2 \lambda_1^{(0)} \bigl(\frac{3 \zeta (2 + \bar{\gamma})}{-1 + \zeta} + \frac{5 \zeta^2 (2 + \bar{\gamma}) (1 - 2 s_2)}{(-1 + \zeta)^2} -  \frac{4 \zeta (2 + \bar{\gamma}) \lambda_1 (1 - 2 s_2)}{(-1 + \zeta)^2}\bigr) + \frac{2 \zeta^2 (2 + \bar{\gamma}) \phi_{0}{} \bar{\delta}_2 \lambda_1^{(1)} (1 - 2 s_2)}{(-1 + \zeta)^2}\Bigr)\Biggr] \nn \\
& \hspace{1cm}+ \frac{\alpha^2 \tilde{G}^2}{c^4 r_{12}{}^4} \Biggl[m_{2}{}^2 \bar{\delta}_2 \lambda_1^{(0)} \bigl(\frac{2 \zeta (n_{12} v_{1})^2}{-1 + \zeta} -  \frac{\zeta v_{1}^2}{-1 + \zeta} -  \frac{4 \zeta (n_{12} v_{1}) (n_{12} v_{2})}{-1 + \zeta} -  \frac{2 \zeta (n_{12} v_{2})^2}{-1 + \zeta}\bigr)\Biggr]+ [1 \leftrightarrow 2]\,, \\
&E_{\rm NNLO}^{(2)}=\alpha^2  \Biggl[- \frac{1}{c^2 r_{12}{}^6}\frac{6 \zeta m_{2}{}^2 \bar{\delta}_2 \mu_1^{(0)}}{-1 + \zeta} + \frac{1}{r_{12}{}^6}\Biggl( m_{2}{}^2 (c_1^{(0)} -  \frac{4 \zeta \nu_1^0 (1 - 2 s_2)}{1 - 2 \zeta  s_2})\bigl(- \tfrac{3}{8} (2 + \bar{\gamma})^2 + \frac{3 \zeta \bar{\delta}_2}{2 (-1 + \zeta)} + \frac{3 \zeta (2 + \bar{\gamma})^2 (1 - 2 s_2)}{4 (-1 + \zeta)}\bigr) \Biggr) \nn \\
& \hspace{1cm}+ \frac{1}{c^6 r_{12}{}^4}\Biggl(\frac{\zeta}{4 (-1 + \zeta)}\Bigl( m_{2}{}^2 \bar{\delta}_2 \lambda_1^{(0)} \bigl(12 (n_{12} v_{1})^2 v_{1}^2 - 3 v_{1}^4 + 336 (n_{12} v_{1})^3 (n_{12} v_{2}) - 200 (n_{12} v_{1}) v_{1}^2 (n_{12} v_{2}) \nn \\
& \hspace{1cm}- 984 (n_{12} v_{1})^2 (n_{12} v_{2})^2 + 196 v_{1}^2 (n_{12} v_{2})^2 + 1056 (n_{12} v_{1}) (n_{12} v_{2})^3 - 336 (n_{12} v_{2})^4 - 120 (n_{12} v_{1})^2 (v_{1} v_{2}) \nn \\
& \hspace{1cm}+ 32 v_{1}^2 (v_{1} v_{2}) + 608 (n_{12} v_{1}) (n_{12} v_{2}) (v_{1} v_{2}) - 504 (n_{12} v_{2})^2 (v_{1} v_{2}) - 64 (v_{1} v_{2})^2 + 120 (n_{12} v_{1})^2 v_{2}^2 - 32 v_{1}^2 v_{2}^2 \nn \\
& \hspace{1cm}- 432 (n_{12} v_{1}) (n_{12} v_{2}) v_{2}^2 + 296 (n_{12} v_{2})^2 v_{2}^2 + 96 (v_{1} v_{2}) v_{2}^2 - 32 v_{2}^4\bigr) \Bigr)\Biggr)\Biggr] + [1 \leftrightarrow 2]\,, \\
&E_{\rm NNLO}^{(3)}=\frac{\alpha^3}{c^6 r_{12}{}^5} \Biggl[m_{1}{} m_{2}{}^2 \bar{\delta}_2 \lambda_1^{(0)} \bigl(\frac{\zeta (93 \bar{\gamma} + 60 \bar{\gamma}^2 + 80 \bar{\beta}_2) (n_{12} v_{1})^2}{(-1 + \zeta) \bar{\gamma}} -  \frac{\zeta (29 \bar{\gamma} + 16 \bar{\gamma}^2 + 16 \bar{\beta}_2) v_{1}^2}{(-1 + \zeta) \bar{\gamma}}\nn \\
& \hspace{1cm} -  \frac{3 \zeta (13 \bar{\gamma} + 16 \bar{\gamma}^2 + 40 \bar{\beta}_2) (n_{12} v_{1}) (n_{12} v_{2})}{(-1 + \zeta) \bar{\gamma}} + \frac{4 \zeta (-2 + \bar{\gamma}) (n_{12} v_{2})^2}{-1 + \zeta} \nn \\
& \hspace{1cm}+ \frac{\zeta (31 \bar{\gamma} + 16 \bar{\gamma}^2 + 8 \bar{\beta}_2) (v_{1} v_{2})}{(-1 + \zeta) \bar{\gamma}}\bigr) \nn \\
& \hspace{1cm}+ m_{2}{}^3 \Biggl(\phi_{0}{} \lambda_1^{(1)} \Bigl((\bar{\delta}_2)^2 (1 - 2 s_1) \bigl(- \frac{4 \zeta^2 (2 + \bar{\gamma}) (n_{12} v_{1})^2}{(-1 + \zeta)^2} + \frac{2 \zeta^2 (2 + \bar{\gamma}) v_{1}^2}{(-1 + \zeta)^2} + \frac{18 \zeta^2 (2 + \bar{\gamma}) (n_{12} v_{1}) (n_{12} v_{2})}{(-1 + \zeta)^2}\nn \\
& \hspace{1cm} -  \frac{4 \zeta^2 (2 + \bar{\gamma}) (n_{12} v_{2})^2}{(-1 + \zeta)^2} -  \frac{2 \zeta^2 (2 + \bar{\gamma}) (v_{1} v_{2})}{(-1 + \zeta)^2} + \frac{2 \zeta^2 (2 + \bar{\gamma}) v_{2}^2}{(-1 + \zeta)^2}\bigr) + \bar{\delta}_2 (1 - 2 s_2) \bigl(- \frac{\zeta^2 (2 + \bar{\gamma}) (2 + 2 \bar{\gamma} + \bar{\gamma}^2) (n_{12} v_{1})^2}{(-1 + \zeta)^2} \nn \\
& \hspace{1cm}+ \frac{\zeta^2 (2 + \bar{\gamma}) (2 + 2 \bar{\gamma} + \bar{\gamma}^2) v_{1}^2}{2 (-1 + \zeta)^2} + \frac{9 \zeta^2 (2 + \bar{\gamma}) (2 + 2 \bar{\gamma} + \bar{\gamma}^2) (n_{12} v_{1}) (n_{12} v_{2})}{2 (-1 + \zeta)^2} -  \frac{\zeta^2 (2 + \bar{\gamma}) (2 + 2 \bar{\gamma} + \bar{\gamma}^2) (n_{12} v_{2})^2}{(-1 + \zeta)^2} \nn \\
& \hspace{1cm}-  \frac{\zeta^2 (2 + \bar{\gamma}) (2 + 2 \bar{\gamma} + \bar{\gamma}^2) (v_{1} v_{2})}{2 (-1 + \zeta)^2} + \frac{\zeta^2 (2 + \bar{\gamma}) (2 + 2 \bar{\gamma} + \bar{\gamma}^2) v_{2}^2}{2 (-1 + \zeta)^2}\bigr)\Bigr) \nn \\
& \hspace{1cm} + \lambda_1^{(0)} \biggl((\bar{\delta}_2)^2 \Bigl(\lambda_1 (1 - 2 s_1) \bigl(\frac{8 \zeta (2 + \bar{\gamma}) (n_{12} v_{1})^2}{(-1 + \zeta)^2}-  \frac{4 \zeta (2 + \bar{\gamma}) v_{1}^2}{(-1 + \zeta)^2} -  \frac{36 \zeta (2 + \bar{\gamma}) (n_{12} v_{1}) (n_{12} v_{2})}{(-1 + \zeta)^2} + \frac{8 \zeta (2 + \bar{\gamma}) (n_{12} v_{2})^2}{(-1 + \zeta)^2}  \nn \\
& \hspace{1cm}+ \frac{4 \zeta (2 + \bar{\gamma}) (v_{1} v_{2})}{(-1 + \zeta)^2} -  \frac{4 \zeta (2 + \bar{\gamma}) v_{2}^2}{(-1 + \zeta)^2}\bigr) + (1 - 2 s_1) \bigl(- \frac{10 \zeta^2 (2 + \bar{\gamma}) (n_{12} v_{1})^2}{(-1 + \zeta)^2} + \frac{5 \zeta^2 (2 + \bar{\gamma}) v_{1}^2}{(-1 + \zeta)^2} + \frac{45 \zeta^2 (2 + \bar{\gamma}) (n_{12} v_{1}) (n_{12} v_{2})}{(-1 + \zeta)^2}  \nn \\
& \hspace{1cm}-  \frac{10 \zeta^2 (2 + \bar{\gamma}) (n_{12} v_{2})^2}{(-1 + \zeta)^2} -  \frac{5 \zeta^2 (2 + \bar{\gamma}) (v_{1} v_{2})}{(-1 + \zeta)^2} + \frac{5 \zeta^2 (2 + \bar{\gamma}) v_{2}^2}{(-1 + \zeta)^2}\bigr)\Bigr) + \bar{\delta}_2 \Bigl(\frac{\zeta (2 + \bar{\gamma}) (n_{12} v_{1})^2}{-1 + \zeta} -  \frac{\zeta (2 + \bar{\gamma}) v_{1}^2}{2 (-1 + \zeta)}  \nn \\
& \hspace{1cm}-  \frac{\zeta (52 + 69 \bar{\gamma}) (n_{12} v_{1}) (n_{12} v_{2})}{2 (-1 + \zeta)} + \frac{3 \zeta (25 + 19 \bar{\gamma}) (n_{12} v_{2})^2}{-1 + \zeta} + \frac{\zeta (52 + 37 \bar{\gamma}) (v_{1} v_{2})}{2 (-1 + \zeta)} -  \frac{11 \zeta (4 + 3 \bar{\gamma}) v_{2}^2}{2 (-1 + \zeta)}  \nn \\
& \hspace{1cm}+ \lambda_1 (1 - 2 s_2) \bigl(\frac{2 \zeta (2 + \bar{\gamma}) (2 + 2 \bar{\gamma} + \bar{\gamma}^2) (n_{12} v_{1})^2}{(-1 + \zeta)^2} -  \frac{\zeta (2 + \bar{\gamma}) (2 + 2 \bar{\gamma} + \bar{\gamma}^2) v_{1}^2}{(-1 + \zeta)^2} -  \frac{9 \zeta (2 + \bar{\gamma}) (2 + 2 \bar{\gamma} + \bar{\gamma}^2) (n_{12} v_{1}) (n_{12} v_{2})}{(-1 + \zeta)^2}  \nn \\
& \hspace{1cm}+ \frac{2 \zeta (2 + \bar{\gamma}) (2 + 2 \bar{\gamma} + \bar{\gamma}^2) (n_{12} v_{2})^2}{(-1 + \zeta)^2} + \frac{\zeta (2 + \bar{\gamma}) (2 + 2 \bar{\gamma} + \bar{\gamma}^2) (v_{1} v_{2})}{(-1 + \zeta)^2} -  \frac{\zeta (2 + \bar{\gamma}) (2 + 2 \bar{\gamma} + \bar{\gamma}^2) v_{2}^2}{(-1 + \zeta)^2}\bigr)  \nn \\
& \hspace{1cm}+ (1 - 2 s_2) \bigl(- \frac{5 \zeta^2 (2 + \bar{\gamma}) (2 + 2 \bar{\gamma} + \bar{\gamma}^2) (n_{12} v_{1})^2}{2 (-1 + \zeta)^2} + \frac{5 \zeta^2 (2 + \bar{\gamma}) (2 + 2 \bar{\gamma} + \bar{\gamma}^2) v_{1}^2}{4 (-1 + \zeta)^2}  \nn \\
& \hspace{1cm}+ \frac{45 \zeta^2 (2 + \bar{\gamma}) (2 + 2 \bar{\gamma} + \bar{\gamma}^2) (n_{12} v_{1}) (n_{12} v_{2})}{4 (-1 + \zeta)^2} -  \frac{5 \zeta^2 (2 + \bar{\gamma}) (2 + 2 \bar{\gamma} + \bar{\gamma}^2) (n_{12} v_{2})^2}{2 (-1 + \zeta)^2} -  \frac{5 \zeta^2 (2 + \bar{\gamma}) (2 + 2 \bar{\gamma} + \bar{\gamma}^2) (v_{1} v_{2})}{4 (-1 + \zeta)^2}  \nn \\
& \hspace{1cm}+ \frac{5 \zeta^2 (2 + \bar{\gamma}) (2 + 2 \bar{\gamma} + \bar{\gamma}^2) v_{2}^2}{4 (-1 + \zeta)^2}\bigr)\Bigr)\biggr)\Biggr)\Biggr]+ [1 \leftrightarrow 2]\,, \\
&E_{\rm NNLO}^{(4)}= \frac{\alpha^4 }{c^6 r_{12}{}^6} \Biggl[ m_{2}{}^4 \Biggl(\frac{4 \zeta^2 \phi_{0}{}^2 (\bar{\delta}_2)^2 \lambda_1^{(2)}}{(-1 + \zeta)^2} + \lambda_1^{(0)} \biggl((\bar{\delta}_2)^2 \Bigl(\frac{\zeta (4 + 21 \zeta)}{(-1 + \zeta)^2} + \frac{40 (\lambda_1)^2}{(-1 + \zeta)^2} -  \frac{8 \lambda_2}{(-1 + \zeta)^2} + \lambda_1 \bigl(- \frac{58 \zeta}{(-1 + \zeta)^2} \nn \\
& \hspace{1cm}+ \frac{12 \zeta (2 + \bar{\gamma})^2 (1 - 2 s_1)}{(-1 + \zeta)^2}\bigr)-  \frac{15 \zeta^2 (2 + \bar{\gamma})^2 (1 - 2 s_1)}{(-1 + \zeta)^2}\Bigr) + \bar{\delta}_2 \bigl(- \frac{13 \zeta (2 + \bar{\gamma})^2}{4 (-1 + \zeta)} -  \frac{15 \zeta^2 (2 + \bar{\gamma})^2 (2 + 2 \bar{\gamma} + \bar{\gamma}^2) (1 - 2 s_2)}{4 (-1 + \zeta)^2}  \nn \\
& \hspace{1cm}+ \frac{3 \zeta (2 + \bar{\gamma})^2 (2 + 2 \bar{\gamma} + \bar{\gamma}^2) \lambda_1 (1 - 2 s_2)}{(-1 + \zeta)^2}\bigr)\biggr)  \nn \\
& \hspace{1cm}+ \phi_{0}{} \lambda_1^{(1)} \Bigl((\bar{\delta}_2)^2 \bigl(\frac{24 \zeta^2}{(-1 + \zeta)^2} -  \frac{20 \zeta \lambda_1}{(-1 + \zeta)^2} -  \frac{6 \zeta^2 (2 + \bar{\gamma})^2 (1 - 2 s_1)}{(-1 + \zeta)^2}\bigr) -  \frac{3 \zeta^2 (2 + \bar{\gamma})^2 (2 + 2 \bar{\gamma} + \bar{\gamma}^2) \bar{\delta}_2 (1 - 2 s_2)}{2 (-1 + \zeta)^2}\Bigr)\Biggr)  \nn \\
& \hspace{1cm}+ m_{1}{} m_{2}{}^3 \Biggl(\phi_{0}{} \lambda_1^{(1)} \bigl(- \frac{6 \zeta^2 (6 + 6 \bar{\gamma} + 2 \bar{\gamma}^2 + 24 \bar{\beta}_2 + 24 \bar{\gamma} \bar{\beta}_2 + 7 \bar{\gamma}^2 \bar{\beta}_2) (\bar{\delta}_2)^2 (1 - 2 s_1)}{(-1 + \zeta)^2}  \nn \\
& \hspace{1cm}-  \frac{3 \zeta^2 (2 + \bar{\gamma})^2 (2 \bar{\gamma} + 2 \bar{\gamma}^2 + 2 \bar{\gamma}^3 - 8 \bar{\beta}_2 + 4 \bar{\gamma} \bar{\beta}_2 + 10 \bar{\gamma}^2 \bar{\beta}_2 + 7 \bar{\gamma}^3 \bar{\beta}_2) \bar{\delta}_2 (1 - 2 s_2)}{2 (-1 + \zeta)^2 \bar{\gamma}}\bigr)  \nn \\
& \hspace{1cm}+ \lambda_1^{(0)} \biggl((\bar{\delta}_2)^2 \bigl(- \frac{15 \zeta^2 (6 + 6 \bar{\gamma} + 2 \bar{\gamma}^2 + 24 \bar{\beta}_2 + 24 \bar{\gamma} \bar{\beta}_2 + 7 \bar{\gamma}^2 \bar{\beta}_2) (1 - 2 s_1)}{(-1 + \zeta)^2} \nn \\
& \hspace{1cm} + \frac{12 \zeta (6 + 6 \bar{\gamma} + 2 \bar{\gamma}^2 + 24 \bar{\beta}_2 + 24 \bar{\gamma} \bar{\beta}_2 + 7 \bar{\gamma}^2 \bar{\beta}_2) \lambda_1 (1 - 2 s_1)}{(-1 + \zeta)^2}\bigr) + \bar{\delta}_2 \Bigl(- \frac{\zeta (79 \bar{\gamma}^2 + 41 \bar{\gamma}^3 + 8 \bar{\gamma}^2 \bar{\beta}_1 - 64 \bar{\gamma} \bar{\beta}_2 - 24 \bar{\gamma}^2 \bar{\beta}_2 + 64 \bar{\beta}_1 \bar{\beta}_2)}{(-1 + \zeta) \bar{\gamma}^2}  \nn \\
& \hspace{1cm}+ \lambda_1 \bigl(- \frac{8 \zeta (1 + \bar{\gamma})^2 (2 + \bar{\gamma})^2 \bar{\beta}_1 (1 - 2 s_1)}{(-1 + \zeta)^2} + \frac{3 \zeta (2 + \bar{\gamma})^2 (2 \bar{\gamma} + 2 \bar{\gamma}^2 + 2 \bar{\gamma}^3 - 8 \bar{\beta}_2 + 4 \bar{\gamma} \bar{\beta}_2 + 10 \bar{\gamma}^2 \bar{\beta}_2 + 7 \bar{\gamma}^3 \bar{\beta}_2) (1 - 2 s_2)}{(-1 + \zeta)^2 \bar{\gamma}}\bigr)  \nn \\
& \hspace{1cm}-  \frac{15 \zeta^2 (2 + \bar{\gamma})^2 (2 \bar{\gamma} + 2 \bar{\gamma}^2 + 2 \bar{\gamma}^3 - 8 \bar{\beta}_2 + 4 \bar{\gamma} \bar{\beta}_2 + 10 \bar{\gamma}^2 \bar{\beta}_2 + 7 \bar{\gamma}^3 \bar{\beta}_2) (1 - 2 s_2)}{4 (-1 + \zeta)^2 \bar{\gamma}}\Bigr) -  \frac{2 \zeta \bar{\gamma} (1 + \bar{\gamma})^2 (2 + \bar{\gamma})^3 \bar{\beta}_1 \lambda_1 (1 - 2 s_2)}{(-1 + \zeta)^2}\biggr)\Biggr)  \nn \\
& \hspace{1cm}+ m_{1}{}^2 m_{2}{}^2 \Biggl( \lambda_1^{(0)} \biggl(- \frac{\zeta \bar{\gamma} (2 + \bar{\gamma})^2 (-4 + 3 \bar{\gamma})}{20 (-1 + \zeta)}  + \bar{\delta}_2 \Bigl(- \frac{\zeta \bigl(22 \bar{\gamma}^2 + 12 \bar{\gamma}^3 + 3 \bar{\gamma}^4 - 160 \bar{\gamma} \bar{\beta}_2 + 20 \bar{\gamma}^2 \bar{\beta}_2 + 160 (\bar{\beta}_2)^2 + 80 \bar{\gamma} \bar{\chi}_2\bigr)}{5 (-1 + \zeta) \bar{\gamma}^2}\nn \\
& \hspace{1cm} + \frac{\zeta (2 + \bar{\gamma})^2 (4 \bar{\gamma} + 12 \bar{\gamma}^2 + 10 \bar{\gamma}^3 + 5 \bar{\gamma}^4 + 24 \bar{\beta}_2) \lambda_1 (1 - 2 s_1)}{2 (-1 + \zeta)^2}\Bigr) \nn \\
& \hspace{1cm}+ \frac{\zeta \bar{\gamma} (2 + \bar{\gamma})^3 (4 \bar{\gamma} + 12 \bar{\gamma}^2 + 10 \bar{\gamma}^3 + 5 \bar{\gamma}^4 + 24 \bar{\beta}_2) \lambda_1 (1 - 2 s_2)}{8 (-1 + \zeta)^2}\biggr)\Biggr)\Biggr] + [1 \leftrightarrow 2]\,,
\end{align}
\end{subequations}

\subsection{Linear momentum}
\label{Pi_harmonic_coordinates}

We display here the tidal correction to the linear momentum in harmonic coordinates up to the NNLO. 

\begin{subequations}
\begin{align}
&P^{i}_{\rm LO}= 0\, ,\\
&P^{i}_{\rm NLO}= \frac{\alpha^2 \tilde{G}^2}{c^4 r_{12}{}^4} \frac{\zeta}{1 -  \zeta} \Biggl(m_{2}{}^2 \bar{\delta}_2 \lambda_1^{(0)} \bigl(8 n_{12}{}^{i} (n_{12} v_{2}) + 2 v_{1}{}^{i}\bigr)  \Biggr)+ [1 \leftrightarrow 2]\, ,\\
&P^{i}_{\rm NNLO}= \frac{\alpha^2 \tilde{G}^2}{c^6 r_{12}{}^4} \frac{\zeta}{1 -  \zeta} \Biggl[ m_{2}{}^2 \bar{\delta}_2 \lambda_1^{(0)} \Bigl(-4 n_{12}{}^{i} \bigl(21 (n_{12} v_{1})^3 - 11 (n_{12} v_{1}) v_{1}^2 - 60 (n_{12} v_{1})^2 (n_{12} v_{2}) + 11 v_{1}^2 (n_{12} v_{2}) + 63 (n_{12} v_{1}) (n_{12} v_{2})^2 \nn \\
& \hspace{1cm}- 18 (n_{12} v_{2})^3 + 22 (n_{12} v_{1}) (v_{1} v_{2}) - 22 (n_{12} v_{2}) (v_{1} v_{2}) - 11 (n_{12} v_{1}) v_{2}^2 + 10 (n_{12} v_{2}) v_{2}^2\bigr) + \bigl(28 (n_{12} v_{1})^2 - 7 v_{1}^2 \nn \\
& \hspace{1cm}- 56 (n_{12} v_{1}) (n_{12} v_{2}) + 28 (n_{12} v_{2})^2 + 16 (v_{1} v_{2}) - 8 v_{2}^2\bigr) v_{1}{}^{i} - 2 \bigl(15 (n_{12} v_{1})^2 - 4 v_{1}^2 - 30 (n_{12} v_{1}) (n_{12} v_{2}) + 13 (n_{12} v_{2})^2 \nn \\
& \hspace{1cm}+ 8 (v_{1} v_{2}) - 4 v_{2}^2\bigr) v_{2}{}^{i}\Bigr) \Biggr]\nn \\
& \hspace{1cm}+ \frac{\alpha^3 \tilde{G}^3}{c^6 r_{12}{}^5} \frac{\zeta}{1 -  \zeta} \Biggl[ m_{1}{} m_{2}{}^2 \bar{\delta}_2 \lambda_1^{(0)} \Bigl(n_{12}{}^{i} \bigl(- \frac{(111 \bar{\gamma} + 56 \bar{\gamma}^2 + 40 \bar{\beta}_2) (n_{12} v_{1})}{\bar{\gamma}} + \frac{5 (11 \bar{\gamma} + 8 \bar{\gamma}^2 + 24 \bar{\beta}_2) (n_{12} v_{2})}{\bar{\gamma}}\bigr) \nn \\
& \hspace{1cm}+ \frac{(27 \bar{\gamma} + 16 \bar{\gamma}^2 + 24 \bar{\beta}_2) v_{1}{}^{i}}{\bar{\gamma}} -  \frac{(31 \bar{\gamma} + 16 \bar{\gamma}^2 + 8 \bar{\beta}_2) v_{2}{}^{i}}{\bar{\gamma}}\Bigr) \nn \\
& \hspace{1cm}+ m_{2}{}^3 \Biggl(\bar{\delta}_2 \lambda_1^{(0)} \biggl(n_{12}{}^{i} \Bigl(\tfrac{1}{2} (44 + 65 \bar{\gamma}) (n_{12} v_{1}) + \tfrac{1}{2} (-176 - 127 \bar{\gamma}) (n_{12} v_{2}) + (1 - 2 s_2) \bigl(- \frac{5 (2 + \bar{\gamma}) (5 \zeta - 4 \lambda_1) (n_{12} v_{1})}{2 (-1 + \zeta)} \nn \\
& \hspace{1cm}-  \frac{5 (2 + \bar{\gamma}) (5 \zeta - 4 \lambda_1) (n_{12} v_{2})}{2 (-1 + \zeta)}\bigr)\Bigr) + \bigl(\tfrac{1}{2} (-48 - 35 \bar{\gamma}) -  \frac{(2 + \bar{\gamma}) (5 \zeta - 4 \lambda_1) (1 - 2 s_2)}{2 (-1 + \zeta)}\bigr) v_{1}{}^{i} \nn \\
& \hspace{1cm}+ \bigl(\tfrac{1}{2} (36 + 29 \bar{\gamma}) -  \frac{(2 + \bar{\gamma}) (5 \zeta - 4 \lambda_1) (1 - 2 s_2)}{2 (-1 + \zeta)}\bigr) v_{2}{}^{i}\biggr) \nn \\
& \hspace{1cm}+ \phi_{0}{} \bar{\delta}_2 \lambda_1^{(1)} \Bigl(n_{12}{}^{i} (1 - 2 s_2) \bigl(- \frac{5 \zeta (2 + \bar{\gamma}) (n_{12} v_{1})}{-1 + \zeta} -  \frac{5 \zeta (2 + \bar{\gamma}) (n_{12} v_{2})}{-1 + \zeta}\bigr) -  \frac{\zeta (2 + \bar{\gamma}) (1 - 2 s_2) v_{1}{}^{i}}{-1 + \zeta} -  \frac{\zeta (2 + \bar{\gamma}) (1 - 2 s_2) v_{2}{}^{i}}{-1 + \zeta}\Bigr)\Biggr) \Biggr]\nn \\
& \hspace{1cm}+ [1 \leftrightarrow 2]\, .
\end{align}
\end{subequations}


\subsection{Angular momentum}
\label{Ji_harmonic_coordinates}

We display here the tidal correction to the angular momentum in harmonic coordinates up to the NNLO. Note that in this section, the vector product is denoted, e.g $(n_{12} \times v_{1})^{i}$. 

\begin{subequations}
\begin{align}
&J^{i}_{\rm LO}= 0\, ,\\
&J^{i}_{\rm NLO}= \frac{\alpha^2 \tilde{G}^2}{c^4 r_{12}{}^4} \frac{\zeta}{1 -  \zeta} \Biggl[ m_{2}{}^2 \bar{\delta}_2 \lambda_1^{(0)} \Bigl(-2 (v_{1} \times y_{1})^{i} + \bigl(- \frac{8 (v_{2} y_{1})}{r_{12}{}^2} + \frac{8 (v_{2} y_{2})}{r_{12}{}^2}\bigr) ( y_{1} \times y_{2})^{i}\Bigr) \Biggr] + [1 \leftrightarrow 2] \, ,\\
&J^{i}_{\rm NNLO}= \frac{\alpha^2 \tilde{G}^2}{c^6 r_{12}{}^3} \frac{\zeta}{1 -  \zeta} \Biggl[m_{2}{}^2 \bar{\delta}_2 \lambda_1^{(0)} \Biggl(-2 (n_{12} v_{1})^2 (n_{12} \times v_{1})^{i} + v_{1}^2 (n_{12} \times v_{1})^{i} + 4 (n_{12} v_{1}) (n_{12} \times v_{1})^{i} (n_{12} v_{2}) - 6 (n_{12} \times v_{1})^{i} (n_{12} v_{2})^2 \nn \\
& \hspace{1cm}+ 14 (n_{12} v_{1})^2(n_{12} \times v_{2})^{i} - 4 v_{1}^2(n_{12} \times v_{2})^{i} - 28 (n_{12} v_{1}) (n_{12} v_{2})(n_{12} \times v_{2})^{i} + 18 (n_{12} v_{2})^2(n_{12} \times v_{2})^{i} + 8 (v_{1} v_{2})(n_{12} \times v_{2})^{i}\nn \\
& \hspace{1cm} - 4 v_{2}^2(n_{12} \times v_{2})^{i} - 4 (n_{12} v_{1}) (v_{1} \times v_{2})^{i} + 4 (n_{12} v_{2}) (v_{1} \times v_{2})^{i} \nn \\
& \hspace{1cm}+ \frac{1}{r_{12}{}}\Bigl(84 (n_{12} v_{1})^3 ( n_{12} \times  y_{2})^{i} - 44 (n_{12} v_{1}) v_{1}^2 ( n_{12} \times  y_{2})^{i} - 240 (n_{12} v_{1})^2 (n_{12} v_{2}) ( n_{12} \times  y_{2})^{i} + 44 v_{1}^2 (n_{12} v_{2}) ( n_{12} \times  y_{2})^{i} \nn \\
& \hspace{1cm}+ 252 (n_{12} v_{1}) (n_{12} v_{2})^2 ( n_{12} \times  y_{2})^{i} - 72 (n_{12} v_{2})^3 ( n_{12} \times  y_{2})^{i} + 88 (n_{12} v_{1}) (v_{1} v_{2}) ( n_{12} \times  y_{2})^{i} - 88 (n_{12} v_{2}) (v_{1} v_{2}) ( n_{12} \times  y_{2})^{i} \nn \\
& \hspace{1cm}- 44 (n_{12} v_{1}) v_{2}^2 ( n_{12} \times  y_{2})^{i} + 40 (n_{12} v_{2}) v_{2}^2 ( n_{12} \times  y_{2})^{i} - 28 (n_{12} v_{1})^2 (v_{1} \times y_{2})^{i} + 7 v_{1}^2 (v_{1} \times y_{2})^{i} \nn \\
& \hspace{1cm}+ 56 (n_{12} v_{1}) (n_{12} v_{2}) (v_{1} \times y_{2})^{i} - 28 (n_{12} v_{2})^2 (v_{1} \times y_{2})^{i} - 16 (v_{1} v_{2}) (v_{1} \times y_{2})^{i} + 8 v_{2}^2 (v_{1} \times y_{2})^{i} + 30 (n_{12} v_{1})^2 (v_{2} \times  y_{2})^{i} \nn \\
& \hspace{1cm}- 8 v_{1}^2 (v_{2} \times  y_{2})^{i} - 60 (n_{12} v_{1}) (n_{12} v_{2}) (v_{2} \times  y_{2})^{i} + 26 (n_{12} v_{2})^2 (v_{2} \times  y_{2})^{i} + 16 (v_{1} v_{2}) (v_{2} \times  y_{2})^{i} - 8 v_{2}^2 (v_{2} \times  y_{2})^{i} \Bigr)\Biggr)  \Biggr]\nn \\
& \hspace{1cm} +\frac{\alpha^3 \tilde{G}^3}{c^6 r_{12}{}^4} \frac{\zeta}{1 -  \zeta} \Biggl[m_{1}{} m_{2}{}^2 \bar{\delta}_2 \lambda_1^{(0)} \Biggl(\frac{4 (9 \bar{\gamma} + 6 \bar{\gamma}^2 + 8 \bar{\beta}_2) (n_{12} \times v_{1})^{i}}{\bar{\gamma}} -  \frac{(27 \bar{\gamma} + 16 \bar{\gamma}^2 + 8 \bar{\beta}_2) (n_{12} \times v_{2})^{i}}{\bar{\gamma}} \nn \\
& \hspace{1cm}+ \frac{1}{r_{12}{}}\Bigl(\frac{(111 \bar{\gamma} + 56 \bar{\gamma}^2 + 40 \bar{\beta}_2) (n_{12} v_{1}) (n_{12}\times y_{2})^{i}}{\bar{\gamma}} -  \frac{5 (11 \bar{\gamma} + 8 \bar{\gamma}^2 + 24 \bar{\beta}_2) (n_{12} v_{2}) (n_{12}\times y_{2})^{i}}{\bar{\gamma}} \nn \\
& \hspace{1cm}-  \frac{(27 \bar{\gamma} + 16 \bar{\gamma}^2 + 24 \bar{\beta}_2) ( v_{1} \times y_{2})^{i}}{\bar{\gamma}} + \frac{(31 \bar{\gamma} + 16 \bar{\gamma}^2 + 8 \bar{\beta}_2) ( v_{2} \times y_{2})^{i}}{\bar{\gamma}} \Bigr)\Biggr) \nn \\
& \hspace{1cm}+ m_{2}{}^3 \Biggl(\phi_{0}{} \bar{\delta}_2 \lambda_1^{(1)} \Bigl((1 - 2 s_2) \bigl(- \frac{2 \zeta (2 + \bar{\gamma}) (n_{12} \times v_{1})^{i}}{-1 + \zeta} + \frac{\zeta (2 + \bar{\gamma}) (n_{12} \times v_{2})^{i}}{-1 + \zeta}\bigr) \nn \\
& \hspace{1cm}+ \frac{1}{r_{12}{}}(1 - 2 s_2) \bigl(\frac{5 \zeta (2 + \bar{\gamma}) (n_{12} v_{1}) (n_{12}\times y_{2})^{i}}{-1 + \zeta} + \frac{5 \zeta (2 + \bar{\gamma}) (n_{12} v_{2}) (n_{12}\times y_{2})^{i}}{-1 + \zeta} \nn \\
& \hspace{1cm}+ \frac{\zeta (2 + \bar{\gamma}) ( v_{1} \times y_{2})^{i}}{-1 + \zeta} + \frac{\zeta (2 + \bar{\gamma}) ( v_{2} \times y_{2})^{i}}{-1 + \zeta}\bigr)\Bigr) \nn \\
& \hspace{1cm}+ \bar{\delta}_2 \lambda_1^{(0)} \biggl((2 + \bar{\gamma}) (n_{12} \times v_{1})^{i} + \tfrac{1}{2} (-32 - 21 \bar{\gamma}) (n_{12} \times v_{2})^{i} + (1 - 2 s_2) \bigl(- \frac{(2 + \bar{\gamma}) (5 \zeta - 4 \lambda_1) (n_{12} \times v_{1})^{i}}{-1 + \zeta} \nn \\
& \hspace{1cm}+ \frac{(2 + \bar{\gamma}) (5 \zeta - 4 \lambda_1) (n_{12} \times v_{2})^{i}}{2 (-1 + \zeta)}\bigr) \nn \\
& \hspace{1cm}+ \frac{1}{r_{12}{}}\Bigl(\tfrac{1}{2} (-44 - 65 \bar{\gamma}) (n_{12} v_{1}) (n_{12}\times y_{2})^{i} + \tfrac{1}{2} (176 + 127 \bar{\gamma}) (n_{12} v_{2}) (n_{12}\times y_{2})^{i} + \tfrac{1}{2} (48 + 35 \bar{\gamma}) ( v_{1} \times y_{2})^{i} \nn \\
& \hspace{1cm}+ \tfrac{1}{2} (-36 - 29 \bar{\gamma}) ( v_{2} \times y_{2})^{i} + (1 - 2 s_2) \bigl(\frac{5 (2 + \bar{\gamma}) (5 \zeta - 4 \lambda_1) (n_{12} v_{1}) (n_{12}\times y_{2})^{i}}{2 (-1 + \zeta)} + \frac{5 (2 + \bar{\gamma}) (5 \zeta - 4 \lambda_1) (n_{12} v_{2}) (n_{12}\times y_{2})^{i}}{2 (-1 + \zeta)} \nn \\
& \hspace{1cm}+ \frac{(2 + \bar{\gamma}) (5 \zeta - 4 \lambda_1) ( v_{1} \times y_{2})^{i}}{2 (-1 + \zeta)} + \frac{(2 + \bar{\gamma}) (5 \zeta - 4 \lambda_1) ( v_{2} \times y_{2})^{i}}{2 (-1 + \zeta)}\bigr) \Bigr)\biggr)\Biggr)\Biggr] + [1 \leftrightarrow 2]\, .
\end{align}
\end{subequations}

\section{NNLO tidal correction to the EOM in EsGB theories in the CM frame} 
In this section, we display the NNLO tidal correction to the equations of motion in the CM frame, in the framework of EsGB theories. The result in split in increasing powers of $G_{12}$.  

\begin{subequations}
\begin{align}
&{a_{\rm CM,NNLO}^{(2),i}}_{|{\rm EsGB}}= G_{12}^2 \Biggl[\frac{M}{c^2 r^7} \biggl((-9 + \frac{\tfrac{9}{2} -  \tfrac{9}{2} m_{-}}{\nu}) \delta_2 {\mu_1^{(0)}}_{|{\rm EsGB}} + (-9 + \frac{\tfrac{9}{2} + \tfrac{9}{2} m_{-}}{\nu}) \delta_1 {\mu_2^{(0)}}_{|{\rm EsGB}}\biggr) n^{i} \nn \\
& \hspace{1cm}+ \frac{M}{r^7} n^{i} \biggl({c_2^{(0)}}_{|{\rm EsGB}} \Bigl(\tfrac{9}{4} (2 + \bar{\gamma}_{12})^2 -  \frac{\tfrac{9}{8} (2 + \bar{\gamma}_{12})^2 + \tfrac{9}{8} (2 + \bar{\gamma}_{12})^2 m_{-}}{\nu} + \bigl(- \tfrac{9}{4} (2 + \bar{\gamma}_{12})^2 \alpha_0^2 \nn \\
& \hspace{1cm}+ \frac{\tfrac{9}{8} (2 + \bar{\gamma}_{12})^2 \alpha_0^2 + \tfrac{9}{8} (2 + \bar{\gamma}_{12})^2 \alpha_0^2 m_{-}}{\nu}\bigr) (\alpha_1^0)^2 + (9 \alpha_0^2 -  \frac{\tfrac{9}{2} \alpha_0^2 + \tfrac{9}{2} \alpha_0^2 m_{-}}{\nu}) \delta_1\Bigr) \nn \\
& \hspace{1cm}+ {c_1^{(0)}}_{|{\rm EsGB}} \Bigl(\tfrac{9}{4} (2 + \bar{\gamma}_{12})^2 -  \frac{\tfrac{9}{8} (2 + \bar{\gamma}_{12})^2 -  \tfrac{9}{8} (2 + \bar{\gamma}_{12})^2 m_{-}}{\nu} + \bigl(- \tfrac{9}{4} (2 + \bar{\gamma}_{12})^2 \alpha_0^2 \nn \\
& \hspace{1cm}+ \frac{\tfrac{9}{8} (2 + \bar{\gamma}_{12})^2 \alpha_0^2 -  \tfrac{9}{8} (2 + \bar{\gamma}_{12})^2 \alpha_0^2 m_{-}}{\nu}\bigr) (\alpha_2^0)^2 + (9 \alpha_0^2 -  \frac{\tfrac{9}{2} \alpha_0^2 -  \tfrac{9}{2} \alpha_0^2 m_{-}}{\nu}) \delta_2\Bigr) \nn \\
& \hspace{1cm}+ {\nu_1^{(0)}}_{|{\rm EsGB}}\Bigl(\bigl(\tfrac{9}{2} (2 + \bar{\gamma}_{12})^2 -  \frac{\tfrac{9}{4} (2 + \bar{\gamma}_{12})^2 -  \tfrac{9}{4} (2 + \bar{\gamma}_{12})^2 m_{-}}{\nu}\bigr) \alpha_2^0 \nn \\
& \hspace{1cm}+ \bigl(\tfrac{9}{2} (2 + \bar{\gamma}_{12})^2 \alpha_0 -  \frac{\tfrac{9}{4} (2 + \bar{\gamma}_{12})^2 \alpha_0 -  \tfrac{9}{4} (2 + \bar{\gamma}_{12})^2 \alpha_0 m_{-}}{\nu}\bigr) (\alpha_2^0)^2 + (-18 \alpha_0 + \frac{9 \alpha_0 - 9 \alpha_0 m_{-}}{\nu}) \delta_2\Bigr)  \nn \\
& \hspace{1cm}+  {\nu_2^{(0)}}_{|{\rm EsGB}}\Bigl(\bigl(\tfrac{9}{2} (2 + \bar{\gamma}_{12})^2 -  \frac{\tfrac{9}{4} (2 + \bar{\gamma}_{12})^2 + \tfrac{9}{4} (2 + \bar{\gamma}_{12})^2 m_{-}}{\nu}\bigr) \alpha_1^0 \nn \\
& \hspace{1cm}+ \bigl(\tfrac{9}{2} (2 + \bar{\gamma}_{12})^2 \alpha_0 -  \frac{\tfrac{9}{4} (2 + \bar{\gamma}_{12})^2 \alpha_0 + \tfrac{9}{4} (2 + \bar{\gamma}_{12})^2 \alpha_0 m_{-}}{\nu}\bigr) (\alpha_1^0)^2+ (-18 \alpha_0 + \frac{9 \alpha_0 + 9 \alpha_0 m_{-}}{\nu}) \delta_1\Bigr)\biggr) \nn \\
& \hspace{1cm}+ \frac{M}{c^6 r^5} \Biggl(n^{i} \biggl(\Bigl(\bigl(84 - 84 m_{-} + (-6 + 6 m_{-}) \nu + 36 \nu^2\bigr) \delta_2 {\lambda_1^{(0)}}_{|{\rm EsGB}} + \bigl(84 + 84 m_{-} + (-6 - 6 m_{-}) \nu + 36 \nu^2\bigr) \delta_1 {\lambda_2^{(0)}}_{|{\rm EsGB}}\Bigr) (n v)^4 \nn \\
& \hspace{1cm}+ \Bigl(\bigl(- \tfrac{261}{4} + \tfrac{261}{4} m_{-} + (\tfrac{9}{2} -  \tfrac{9}{2} m_{-}) \nu - 24 \nu^2\bigr) \delta_2 {\lambda_1^{(0)}}_{|{\rm EsGB}} + \bigl(- \tfrac{261}{4} -  \tfrac{261}{4} m_{-} + (\tfrac{9}{2} + \tfrac{9}{2} m_{-}) \nu - 24 \nu^2\bigr) \delta_1 {\lambda_2^{(0)}}_{|{\rm EsGB}}\Bigr) (n v)^2 v^2 \nn \\
& \hspace{1cm}+ \Bigl(\bigl(\tfrac{53}{8} -  \tfrac{53}{8} m_{-} + (- \tfrac{11}{2} + \tfrac{7}{2} m_{-}) \nu + 8 \nu^2\bigr) \delta_2 {\lambda_1^{(0)}}_{|{\rm EsGB}} + \bigl(\tfrac{53}{8} + \tfrac{53}{8} m_{-} + (- \tfrac{11}{2} -  \tfrac{7}{2} m_{-}) \nu + 8 \nu^2\bigr) \delta_1 {\lambda_2^{(0)}}_{|{\rm EsGB}}\Bigr) v^4 \biggr) \nn \\
& \hspace{1cm}+ v^{i}\biggl(\Bigl(\bigl(- \tfrac{69}{2} + \tfrac{69}{2} m_{-} + (3 - 3 m_{-}) \nu - 12 \nu^2\bigr) \delta_2 {\lambda_1^{(0)}}_{|{\rm EsGB}} + \bigl(- \tfrac{69}{2} -  \tfrac{69}{2} m_{-} + (3 + 3 m_{-}) \nu - 12 \nu^2\bigr) \delta_1 {\lambda_2^{(0)}}_{|{\rm EsGB}}\Bigr) (n v)^3 \nn \\
& \hspace{1cm}+ \Bigl(\bigl(17 - 17 m_{-} + (1 + m_{-}) \nu + 4 \nu^2\bigr) \delta_2 {\lambda_1^{(0)}}_{|{\rm EsGB}} + \bigl(17 + 17 m_{-} + (1 -  m_{-}) \nu + 4 \nu^2\bigr) \delta_1 {\lambda_2^{(0)}}_{|{\rm EsGB}}\Bigr) (n v) v^2\biggr) \Biggr)\Biggr] \,, \\
& {a_{\rm CM,NNLO}^{(3),i}}_{|{\rm EsGB}}=\frac{G_{12}^3 M^2}{c^6 r^6} \Biggl[n^{i} \Biggl(\biggl(\Bigl(\frac{\tfrac{1}{8} (32 + 25 \bar{\gamma}_{12}) + \tfrac{1}{8} (-32 - 25 \bar{\gamma}_{12}) m_{-}}{\nu} + \tfrac{83}{4} \nu^2 + \frac{m_{-} (1114 \bar{\gamma}_{12} + 1265 \bar{\gamma}_{12}^2 + 1120 \bar{\beta}_2)}{16 \bar{\gamma}_{12}} \nn \\
& \hspace{1cm}-  \frac{1242 \bar{\gamma}_{12} + 1365 \bar{\gamma}_{12}^2 + 1120 \bar{\beta}_2}{16 \bar{\gamma}_{12}} + \nu \bigl(\frac{m_{-} (204 \bar{\gamma}_{12} + 105 \bar{\gamma}_{12}^2 + 560 \bar{\beta}_2)}{16 \bar{\gamma}_{12}} + \frac{3 (888 \bar{\gamma}_{12} + 695 \bar{\gamma}_{12}^2 + 560 \bar{\beta}_2)}{16 \bar{\gamma}_{12}}\bigr)\Bigr) \delta_2 {\lambda_1^{(0)}}_{|{\rm EsGB}} \nn \\
& \hspace{1cm}+ \Bigl(- \tfrac{7}{16} (2 + \bar{\gamma}_{12}) -  \tfrac{21}{16} (2 + \bar{\gamma}_{12}) m_{-} + \frac{\tfrac{7}{8} (2 + \bar{\gamma}_{12}) -  \tfrac{7}{8} (2 + \bar{\gamma}_{12}) m_{-}}{\nu} + \bigl(- \tfrac{105}{16} (2 + \bar{\gamma}_{12}) + \tfrac{35}{16} (2 + \bar{\gamma}_{12}) m_{-}\bigr) \nu\Bigr) \alpha_2^0 \delta_2 {\lambda_1^{(1)}}_{|{\rm EsGB}} \nn \\
& \hspace{1cm}+ \Bigl(\frac{\tfrac{1}{8} (32 + 25 \bar{\gamma}_{12}) + \tfrac{1}{8} (32 + 25 \bar{\gamma}_{12}) m_{-}}{\nu} + \tfrac{83}{4} \nu^2 -  \frac{m_{-} (1114 \bar{\gamma}_{12} + 1265 \bar{\gamma}_{12}^2 + 1120 \bar{\beta}_1)}{16 \bar{\gamma}_{12}} -  \frac{1242 \bar{\gamma}_{12} + 1365 \bar{\gamma}_{12}^2 + 1120 \bar{\beta}_1}{16 \bar{\gamma}_{12}} \nn \\
& \hspace{1cm}+ \nu \bigl(- \frac{m_{-} (204 \bar{\gamma}_{12} + 105 \bar{\gamma}_{12}^2 + 560 \bar{\beta}_1)}{16 \bar{\gamma}_{12}} + \frac{3 (888 \bar{\gamma}_{12} + 695 \bar{\gamma}_{12}^2 + 560 \bar{\beta}_1)}{16 \bar{\gamma}_{12}}\bigr)\Bigr) \delta_1 {\lambda_2^{(0)}}_{|{\rm EsGB}} \nn \\
& \hspace{1cm}+ \Bigl(- \tfrac{7}{16} (2 + \bar{\gamma}_{12}) + \tfrac{21}{16} (2 + \bar{\gamma}_{12}) m_{-} + \frac{\tfrac{7}{8} (2 + \bar{\gamma}_{12}) + \tfrac{7}{8} (2 + \bar{\gamma}_{12}) m_{-}}{\nu} \nn \\
& \hspace{1cm}+ \bigl(- \tfrac{105}{16} (2 + \bar{\gamma}_{12}) -  \tfrac{35}{16} (2 + \bar{\gamma}_{12}) m_{-}\bigr) \nu\Bigr) \alpha_1^0 \delta_1 {\lambda_2^{(1)}}_{|{\rm EsGB}}\biggr) (n v)^2 \nn \\
& \hspace{1cm}+ \biggl(\Bigl(\frac{\tfrac{1}{8} (8 + 5 \bar{\gamma}_{12}) + \tfrac{1}{8} (-8 - 5 \bar{\gamma}_{12}) m_{-}}{\nu} -  \tfrac{7}{2} \nu^2 -  \frac{51 \bar{\gamma}_{12} + \bar{\gamma}_{12}^2 - 20 \bar{\beta}_2}{4 \bar{\gamma}_{12}} -  \frac{m_{-} (-43 \bar{\gamma}_{12} + 4 \bar{\gamma}_{12}^2 + 20 \bar{\beta}_2)}{4 \bar{\gamma}_{12}} \nn \\
& \hspace{1cm}+ \nu \bigl(- \frac{15 m_{-} (4 \bar{\gamma}_{12} + 3 \bar{\gamma}_{12}^2 + 16 \bar{\beta}_2)}{8 \bar{\gamma}_{12}} + \frac{160 \bar{\gamma}_{12} + 59 \bar{\gamma}_{12}^2 + 80 \bar{\beta}_2}{8 \bar{\gamma}_{12}}\bigr)\Bigr) \delta_2 {\lambda_1^{(0)}}_{|{\rm EsGB}} \nn \\
& \hspace{1cm}+ \Bigl(-3 (2 + \bar{\gamma}_{12}) + \tfrac{9}{4} (2 + \bar{\gamma}_{12}) m_{-} + \frac{\tfrac{3}{8} (2 + \bar{\gamma}_{12}) -  \tfrac{3}{8} (2 + \bar{\gamma}_{12}) m_{-}}{\nu} + \bigl(\tfrac{45}{8} (2 + \bar{\gamma}_{12}) -  \tfrac{15}{8} (2 + \bar{\gamma}_{12}) m_{-}\bigr) \nu\Bigr) \alpha_2^0 \delta_2 {\lambda_1^{(1)}}_{|{\rm EsGB}} \nn \\
& \hspace{1cm}+ \Bigl(\frac{\tfrac{1}{8} (8 + 5 \bar{\gamma}_{12}) + \tfrac{1}{8} (8 + 5 \bar{\gamma}_{12}) m_{-}}{\nu} -  \tfrac{7}{2} \nu^2 -  \frac{51 \bar{\gamma}_{12} + \bar{\gamma}_{12}^2 - 20 \bar{\beta}_1}{4 \bar{\gamma}_{12}} + \frac{m_{-} (-43 \bar{\gamma}_{12} + 4 \bar{\gamma}_{12}^2 + 20 \bar{\beta}_1)}{4 \bar{\gamma}_{12}} \nn \\
& \hspace{1cm}+ \nu \bigl(\frac{15 m_{-} (4 \bar{\gamma}_{12} + 3 \bar{\gamma}_{12}^2 + 16 \bar{\beta}_1)}{8 \bar{\gamma}_{12}} + \frac{160 \bar{\gamma}_{12} + 59 \bar{\gamma}_{12}^2 + 80 \bar{\beta}_1}{8 \bar{\gamma}_{12}}\bigr)\Bigr) \delta_1 {\lambda_2^{(0)}}_{|{\rm EsGB}} \nn \\
& \hspace{1cm}+ \Bigl(-3 (2 + \bar{\gamma}_{12}) -  \tfrac{9}{4} (2 + \bar{\gamma}_{12}) m_{-} + \frac{\tfrac{3}{8} (2 + \bar{\gamma}_{12}) + \tfrac{3}{8} (2 + \bar{\gamma}_{12}) m_{-}}{\nu} + \bigl(\tfrac{45}{8} (2 + \bar{\gamma}_{12}) + \tfrac{15}{8} (2 + \bar{\gamma}_{12}) m_{-}\bigr) \nu\Bigr) \alpha_1^0 \delta_1 {\lambda_2^{(1)}}_{|{\rm EsGB}}\biggr) v^2\Biggr) \nn \\
& \hspace{1cm}+ \biggl(\Bigl(-13 \nu^2 + \frac{116 \bar{\gamma}_{12} + 107 \bar{\gamma}_{12}^2 + 40 \bar{\beta}_2}{4 \bar{\gamma}_{12}} -  \frac{m_{-} (116 \bar{\gamma}_{12} + 107 \bar{\gamma}_{12}^2 + 40 \bar{\beta}_2)}{4 \bar{\gamma}_{12}} \nn \\
& \hspace{1cm}+ \nu \bigl(- \frac{m_{-} (32 \bar{\gamma}_{12} + 15 \bar{\gamma}_{12}^2 + 80 \bar{\beta}_2)}{4 \bar{\gamma}_{12}} -  \frac{276 \bar{\gamma}_{12} + 199 \bar{\gamma}_{12}^2 + 80 \bar{\beta}_2}{4 \bar{\gamma}_{12}}\bigr)\Bigr) \delta_2 {\lambda_1^{(0)}}_{|{\rm EsGB}} \nn \\
& \hspace{1cm}+ \Bigl(- \tfrac{5}{4} (2 + \bar{\gamma}_{12}) + \tfrac{5}{4} (2 + \bar{\gamma}_{12}) m_{-} + \bigl(\tfrac{15}{4} (2 + \bar{\gamma}_{12}) -  \tfrac{5}{4} (2 + \bar{\gamma}_{12}) m_{-}\bigr) \nu\Bigr) \alpha_2^0 \delta_2 {\lambda_1^{(1)}}_{|{\rm EsGB}} \nn \\
& \hspace{1cm}+ \Bigl(-13 \nu^2 + \frac{116 \bar{\gamma}_{12} + 107 \bar{\gamma}_{12}^2 + 40 \bar{\beta}_1}{4 \bar{\gamma}_{12}} + \frac{m_{-} (116 \bar{\gamma}_{12} + 107 \bar{\gamma}_{12}^2 + 40 \bar{\beta}_1)}{4 \bar{\gamma}_{12}} \nn \\
& \hspace{1cm}+ \nu \bigl(\frac{m_{-} (32 \bar{\gamma}_{12} + 15 \bar{\gamma}_{12}^2 + 80 \bar{\beta}_1)}{4 \bar{\gamma}_{12}} -  \frac{276 \bar{\gamma}_{12} + 199 \bar{\gamma}_{12}^2 + 80 \bar{\beta}_1}{4 \bar{\gamma}_{12}}\bigr)\Bigr) \delta_1 {\lambda_2^{(0)}}_{|{\rm EsGB}} \nn \\
& \hspace{1cm}+ \Bigl(- \tfrac{5}{4} (2 + \bar{\gamma}_{12}) -  \tfrac{5}{4} (2 + \bar{\gamma}_{12}) m_{-} + \bigl(\tfrac{15}{4} (2 + \bar{\gamma}_{12}) + \tfrac{5}{4} (2 + \bar{\gamma}_{12}) m_{-}\bigr) \nu\Bigr) \alpha_1^0 \delta_1 {\lambda_2^{(1)}}_{|{\rm EsGB}}\biggr) (n v) v^{i}\Biggr]  \,,\\
& {a_{\rm CM,NNLO}^{(4),i}}_{|{\rm EsGB}}= \frac{G_{12}^4 M^3}{c^6 r^7} n^{i} \Biggl[\alpha_2^0 \Bigl(\frac{\tfrac{1}{8} (2 + \bar{\gamma}_{12}) (34 + 19 \bar{\gamma}_{12}) -  \tfrac{1}{8} (2 + \bar{\gamma}_{12}) (34 + 19 \bar{\gamma}_{12}) m_{-}}{\nu} + \frac{(2 + \bar{\gamma}_{12}) m_{-} (9 \bar{\gamma}_{12} + 14 \bar{\gamma}_{12}^2 + 36 \bar{\beta}_2)}{4 \bar{\gamma}_{12}} \nn \\
& \hspace{1cm}-  \frac{(2 + \bar{\gamma}_{12}) (43 \bar{\gamma}_{12} + 33 \bar{\gamma}_{12}^2 + 36 \bar{\beta}_2)}{4 \bar{\gamma}_{12}} + \nu \bigl(2 (2 + \bar{\gamma}_{12}) m_{-} + \frac{3 (2 + \bar{\gamma}_{12}) (-8 \bar{\gamma}_{12} + 3 \bar{\gamma}_{12}^2 + 24 \bar{\beta}_2)}{4 \bar{\gamma}_{12}}\bigr)\Bigr) \delta_2 {\lambda_1^{(1)}}_{|{\rm EsGB}} \nn \\
& \hspace{1cm}+ \bigl(- \tfrac{3}{4} (2 + \bar{\gamma}_{12})^2 + \tfrac{3}{8} (2 + \bar{\gamma}_{12})^2 m_{-} + \frac{\tfrac{3}{16} (2 + \bar{\gamma}_{12})^2 -  \tfrac{3}{16} (2 + \bar{\gamma}_{12})^2 m_{-}}{\nu} + \tfrac{3}{8} (2 + \bar{\gamma}_{12})^2 \nu\bigr) (\alpha_2^0)^2 \delta_2 {\lambda_1^{(2)}}_{|{\rm EsGB}} \nn \\
& \hspace{1cm}+ \alpha_1^0 \Bigl(\frac{\tfrac{1}{8} (2 + \bar{\gamma}_{12}) (34 + 19 \bar{\gamma}_{12}) + \tfrac{1}{8} (2 + \bar{\gamma}_{12}) (34 + 19 \bar{\gamma}_{12}) m_{-}}{\nu} -  \frac{(2 + \bar{\gamma}_{12}) m_{-} (9 \bar{\gamma}_{12} + 14 \bar{\gamma}_{12}^2 + 36 \bar{\beta}_1)}{4 \bar{\gamma}_{12}} \nn \\
& \hspace{1cm}-  \frac{(2 + \bar{\gamma}_{12}) (43 \bar{\gamma}_{12} + 33 \bar{\gamma}_{12}^2 + 36 \bar{\beta}_1)}{4 \bar{\gamma}_{12}} + \nu \bigl(-2 (2 + \bar{\gamma}_{12}) m_{-} + \frac{3 (2 + \bar{\gamma}_{12}) (-8 \bar{\gamma}_{12} + 3 \bar{\gamma}_{12}^2 + 24 \bar{\beta}_1)}{4 \bar{\gamma}_{12}}\bigr)\Bigr) \delta_1 {\lambda_2^{(1)}}_{|{\rm EsGB}} \nn \\
& \hspace{1cm}+ \bigl(- \tfrac{3}{4} (2 + \bar{\gamma}_{12})^2 -  \tfrac{3}{8} (2 + \bar{\gamma}_{12})^2 m_{-} + \frac{\tfrac{3}{16} (2 + \bar{\gamma}_{12})^2 + \tfrac{3}{16} (2 + \bar{\gamma}_{12})^2 m_{-}}{\nu} + \tfrac{3}{8} (2 + \bar{\gamma}_{12})^2 \nu\bigr) (\alpha_1^0)^2 \delta_1 {\lambda_2^{(2)}}_{|{\rm EsGB}} \nn \\
& \hspace{1cm}+ \delta_1 {\lambda_2^{(0)}}_{|{\rm EsGB}} \Bigl(\bigl(-3 (2 + \bar{\gamma}_{12})^2 -  \tfrac{3}{2} (2 + \bar{\gamma}_{12})^2 m_{-} + \frac{\tfrac{3}{4} (2 + \bar{\gamma}_{12})^2 + \tfrac{3}{4} (2 + \bar{\gamma}_{12})^2 m_{-}}{\nu} + \tfrac{3}{2} (2 + \bar{\gamma}_{12})^2 \nu\bigr) (\alpha_1^0)^2 \nn \\
& \hspace{1cm}+ \frac{3 (2 + \bar{\gamma}_{12})^2 (-4 + 3 \bar{\gamma}_{12}) \nu (\alpha_2^0)^2}{10 \bar{\gamma}_{12}} + \frac{\tfrac{1}{16} (432 + 476 \bar{\gamma}_{12} + 131 \bar{\gamma}_{12}^2 - 8 \bar{\beta}_2) + \tfrac{1}{16} m_{-} (432 + 476 \bar{\gamma}_{12} + 131 \bar{\gamma}_{12}^2 - 8 \bar{\beta}_2)}{\nu}\nn \\
& \hspace{1cm} -  \frac{m_{-} (-334 \bar{\gamma}_{12} + 14 \bar{\gamma}_{12}^2 + 69 \bar{\gamma}_{12}^3 + 48 \zeta_{12} + 752 \bar{\beta}_1 + 300 \bar{\gamma}_{12} \bar{\beta}_1 - 92 \bar{\gamma}_{12} \bar{\beta}_2)}{8 \bar{\gamma}_{12}} \nn \\
& \hspace{1cm}-  \frac{49 \bar{\gamma}_{12} + 245 \bar{\gamma}_{12}^2 + 100 \bar{\gamma}_{12}^3 + 24 \zeta_{12} + 376 \bar{\beta}_1 + 150 \bar{\gamma}_{12} \bar{\beta}_1 - 50 \bar{\gamma}_{12} \bar{\beta}_2}{4 \bar{\gamma}_{12}} \nn \\
& \hspace{1cm}+ \nu \bigl(- \frac{m_{-} (11 \bar{\gamma}_{12} + 12 \bar{\gamma}_{12}^2 + 64 \bar{\beta}_1 - 11 \bar{\gamma}_{12} \bar{\beta}_1 + 11 \bar{\gamma}_{12} \bar{\beta}_2)}{2 \bar{\gamma}_{12}} \nn \\
& \hspace{1cm}-  \frac{4616 \bar{\gamma}_{12}^2 + 1696 \bar{\gamma}_{12}^3 - 231 \bar{\gamma}_{12}^4 - 480 \bar{\gamma}_{12} \zeta_{12} - 1920 \bar{\gamma}_{12} \bar{\beta}_1 - 1700 \bar{\gamma}_{12}^2 \bar{\beta}_1 - 1920 (\bar{\beta}_1)^2 + 1100 \bar{\gamma}_{12}^2 \bar{\beta}_2 + 240 \bar{\gamma}_{12} \epsilon_1}{40 \bar{\gamma}_{12}^2}\bigr)\Bigr) \nn \\
& \hspace{1cm}+ \delta_2 {\lambda_1^{(0)}}_{|{\rm EsGB}} \Bigl(\frac{3 (2 + \bar{\gamma}_{12})^2 (-4 + 3 \bar{\gamma}_{12}) \nu (\alpha_1^0)^2}{10 \bar{\gamma}_{12}} + \bigl(-3 (2 + \bar{\gamma}_{12})^2 + \tfrac{3}{2} (2 + \bar{\gamma}_{12})^2 m_{-} + \frac{\tfrac{3}{4} (2 + \bar{\gamma}_{12})^2 -  \tfrac{3}{4} (2 + \bar{\gamma}_{12})^2 m_{-}}{\nu} \nn \\
& \hspace{1cm}+ \tfrac{3}{2} (2 + \bar{\gamma}_{12})^2 \nu\bigr) (\alpha_2^0)^2 + \frac{\tfrac{1}{16} (432 + 476 \bar{\gamma}_{12} + 131 \bar{\gamma}_{12}^2 - 8 \bar{\beta}_1) + \tfrac{1}{16} m_{-} (-432 - 476 \bar{\gamma}_{12} - 131 \bar{\gamma}_{12}^2 + 8 \bar{\beta}_1)}{\nu} \nn \\
& \hspace{1cm}-  \frac{49 \bar{\gamma}_{12} + 245 \bar{\gamma}_{12}^2 + 100 \bar{\gamma}_{12}^3 + 24 \zeta_{12} - 50 \bar{\gamma}_{12} \bar{\beta}_1 + 376 \bar{\beta}_2 + 150 \bar{\gamma}_{12} \bar{\beta}_2}{4 \bar{\gamma}_{12}} \nn \\
& \hspace{1cm}+ \frac{m_{-} (-334 \bar{\gamma}_{12} + 14 \bar{\gamma}_{12}^2 + 69 \bar{\gamma}_{12}^3 + 48 \zeta_{12} - 92 \bar{\gamma}_{12} \bar{\beta}_1 + 752 \bar{\beta}_2 + 300 \bar{\gamma}_{12} \bar{\beta}_2)}{8 \bar{\gamma}_{12}} \nn \\
& \hspace{1cm}+ \nu \bigl(\frac{m_{-} (11 \bar{\gamma}_{12} + 12 \bar{\gamma}_{12}^2 + 11 \bar{\gamma}_{12} \bar{\beta}_1 + 64 \bar{\beta}_2 - 11 \bar{\gamma}_{12} \bar{\beta}_2)}{2 \bar{\gamma}_{12}} \nn \\
& \hspace{1cm}-  \frac{4616 \bar{\gamma}_{12}^2 + 1696 \bar{\gamma}_{12}^3 - 231 \bar{\gamma}_{12}^4 - 480 \bar{\gamma}_{12} \zeta_{12} + 1100 \bar{\gamma}_{12}^2 \bar{\beta}_1 - 1920 \bar{\gamma}_{12} \bar{\beta}_2 - 1700 \bar{\gamma}_{12}^2 \bar{\beta}_2 - 1920 (\bar{\beta}_2)^2 + 240 \bar{\gamma}_{12} \epsilon_2}{40 \bar{\gamma}_{12}^2}\bigr)\Bigr)\Biggr] \,.
\end{align}
\end{subequations}

\section{Computation of diagrams at NNLO order in the PNEFT formalism}


\subsection{$G^2 m_1^2\lambda_{2}^{(0)}v^4$}
The values of the diagrams are normalized with $2\frac{G^2m_1^2\lambda_{2}^{(0)}}{r^4}\tilde{f}_0^{(2)}(\tilde{d}_1^{(1)})^2$. We have to take into account also the mirror images $(1\leftrightarrow 2)$.

\begin{figure}[H]
\centering

\begin{tikzpicture}[scale=0.6]
\begin{feynman}
\vertex  (a) at (0,0);
\vertex (b) at (-1,1);
\vertex (c) at (1,1);
\vertex (d) at (-1,-1);
\vertex  (e) at (1,-1);
\vertex (g) at (0,-1);
\vertex (f) at (0,1);
\diagram{
(d)--[horizontal,very thick](e);
(b)--[horizontal,very thick](c);
(d)--[horizontal,red](f);
(e)--[horizontal,red](f);
};
\end{feynman}
\node[below=0.7cm,scale=0.8]{(a)};
\end{tikzpicture}
\begin{tikzpicture}[scale=0.6]
\begin{feynman}
\vertex  (a) at (0,0);
\vertex (b) at (-1,1);
\vertex (c) at (1,1);
\vertex (d) at (-1,-1);
\vertex  (e) at (1,-1);
\vertex (g) at (0,-1);
\vertex (f) at (0,1);
\diagram{
(d)--[horizontal,very thick](e);
(b)--[horizontal,very thick](c);
(d)--[horizontal](f);
(e)--[horizontal](f);
};
\end{feynman}
\node[below=0.7cm,scale=0.8]{(b)};
\end{tikzpicture}
\begin{tikzpicture}[scale=0.6]
\begin{feynman}
\vertex  (a) at (0,0);
\vertex (b) at (-1,1);
\vertex (c) at (1,1);
\vertex (d) at (-1,-1);
\vertex  (e) at (1,-1);
\vertex (g) at (0,-1);
\vertex (f) at (0,1);
\diagram{
(d)--[horizontal,very thick](e);
(b)--[horizontal,very thick](c);
(d)--[horizontal,red](f);
(e)--[horizontal](f);
};
\end{feynman}
\node[below=0.7cm,scale=0.8]{(c)};
\end{tikzpicture}

\caption{Diagrams contributing at order $G^2m_1^2\lambda_2^{(0)}(1+v^2+v^4)$}
\end{figure}

\begin{align}
Fig.(1a)=&-1+\nn\\
&+nv_1(nv_1+2nv_2)-nv_2^2+\frac{v_2^2}{2}\nn\\
&+r^2\Big(2na_1\ na_2-\frac{a_1a_2}{2}-na_1^2\Big)+\nn\\
&+2r\Big((a_1v_2-a_1v_1)(nv_1-nv_2)-na_1(v_2^2+v_1^2-2v_1v_2-\frac{7}{2}(nv_1^2+nv_2^2)+6nv_1\ nv_2)\Big)\nn\\
&+\frac{v_2^4}{8}+nv_2^2(3nv_1^2-\frac{v_2^2}{2})+nv_1^2(2v_1v_2-3\frac{v_2^2}{2})+nv_1\ nv_2(v_2^2+2v_1^2-2v_1v_2-6nv_1^2)-\nn\\
&-\frac{3}{4r^2}(4\mu_2^{(0)}+4\nu_2^{(0)}-c_2^{(0)})\nn\\
Fig.(1b)=&\frac{3}{4r^2}\frac{c_2^{(0)}}{\tilde{f}_0^{(2)}(\tilde{d}_1^{(1)})^2}\nn\\
Fig.(1c)=&-\frac{3}{2r^2}\frac{(c_2^{(0)}-2\nu_2^{(0)})}{\tilde{f}_0^{(2)}\tilde{d}_1^{(1)}}\frac{\sqrt{2}}{\sqrt{3+2\omega_0}}\nn\\
\end{align}


\subsection{$G^3 m_1^2m_2\lambda_{(2)}^0\tilde{f}^{(2)}_nv^2$}

The values of the diagrams are normalized with $2\frac{G^3m_1^2m_2\lambda_{(2)}^0}{r^5}\tilde{f}_0^{(2)}(\tilde{d}_1^{(1)})^2$. We have to take into account also the mirror images $(1\leftrightarrow 2)$.
\begin{figure}[H]
\centering

\begin{tikzpicture}[scale=0.6]
\begin{feynman}
\vertex  (a) at (0,0);
\vertex (b) at (-1,1);
\vertex (c) at (1,1);
\vertex (d) at (-1,-1);
\vertex  (e) at (1,-1);
\vertex (g) at (0,-1);
\vertex (f) at (0,1);
\diagram{
(d)--[horizontal,very thick](e);
(b)--[horizontal,very thick](c);
(d)--[horizontal,red](b);
(d)--[horizontal,red](c);
(e)--[horizontal,red](c);
};
\end{feynman}
\node[below=0.7cm,scale=0.8]{(1a)};
\end{tikzpicture}
\begin{tikzpicture}[scale=0.6]
\begin{feynman}
\vertex  (a) at (0,0);
\vertex (b) at (-1,1);
\vertex (c) at (1,1);
\vertex (d) at (-1,-1);
\vertex  (e) at (1,-1);
\vertex (g) at (0,-1);
\vertex (f) at (0,1);
\diagram{
(d)--[horizontal,very thick](e);
(b)--[horizontal,very thick](c);
(d)--[horizontal,red](b);
(d)--[horizontal,red](c);
(e)--[horizontal](c);
};
\end{feynman}
\node[below=0.7cm,scale=0.8]{(1b)};
\end{tikzpicture}
\begin{tikzpicture}[scale=0.6]
\begin{feynman}
\vertex  (a) at (0,0);
\vertex (b) at (-1,1);
\vertex (c) at (1,1);
\vertex (d) at (-1,-1);
\vertex  (e) at (1,-1);
\vertex (g) at (0,-1);
\vertex (f) at (0,1);
\diagram{
(d)--[horizontal,very thick](e);
(b)--[horizontal,very thick](c);
(d)--[horizontal,red](b);
(d)--[horizontal,red](c);
(e)--[horizontal,dashed](c);
};
\end{feynman}
\node[below=0.7cm,scale=0.8]{(1c)};
\end{tikzpicture}

\begin{tikzpicture}[scale=0.6]
\begin{feynman}
\vertex  (a) at (0,0);
\vertex (b) at (-1,1);
\vertex (c) at (1,1);
\vertex (d) at (-1,-1);
\vertex  (e) at (1,-1);
\vertex (g) at (0,-1);
\vertex (f) at (0,1);
\diagram{
(d)--[horizontal,very thick](e);
(b)--[horizontal,very thick](c);
(d)--[horizontal,red](b);
(b)--[horizontal,red](a);
(c)--[horizontal,red](a);
(g)--[horizontal,red](a);
};
\end{feynman}
\node[below=0.7cm,scale=0.8]{(2a)};
\end{tikzpicture}
\begin{tikzpicture}[scale=0.6]
\begin{feynman}
\vertex  (a) at (0,0);
\vertex (b) at (-1,1);
\vertex (c) at (1,1);
\vertex (d) at (-1,-1);
\vertex  (e) at (1,-1);
\vertex (g) at (0,-1);
\vertex (f) at (0,1);
\diagram{
(d)--[horizontal,very thick](e);
(b)--[horizontal,very thick](c);
(d)--[horizontal,red](b);
(b)--[horizontal,red](a);
(c)--[horizontal](a);
(g)--[horizontal,red](a);
};
\end{feynman}
\node[below=0.7cm,scale=0.8]{(2b)};
\end{tikzpicture}
\begin{tikzpicture}[scale=0.6]
\begin{feynman}
\vertex  (a) at (0,0);
\vertex (b) at (-1,1);
\vertex (c) at (1,1);
\vertex (d) at (-1,-1);
\vertex  (e) at (1,-1);
\vertex (g) at (0,-1);
\vertex (f) at (0,1);
\diagram{
(d)--[horizontal,very thick](e);
(b)--[horizontal,very thick](c);
(d)--[horizontal,red](b);
(b)--[horizontal,red](a);
(c)--[horizontal,red](a);
(g)--[horizontal](a);
};
\end{feynman}
\node[below=0.7cm,scale=0.8]{(2c)};
\end{tikzpicture}
\begin{tikzpicture}[scale=0.6]
\begin{feynman}
\vertex  (a) at (0,0);
\vertex (b) at (-1,1);
\vertex (c) at (1,1);
\vertex (d) at (-1,-1);
\vertex  (e) at (1,-1);
\vertex (g) at (0,-1);
\vertex (f) at (0,1);
\diagram{
(d)--[horizontal,very thick](e);
(b)--[horizontal,very thick](c);
(d)--[horizontal,red](b);
(b)--[horizontal,red](a);
(c)--[horizontal,dashed](a);
(g)--[horizontal,red](a);
};
\end{feynman}
\node[below=0.7cm,scale=0.8]{(2d)};
\end{tikzpicture}
\begin{tikzpicture}[scale=0.6]
\begin{feynman}
\vertex  (a) at (0,0);
\vertex (b) at (-1,1);
\vertex (c) at (1,1);
\vertex (d) at (-1,-1);
\vertex  (e) at (1,-1);
\vertex (g) at (0,-1);
\vertex (f) at (0,1);
\diagram{
(d)--[horizontal,very thick](e);
(b)--[horizontal,very thick](c);
(d)--[horizontal,red](b);
(b)--[horizontal,red](a);
(c)--[horizontal,red](a);
(g)--[horizontal,dashed](a);
};
\end{feynman}
\node[below=0.7cm,scale=0.8]{(2e)};
\end{tikzpicture}
\begin{tikzpicture}[scale=0.6]
\begin{feynman}
\vertex  (a) at (0,0);
\vertex (b) at (-1,1);
\vertex (c) at (1,1);
\vertex (d) at (-1,-1);
\vertex  (e) at (1,-1);
\vertex (g) at (0,-1);
\vertex (f) at (0,1);
\diagram{
(d)--[horizontal,very thick](e);
(b)--[horizontal,very thick](c);
(d)--[horizontal,red](b);
(b)--[horizontal,red](a);
(c)--[horizontal,double](a);
(g)--[horizontal,red](a);
};
\end{feynman}
\node[below=0.7cm,scale=0.8]{(2f)};
\end{tikzpicture}
\begin{tikzpicture}[scale=0.6]
\begin{feynman}
\vertex  (a) at (0,0);
\vertex (b) at (-1,1);
\vertex (c) at (1,1);
\vertex (d) at (-1,-1);
\vertex  (e) at (1,-1);
\vertex (g) at (0,-1);
\vertex (f) at (0,1);
\diagram{
(d)--[horizontal,very thick](e);
(b)--[horizontal,very thick](c);
(d)--[horizontal,red](b);
(b)--[horizontal,red](a);
(c)--[horizontal,red](a);
(g)--[horizontal,double](a);
};
\end{feynman}
\node[below=0.7cm,scale=0.8]{(2g)};
\end{tikzpicture}

\caption{Diagrams contributing at order $G^3m_1^2m_2\lambda_2^{(0)}(1+v^2)$}
\end{figure}

\begin{align}
Fig.(2.1a)=&2\frac{\tilde{d}_1^{(2)}\tilde{d}_2^{(1)}}{\tilde{d}_1^{(1)}}\Big[2+\nn\\
&+\Big(v_1v_2-2v_2^2+5nv_2(2nv_2-3nv_1)\Big)\Big]\nn\\
Fig.(2.1b)=&2+\nn\\
&+v_1v_2+4v_1^2+2v_2^2+5nv_2(2nv_2-3nv_1) \nn\\
Fig.(2.1c)=&-8v_1v_2 \nn\\
Fig.(2.2a)=&-\tilde{d}_1^{(2)}x_1\Big[2+\nn\\
&+\Big(v_1v_2-2v_2^2+5nv_2(2nv_2-3nv_1)\Big)\Big] \nn\\
Fig.(2.2b)=&-2\Big((v_1v_2-9nv_1\ nv_2)+2(v_1^2+nv_1^2)\Big) \nn\\
Fig.(2.2c)=&2\frac{\tilde{d}_1^{(2)}}{\tilde{d}_1^{(1)}}(v_1v_2-nv_1\ nv_2) \nn\\
Fig.(2.2d)=&2\Big(5(v_1v_2-nv_1\ nv_2)+(v_2^2-9nv_2^2)\Big) \nn\\
Fig.(2.2e)=&2\frac{\tilde{d}_1^{(2)}}{\tilde{d}_1^{(1)}}\Big((v_1^2-nv_1^2)-(v_1v_2-nv_1\ nv_2)\Big) \nn\\
Fig.(2.2f)=&-2(3v_2^2-7nv_2^2) \nn\\
Fig.(2.2g)=&-2\frac{\tilde{d}_1^{(2)}}{\tilde{d}_1^{(1)}}(v_1^2-nv_1^2)\nn\\
\end{align}

\subsection{$G^3 m_1^3\lambda_{(2)}^0\tilde{f}^{(2)}_nv^2$}

The values of the diagrams are normalized with $2\frac{G^3m_1^3\lambda_{(2)}^0}{r^5}\tilde{f}_0^{(2)}(\tilde{d}_1^{(1)})^2$. We have to take into account also the mirror images $(1\leftrightarrow 2)$.

\begin{figure}[H]
\centering

\begin{tikzpicture}[scale=0.6]
\begin{feynman}
\vertex  (a) at (0,0);
\vertex (b) at (-1,1);
\vertex (c) at (1,1);
\vertex (d) at (-1,-1);
\vertex  (e) at (1,-1);
\vertex (g) at (0,-1);
\vertex (f) at (0,1);
\diagram{
(d)--[horizontal,very thick](e);
(b)--[horizontal,very thick](c);
(d)--[horizontal,red](f);
(e)--[horizontal,red](f);
(g)--[horizontal,red](f);
};
\end{feynman}
\node[below=0.7cm,scale=0.8]{(1a)};
\end{tikzpicture}
\begin{tikzpicture}[scale=0.6]
\begin{feynman}
\vertex  (a) at (0,0);
\vertex (b) at (-1,1);
\vertex (c) at (1,1);
\vertex (d) at (-1,-1);
\vertex  (e) at (1,-1);
\vertex (g) at (0,-1);
\vertex (f) at (0,1);
\diagram{
(d)--[horizontal,very thick](e);
(b)--[horizontal,very thick](c);
(d)--[horizontal](f);
(e)--[horizontal,red](f);
(g)--[horizontal,red](f);
};
\end{feynman}
\node[below=0.7cm,scale=0.8]{(1b)};
\end{tikzpicture}
\begin{tikzpicture}[scale=0.6]
\begin{feynman}
\vertex  (a) at (0,0);
\vertex (b) at (-1,1);
\vertex (c) at (1,1);
\vertex (d) at (-1,-1);
\vertex  (e) at (1,-1);
\vertex (g) at (0,-1);
\vertex (f) at (0,1);
\diagram{
(d)--[horizontal,very thick](e);
(b)--[horizontal,very thick](c);
(d)--[horizontal,dashed](f);
(e)--[horizontal,red](f);
(g)--[horizontal,red](f);
};
\end{feynman}
\node[below=0.7cm,scale=0.8]{(1c)};
\end{tikzpicture}
\begin{tikzpicture}[scale=0.6]
\begin{feynman}
\vertex  (a) at (0,0);
\vertex (b) at (-1,1);
\vertex (c) at (1,1);
\vertex (d) at (-1,-1);
\vertex  (e) at (1,-1);
\vertex (g) at (0,-1);
\vertex (f) at (0,1);
\diagram{
(d)--[horizontal,very thick](e);
(b)--[horizontal,very thick](c);
(d)--[horizontal,red](f);
(e)--[horizontal,double](f);
(g)--[horizontal,red](f);
};
\end{feynman}
\node[below=0.7cm,scale=0.8]{(1d)};
\end{tikzpicture}

\begin{tikzpicture}[scale=0.6]
\begin{feynman}
\vertex  (a) at (0,0);
\vertex (b) at (-1,1);
\vertex (c) at (1,1);
\vertex (d) at (-1,-1);
\vertex  (e) at (1,-1);
\vertex (g) at (0,-1);
\vertex (f) at (0,1);
\vertex (i) at (0.5,0);
\diagram{
(d)--[horizontal,very thick](e);
(b)--[horizontal,very thick](c);
(d)--[horizontal,red](f);
(e)--[horizontal,red](f);
(g)--[horizontal,red](i);
};
\end{feynman}
\node[below=0.7cm,scale=0.8]{(2a)};
\end{tikzpicture}
\begin{tikzpicture}[scale=0.6]
\begin{feynman}
\vertex  (a) at (0,0);
\vertex (b) at (-1,1);
\vertex (c) at (1,1);
\vertex (d) at (-1,-1);
\vertex  (e) at (1,-1);
\vertex (g) at (0,-1);
\vertex (f) at (0,1);
\vertex (i) at (0.5,0);
\diagram{
(d)--[horizontal,very thick](e);
(b)--[horizontal,very thick](c);
(d)--[horizontal,red](f);
(e)--[horizontal,red](f);
(g)--[horizontal](i);
};
\end{feynman}
\node[below=0.7cm,scale=0.8]{(2b)};
\end{tikzpicture}
\begin{tikzpicture}[scale=0.6]
\begin{feynman}
\vertex  (a) at (0,0);
\vertex (b) at (-1,1);
\vertex (c) at (1,1);
\vertex (d) at (-1,-1);
\vertex  (e) at (1,-1);
\vertex (g) at (0,-1);
\vertex (f) at (0,1);
\vertex (i) at (0.5,0);
\diagram{
(d)--[horizontal,very thick](e);
(b)--[horizontal,very thick](c);
(d)--[horizontal,red](f);
(e)--[horizontal,red](f);
(g)--[horizontal,dashed](i);
};
\end{feynman}
\node[below=0.7cm,scale=0.8]{(2c)};
\end{tikzpicture}
\begin{tikzpicture}[scale=0.6]
\begin{feynman}
\vertex  (a) at (0,0);
\vertex (b) at (-1,1);
\vertex (c) at (1,1);
\vertex (d) at (-1,-1);
\vertex  (e) at (1,-1);
\vertex (g) at (0,-1);
\vertex (f) at (0,1);
\vertex (i) at (0.5,0);
\diagram{
(d)--[horizontal,very thick](e);
(b)--[horizontal,very thick](c);
(d)--[horizontal,red](f);
(e)--[horizontal,red](f);
(g)--[horizontal,double](i);
};
\end{feynman}
\node[below=0.7cm,scale=0.8]{(2d)};
\end{tikzpicture}

\caption{Diagrams contributing at order $G^3m_1^3\lambda_2^{(0)}(1+v^2)$}
\end{figure}

\begin{align}
Fig.(3.1a)=&\frac{\tilde{f}_1^{(2)}}{\tilde{f}_0^{(2)}}\tilde{d}_1^{(1)}\Big[2+\nn\\
&+\Big(2nv_1^2+2nv_2^2-9nv_1\ nv_2-v_1^2-v_2^2+v_1v_2\Big)\Big]\nn\\
Fig.(3.1b)=&3+\nn\\
&+\frac{1}{2}\Big(v_2^2-2nv_2^2+9v_1^2+6nv_1^2+3v_1v_2-11nv_1\ nv_2\Big) \nn\\
Fig.(3.1c)=&-4\Big(v_1v_2+2nv_1^2\Big) \nn\\
Fig.(3.1d)=&-4\Big(v_1^2-nv_1^2\Big) \nn\\
Fig.(3.2a)=&-\tilde{d}_1^{(1)}x_1\Big[2+\nn\\
&+\Big(2nv_1^2+2nv_2^2-9nv_1\ nv_2-v_1^2-v_2^2+v_1v_2\Big)\Big] \nn\\
Fig.(3.2b)=&-4\Big(v_1^2+nv_1^2+v_1v_2-5nv_1\ nv_2\Big) \nn\\
Fig.(3.2c)=&4\Big(3(v_1^2-nv_1^2)+v_1v_2-5nv_1\ nv_2\Big) \nn\\
Fig.(3.2d)=&-8 \Big(v_1^2-2nv_1^2\Big) \nn\\
\end{align}

\subsection{$G^4m_1^4\lambda_{(2)}^0\tilde{f}^{(2)}_n$}

The values of the diagrams are normalized with $2\frac{G^4m_1^4\lambda_{(2)}^0}{r^6}\tilde{f}_0^{(2)}(\tilde{d}_1^{(1)})^2$. We have to take into account also the mirror images $(1\leftrightarrow 2)$.

\begin{figure}[H]
\centering

\begin{tikzpicture}[scale=0.6]
\begin{feynman}
\vertex  (a) at (0,0);
\vertex (b) at (-1,1);
\vertex (c) at (1,1);
\vertex (d) at (-1,-1);
\vertex  (e) at (1,-1);
\vertex (g) at (-0.33,-1);
\vertex (h) at (0.33,-1);
\vertex (f) at (0,1);
\diagram{
(d)--[horizontal,very thick](e);
(b)--[horizontal,very thick](c);
(d)--[horizontal,red](f);
(e)--[horizontal,red](f);
(h)--[horizontal,red](f);
(g)--[horizontal,red](f);
};
\end{feynman}
\node[below=0.7cm,scale=0.8]{(a)};
\end{tikzpicture}
\begin{tikzpicture}[scale=0.6]
\begin{feynman}
\vertex  (a) at (0,0);
\vertex (b) at (-1,1);
\vertex (c) at (1,1);
\vertex (d) at (-1,-1);
\vertex  (e) at (1,-1);
\vertex (g) at (-0.33,-1);
\vertex (h) at (0.33,-1);
\vertex (f) at (0,1);
\diagram{
(d)--[horizontal,very thick](e);
(b)--[horizontal,very thick](c);
(d)--[horizontal,red](f);
(e)--[horizontal,red](f);
(h)--[horizontal,red](f);
(g)--[horizontal](f);
};
\end{feynman}
\node[below=0.7cm,scale=0.8]{(b)};
\end{tikzpicture}
\begin{tikzpicture}[scale=0.6]
\begin{feynman}
\vertex  (a) at (0,0);
\vertex (b) at (-1,1);
\vertex (c) at (1,1);
\vertex (d) at (-1,-1);
\vertex  (e) at (1,-1);
\vertex (g) at (-0.33,-1);
\vertex (h) at (0.33,-1);
\vertex (f) at (0,1);
\diagram{
(d)--[horizontal,very thick](e);
(b)--[horizontal,very thick](c);
(d)--[horizontal,red](f);
(e)--[horizontal,red](f);
(h)--[horizontal](f);
(g)--[horizontal](f);
};
\end{feynman}
\node[below=0.7cm,scale=0.8]{(c)};
\end{tikzpicture}
\begin{tikzpicture}[scale=0.6]
\begin{feynman}
\vertex  (a) at (0,0);
\vertex (b) at (-1,1);
\vertex (c) at (1,1);
\vertex (d) at (-1,-1);
\vertex  (e) at (1,-1);
\vertex (g) at (-0.33,-1);
\vertex (h) at (0.33,-1);
\vertex (f) at (0,1);
\diagram{
(d)--[horizontal,very thick](e);
(b)--[horizontal,very thick](c);
(d)--[horizontal,red](f);
(e)--[horizontal,red](f);
(h)--[horizontal,red](a);
(g)--[horizontal,red](a);
(f)--[horizontal,red](a);
};
\end{feynman}
\node[below=0.7cm,scale=0.8]{(d)};
\end{tikzpicture}
\begin{tikzpicture}[scale=0.6]
\begin{feynman}
\vertex  (a) at (0,0);
\vertex (b) at (-1,1);
\vertex (c) at (1,1);
\vertex (d) at (-1,-1);
\vertex  (e) at (1,-1);
\vertex (g) at (-0.33,-1);
\vertex (h) at (0.33,-1);
\vertex (f) at (0,1);
\diagram{
(d)--[horizontal,very thick](e);
(b)--[horizontal,very thick](c);
(d)--[horizontal](f);
(e)--[horizontal,red](f);
(h)--[horizontal,red](a);
(g)--[horizontal,red](a);
(f)--[horizontal,red](a);
};
\end{feynman}
\node[below=0.7cm,scale=0.8]{(e)};
\end{tikzpicture}
\begin{tikzpicture}[scale=0.6]
\begin{feynman}
\vertex  (a) at (0,0);
\vertex (b) at (-1,1);
\vertex (c) at (1,1);
\vertex (d) at (-1,-1);
\vertex  (e) at (1,-1);
\vertex (g) at (-0.33,-1);
\vertex (h) at (0.33,-1);
\vertex (f) at (0,1);
\vertex (j) at (-0.51,0);
\vertex (i) at (0.51,0);
\diagram{
(d)--[horizontal,very thick](e);
(b)--[horizontal,very thick](c);
(d)--[horizontal,red](f);
(e)--[horizontal,red](f);
(h)--[horizontal,red](i);
(g)--[horizontal,red](j);
};
\end{feynman}
\node[below=0.7cm,scale=0.8]{(f)};
\end{tikzpicture}
\begin{tikzpicture}[scale=0.6]
\begin{feynman}
\vertex  (a) at (0,0);
\vertex (b) at (-1,1);
\vertex (c) at (1,1);
\vertex (d) at (-1,-1);
\vertex  (e) at (1,-1);
\vertex (g) at (-0.33,-1);
\vertex (h) at (0.33,-1);
\vertex (f) at (0,1);
\vertex (j) at (-0.51,0);
\vertex (i) at (0.51,0);
\diagram{
(d)--[horizontal,very thick](e);
(b)--[horizontal,very thick](c);
(d)--[horizontal,red](f);
(e)--[horizontal,red](f);
(h)--[horizontal,red](i);
(g)--[horizontal,red](i);
};
\end{feynman}
\node[below=0.7cm,scale=0.8]{(g)};
\end{tikzpicture}
\begin{tikzpicture}[scale=0.6]
\begin{feynman}
\vertex  (a) at (0,0);
\vertex (b) at (-1,1);
\vertex (c) at (1,1);
\vertex (d) at (-1,-1);
\vertex  (e) at (1,-1);
\vertex (g) at (-0.33,-1);
\vertex (h) at (0.33,-1);
\vertex (f) at (0,1);
\vertex (j) at (-0.51,0);
\vertex (i) at (0.51,0);
\vertex (k) at (0,-0.33);
\diagram{
(d)--[horizontal,very thick](e);
(b)--[horizontal,very thick](c);
(d)--[horizontal,red](f);
(e)--[horizontal,red](f);
(h)--[horizontal,red](k);
(g)--[horizontal,red](k);
(i)--[horizontal,red](k);
};
\end{feynman}
\node[below=0.7cm,scale=0.8]{(h)};
\end{tikzpicture}
\begin{tikzpicture}[scale=0.6]
\begin{feynman}
\vertex  (a) at (0,0);
\vertex (b) at (-1,1);
\vertex (c) at (1,1);
\vertex (d) at (-1,-1);
\vertex  (e) at (1,-1);
\vertex (g) at (-0.33,-1);
\vertex (h) at (0.33,-1);
\vertex (f) at (0,1);
\vertex (j) at (-0.51,0);
\vertex (i) at (0.51,0);
\vertex (k) at (0,-0.33);
\diagram{
(d)--[horizontal,very thick](e);
(b)--[horizontal,very thick](c);
(d)--[horizontal,red](f);
(e)--[horizontal,red](f);
(h)--[horizontal,red](k);
(g)--[horizontal,red](k);
(i)--[horizontal,double](k);
};
\end{feynman}
\node[below=0.7cm,scale=0.8]{(i)};
\end{tikzpicture}
\begin{tikzpicture}[scale=0.6]
\begin{feynman}
\vertex  (a) at (0,0);
\vertex (b) at (-1,1);
\vertex (c) at (1,1);
\vertex (d) at (-1,-1);
\vertex  (e) at (1,-1);
\vertex (g) at (-0.33,-1);
\vertex (h) at (0.33,-1);
\vertex (f) at (0,1);
\vertex (j) at (-0.51,0);
\vertex (i) at (0.51,0);
\vertex (k) at (0,-0.33);
\diagram{
(d)--[horizontal,very thick](e);
(b)--[horizontal,very thick](c);
(d)--[horizontal,red](f);
(e)--[horizontal,red](f);
(h)--[horizontal](k);
(g)--[horizontal](k);
(i)--[horizontal,double](k);
};
\end{feynman}
\node[below=0.7cm,scale=0.8]{(j)};
\end{tikzpicture}

\caption{Diagrams contributing at order $G^4m_1^4\lambda_2^{(0)}$}
\end{figure}
\begin{align}
Fig.(4a)=&-2\frac{\tilde{f}_2^{(2)}}{\tilde{f}_0^{(2)}}(\tilde{d}_1^{(1)})^2\nn\\
Fig.(4b)=&-6\frac{\tilde{f}_1^{(2)}}{\tilde{f}_0^{(2)}}\tilde{d}_1^{(1)}\nn\\
Fig.(4c)=&-\frac{9}{2}\nn\\
Fig.(4d)=&5\frac{\tilde{f}_1^{(2)}}{\tilde{f}_0^{(2)}}(\tilde{d}_1^{(1)})^2x_1\nn\\
Fig.(4e)=&6\tilde{d}_1^{(1)}x_1\nn\\
Fig.(4f)=&-(\tilde{d}_1^{(1)})^2(x_1)^2\nn\\
Fig.(4g)=&2(\tilde{d}_1^{(1)})^2x_2\nn\\
Fig.(4h)=&-4(\tilde{d}_1^{(1)})^2(x_1)^2\nn\\
Fig.(4i)=&-4(\tilde{d}_1^{(1)})^2\nn\\
Fig.(4j)=&-2\nn\\
\end{align}

\subsection{$G^4m_1^3m_2\lambda_{(2)}^0\tilde{f}^{(2)}_n$}

The values of the diagrams are normalized with $2\frac{G^4m_1^3m_2\lambda_{(2)}^0}{r^6}\tilde{f}_0^{(2)}(\tilde{d}_1^{(1)})^2$. We have to take into account also the mirror images $(1\leftrightarrow 2)$.

\begin{figure}[H]
\centering

\begin{tikzpicture}[scale=0.6]
\begin{feynman}
\vertex  (a) at (0,0);
\vertex (b) at (-1,1);
\vertex (c) at (1,1);
\vertex (d) at (-1,-1);
\vertex  (e) at (1,-1);
\vertex (g) at (-0.33,-1);
\vertex (h) at (0.33,-1);
\vertex (f) at (0,1);
\diagram{
(d)--[horizontal,very thick](e);
(b)--[horizontal,very thick](c);
(g)--[horizontal,red](b);
(e)--[horizontal,red](f);
(h)--[horizontal,red](f);
(g)--[horizontal,red](f);
};
\end{feynman}
\node[below=0.7cm,scale=0.8]{(1a)};
\end{tikzpicture}
\begin{tikzpicture}[scale=0.6]
\begin{feynman}
\vertex  (a) at (0,0);
\vertex (b) at (-1,1);
\vertex (c) at (1,1);
\vertex (d) at (-1,-1);
\vertex  (e) at (1,-1);
\vertex (g) at (-0.33,-1);
\vertex (h) at (0.33,-1);
\vertex (f) at (0,1);
\diagram{
(d)--[horizontal,very thick](e);
(b)--[horizontal,very thick](c);
(g)--[horizontal](b);
(e)--[horizontal,red](f);
(h)--[horizontal,red](f);
(g)--[horizontal,red](f);
};
\end{feynman}
\node[below=0.7cm,scale=0.8]{(1b)};
\end{tikzpicture}
\begin{tikzpicture}[scale=0.6]
\begin{feynman}
\vertex  (a) at (0,0);
\vertex (b) at (-1,1);
\vertex (c) at (1,1);
\vertex (d) at (-1,-1);
\vertex  (e) at (1,-1);
\vertex (g) at (-0.33,-1);
\vertex (h) at (0.33,-1);
\vertex (f) at (0,1);
\diagram{
(d)--[horizontal,very thick](e);
(b)--[horizontal,very thick](c);
(g)--[horizontal,red](b);
(e)--[horizontal](f);
(h)--[horizontal,red](f);
(g)--[horizontal,red](f);
};
\end{feynman}
\node[below=0.7cm,scale=0.8]{(1c)};
\end{tikzpicture}
\begin{tikzpicture}[scale=0.6]
\begin{feynman}
\vertex  (a) at (0,0);
\vertex (b) at (-1,1);
\vertex (c) at (1,1);
\vertex (d) at (-1,-1);
\vertex  (e) at (1,-1);
\vertex (g) at (-0.33,-1);
\vertex (h) at (0.33,-1);
\vertex (f) at (0,1);
\diagram{
(d)--[horizontal,very thick](e);
(b)--[horizontal,very thick](c);
(g)--[horizontal,red](b);
(e)--[horizontal,red](f);
(h)--[horizontal,red](f);
(g)--[horizontal](f);
};
\end{feynman}
\node[below=0.7cm,scale=0.8]{(1d)};
\end{tikzpicture}
\begin{tikzpicture}[scale=0.6]
\begin{feynman}
\vertex  (a) at (0,0);
\vertex (b) at (-1,1);
\vertex (c) at (1,1);
\vertex (d) at (-1,-1);
\vertex  (e) at (1,-1);
\vertex (g) at (-0.33,-1);
\vertex (h) at (0.33,-1);
\vertex (f) at (0,1);
\diagram{
(d)--[horizontal,very thick](e);
(b)--[horizontal,very thick](c);
(g)--[horizontal](b);
(e)--[horizontal](f);
(h)--[horizontal,red](f);
(g)--[horizontal,red](f);
};
\end{feynman}
\node[below=0.7cm,scale=0.8]{(1e)};
\end{tikzpicture}
\begin{tikzpicture}[scale=0.6]
\begin{feynman}
\vertex  (a) at (0,0);
\vertex (b) at (-1,1);
\vertex (c) at (1,1);
\vertex (d) at (-1,-1);
\vertex  (e) at (1,-1);
\vertex (g) at (-0.33,-1);
\vertex (h) at (0.33,-1);
\vertex (f) at (0,1);
\diagram{
(d)--[horizontal,very thick](e);
(b)--[horizontal,very thick](c);
(g)--[horizontal](b);
(e)--[horizontal,red](f);
(h)--[horizontal,red](f);
(g)--[horizontal](f);
};
\end{feynman}
\node[below=0.7cm,scale=0.8]{(1f)};
\end{tikzpicture}
\begin{tikzpicture}[scale=0.6]
\begin{feynman}
\vertex  (a) at (0,0);
\vertex (b) at (-1,1);
\vertex (c) at (1,1);
\vertex (d) at (-1,-1);
\vertex  (e) at (1,-1);
\vertex (g) at (-0.33,-1);
\vertex (h) at (0.33,-1);
\vertex (f) at (0,1);
\diagram{
(d)--[horizontal,very thick](e);
(b)--[horizontal,very thick](c);
(g)--[horizontal,red](b);
(e)--[horizontal](f);
(h)--[horizontal](f);
(g)--[horizontal,red](f);
};
\end{feynman}
\node[below=0.7cm,scale=0.8]{(1g)};
\end{tikzpicture}

\begin{tikzpicture}[scale=0.6]
\begin{feynman}
\vertex  (a) at (0,0);
\vertex (b) at (-1,1);
\vertex (c) at (1,1);
\vertex (d) at (-1,-1);
\vertex  (e) at (1,-1);
\vertex (g) at (-0.33,-1);
\vertex (h) at (0.33,-1);
\vertex (f) at (0,1);
\vertex (r) at (-0.21,-0.3);
\diagram{
(d)--[horizontal,very thick](e);
(b)--[horizontal,very thick](c);
(r)--[horizontal,red](b);
(e)--[horizontal,red](f);
(h)--[horizontal,red](f);
(g)--[horizontal,red](f);
};
\end{feynman}
\node[below=0.7cm,scale=0.8]{(2a)};
\end{tikzpicture}
\begin{tikzpicture}[scale=0.6]
\begin{feynman}
\vertex  (a) at (0,0);
\vertex (b) at (-1,1);
\vertex (c) at (1,1);
\vertex (d) at (-1,-1);
\vertex  (e) at (1,-1);
\vertex (g) at (-0.33,-1);
\vertex (h) at (0.33,-1);
\vertex (f) at (0,1);
\vertex (r) at (-0.21,-0.3);
\diagram{
(d)--[horizontal,very thick](e);
(b)--[horizontal,very thick](c);
(r)--[horizontal,red](b);
(e)--[horizontal,red](f);
(h)--[horizontal](f);
(g)--[horizontal,red](f);
};
\end{feynman}
\node[below=0.7cm,scale=0.8]{(2b)};
\end{tikzpicture}
\begin{tikzpicture}[scale=0.6]
\begin{feynman}
\vertex  (a) at (0,0);
\vertex (b) at (-1,1);
\vertex (c) at (1,1);
\vertex (d) at (-1,-1);
\vertex  (e) at (1,-1);
\vertex (g) at (-0.33,-1);
\vertex (h) at (0.33,-1);
\vertex (f) at (0,1);
\vertex (r) at (-0.21,-0.3);
\diagram{
(d)--[horizontal,very thick](e);
(b)--[horizontal,very thick](c);
(r)--[horizontal,red](b);
(e)--[horizontal,red](f);
(h)--[horizontal,red](f);
(f)--[horizontal,double](r);
(r)--[horizontal,red](g);
};
\end{feynman}
\node[below=0.7cm,scale=0.8]{(2c)};
\end{tikzpicture}
\begin{tikzpicture}[scale=0.6]
\begin{feynman}
\vertex  (a) at (0,0);
\vertex (b) at (-1,1);
\vertex (c) at (1,1);
\vertex (d) at (-1,-1);
\vertex  (e) at (1,-1);
\vertex (g) at (-0.33,-1);
\vertex (h) at (0.33,-1);
\vertex (f) at (0,1);
\vertex (r) at (-0.21,-0.3);
\diagram{
(d)--[horizontal,very thick](e);
(b)--[horizontal,very thick](c);
(r)--[horizontal](b);
(e)--[horizontal,red](f);
(h)--[horizontal,red](f);
(f)--[horizontal,double](r);
(r)--[horizontal](g);
};
\end{feynman}
\node[below=0.7cm,scale=0.8]{(2d)};
\end{tikzpicture}

\begin{tikzpicture}[scale=0.6]
\begin{feynman}
\vertex  (a) at (0,0);
\vertex (b) at (-1,1);
\vertex (c) at (1,1);
\vertex (d) at (-1,-1);
\vertex  (e) at (1,-1);
\vertex (g) at (-0.33,-1);
\vertex (h) at (0.33,-1);
\vertex (f) at (0,1);
\diagram{
(d)--[horizontal,very thick](e);
(b)--[horizontal,very thick](c);
(g)--[horizontal,red](b);
(h)--[horizontal,red](c);
(h)--[horizontal,red](f);
(g)--[horizontal,red](f);
};
\end{feynman}
\node[below=0.7cm,scale=0.8]{(3a)};
\end{tikzpicture}
\begin{tikzpicture}[scale=0.6]
\begin{feynman}
\vertex  (a) at (0,0);
\vertex (b) at (-1,1);
\vertex (c) at (1,1);
\vertex (d) at (-1,-1);
\vertex  (e) at (1,-1);
\vertex (g) at (-0.33,-1);
\vertex (h) at (0.33,-1);
\vertex (f) at (0,1);
\diagram{
(d)--[horizontal,very thick](e);
(b)--[horizontal,very thick](c);
(g)--[horizontal,red](b);
(h)--[horizontal,red](c);
(h)--[horizontal,red](f);
(g)--[horizontal](f);
};
\end{feynman}
\node[below=0.7cm,scale=0.8]{(3b)};
\end{tikzpicture}
\begin{tikzpicture}[scale=0.6]
\begin{feynman}
\vertex  (a) at (0,0);
\vertex (b) at (-1,1);
\vertex (c) at (1,1);
\vertex (d) at (-1,-1);
\vertex  (e) at (1,-1);
\vertex (g) at (-0.33,-1);
\vertex (h) at (0.33,-1);
\vertex (f) at (0,1);
\diagram{
(d)--[horizontal,very thick](e);
(b)--[horizontal,very thick](c);
(g)--[horizontal,red](b);
(h)--[horizontal](c);
(h)--[horizontal,red](f);
(g)--[horizontal,red](f);
};
\end{feynman}
\node[below=0.7cm,scale=0.8]{(3c)};
\end{tikzpicture}
\begin{tikzpicture}[scale=0.6]
\begin{feynman}
\vertex  (a) at (0,0);
\vertex (b) at (-1,1);
\vertex (c) at (1,1);
\vertex (d) at (-1,-1);
\vertex  (e) at (1,-1);
\vertex (g) at (-0.33,-1);
\vertex (h) at (0.33,-1);
\vertex (f) at (0,1);
\diagram{
(d)--[horizontal,very thick](e);
(b)--[horizontal,very thick](c);
(g)--[horizontal](b);
(h)--[horizontal](c);
(h)--[horizontal,red](f);
(g)--[horizontal,red](f);
};
\end{feynman}
\node[below=0.7cm,scale=0.8]{(3d)};
\end{tikzpicture}
\begin{tikzpicture}[scale=0.6]
\begin{feynman}
\vertex  (a) at (0,0);
\vertex (b) at (-1,1);
\vertex (c) at (1,1);
\vertex (d) at (-1,-1);
\vertex  (e) at (1,-1);
\vertex (g) at (-0.33,-1);
\vertex (h) at (0.33,-1);
\vertex (f) at (0,1);
\diagram{
(d)--[horizontal,very thick](e);
(b)--[horizontal,very thick](c);
(g)--[horizontal](b);
(h)--[horizontal,red](c);
(h)--[horizontal,red](f);
(g)--[horizontal](f);
};
\end{feynman}
\node[below=0.7cm,scale=0.8]{(3e)};
\end{tikzpicture}

\begin{tikzpicture}[scale=0.6]
\begin{feynman}
\vertex  (a) at (0,0);
\vertex (b) at (-1,1);
\vertex (c) at (1,1);
\vertex (d) at (-1,-1);
\vertex  (e) at (1,-1);
\vertex (g) at (-0.33,-1);
\vertex (h) at (0.33,-1);
\vertex (f) at (0,1);
\diagram{
(d)--[horizontal,very thick](e);
(b)--[horizontal,very thick](c);
(d)--[horizontal,red](b);
(e)--[horizontal,red](c);
(d)--[horizontal,red](a);
(e)--[horizontal,red](a);
(f)--[horizontal,red](a);
};
\end{feynman}
\node[below=0.7cm,scale=0.8]{(4a)};
\end{tikzpicture}
\begin{tikzpicture}[scale=0.6]
\begin{feynman}
\vertex  (a) at (0,0);
\vertex (b) at (-1,1);
\vertex (c) at (1,1);
\vertex (d) at (-1,-1);
\vertex  (e) at (1,-1);
\vertex (g) at (-0.33,-1);
\vertex (h) at (0.33,-1);
\vertex (f) at (0,1);
\diagram{
(d)--[horizontal,very thick](e);
(b)--[horizontal,very thick](c);
(d)--[horizontal](b);
(e)--[horizontal,red](c);
(d)--[horizontal,red](a);
(e)--[horizontal,red](a);
(f)--[horizontal,red](a);
};
\end{feynman}
\node[below=0.7cm,scale=0.8]{(4b)};
\end{tikzpicture}
\begin{tikzpicture}[scale=0.6]
\begin{feynman}
\vertex  (a) at (0,0);
\vertex (b) at (-1,1);
\vertex (c) at (1,1);
\vertex (d) at (-1,-1);
\vertex  (e) at (1,-1);
\vertex (g) at (-0.33,-1);
\vertex (h) at (0.33,-1);
\vertex (f) at (0,1);
\vertex (r) at (0.5,0);
\diagram{
(d)--[horizontal,very thick](e);
(b)--[horizontal,very thick](c);
(d)--[horizontal,red](b);
(r)--[horizontal,red](c);
(d)--[horizontal,red](f);
(e)--[horizontal,red](r);
(f)--[horizontal,red](r);
};
\end{feynman}
\node[below=0.7cm,scale=0.8]{(5a)};
\end{tikzpicture}
\begin{tikzpicture}[scale=0.6]
\begin{feynman}
\vertex  (a) at (0,0);
\vertex (b) at (-1,1);
\vertex (c) at (1,1);
\vertex (d) at (-1,-1);
\vertex  (e) at (1,-1);
\vertex (g) at (-0.33,-1);
\vertex (h) at (0.33,-1);
\vertex (f) at (0,1);
\vertex (r) at (0.5,0);
\diagram{
(d)--[horizontal,very thick](e);
(b)--[horizontal,very thick](c);
(d)--[horizontal](b);
(r)--[horizontal,red](c);
(d)--[horizontal,red](f);
(e)--[horizontal,red](r);
(f)--[horizontal,red](r);
};
\end{feynman}
\node[below=0.7cm,scale=0.8]{(5b)};
\end{tikzpicture}
\begin{tikzpicture}[scale=0.6]
\begin{feynman}
\vertex  (a) at (0,0);
\vertex (b) at (-1,1);
\vertex (c) at (1,1);
\vertex (d) at (-1,-1);
\vertex  (e) at (1,-1);
\vertex (g) at (-0.33,-1);
\vertex (h) at (0.33,-1);
\vertex (f) at (0,1);
\vertex (r) at (0.5,0);
\diagram{
(d)--[horizontal,very thick](e);
(b)--[horizontal,very thick](c);
(d)--[horizontal,red](b);
(r)--[horizontal,red](c);
(b)--[horizontal,red](h);
(h)--[horizontal,red](r);
(f)--[horizontal,red](r);
};
\end{feynman}
\node[below=0.7cm,scale=0.8]{(6a)};
\end{tikzpicture}
\begin{tikzpicture}[scale=0.6]
\begin{feynman}
\vertex  (a) at (0,0);
\vertex (b) at (-1,1);
\vertex (c) at (1,1);
\vertex (d) at (-1,-1);
\vertex  (e) at (1,-1);
\vertex (g) at (-0.33,-1);
\vertex (h) at (0.33,-1);
\vertex (f) at (0,1);
\vertex (r) at (0.5,0);
\diagram{
(d)--[horizontal,very thick](e);
(b)--[horizontal,very thick](c);
(d)--[horizontal](b);
(r)--[horizontal,red](c);
(b)--[horizontal,red](h);
(h)--[horizontal,red](r);
(f)--[horizontal,red](r);
};
\end{feynman}
\node[below=0.7cm,scale=0.8]{(6b)};
\end{tikzpicture}

\begin{tikzpicture}[scale=0.6]
\begin{feynman}
\vertex  (a) at (0,0);
\vertex (b) at (-1,1);
\vertex (c) at (1,1);
\vertex (d) at (-1,-1);
\vertex  (e) at (1,-1);
\vertex (g) at (-0.33,-1);
\vertex (h) at (0.33,-1);
\vertex (f) at (0,1);
\vertex (r) at (0.17,0);
\diagram{
(d)--[horizontal,very thick](e);
(b)--[horizontal,very thick](c);
(d)--[horizontal,red](f);
(e)--[horizontal,red](c);
(f)--[horizontal,red](h);
(e)--[horizontal,red](r);
};
\end{feynman}
\node[below=0.7cm,scale=0.8]{(7a)};
\end{tikzpicture}
\begin{tikzpicture}[scale=0.6]
\begin{feynman}
\vertex  (a) at (0,0);
\vertex (b) at (-1,1);
\vertex (c) at (1,1);
\vertex (d) at (-1,-1);
\vertex  (e) at (1,-1);
\vertex (g) at (-0.33,-1);
\vertex (h) at (0.33,-1);
\vertex (f) at (0,1);
\vertex (r) at (0.17,0);
\diagram{
(d)--[horizontal,very thick](e);
(b)--[horizontal,very thick](c);
(d)--[horizontal,red](f);
(e)--[horizontal](c);
(f)--[horizontal,red](h);
(e)--[horizontal,red](r);
};
\end{feynman}
\node[below=0.7cm,scale=0.8]{(7b)};
\end{tikzpicture}
\begin{tikzpicture}[scale=0.6]
\begin{feynman}
\vertex  (a) at (0,0);
\vertex (b) at (-1,1);
\vertex (c) at (1,1);
\vertex (d) at (-1,-1);
\vertex  (e) at (1,-1);
\vertex (g) at (-0.33,-1);
\vertex (h) at (0.33,-1);
\vertex (f) at (0,1);
\vertex (r) at (0.17,0);
\diagram{
(d)--[horizontal,very thick](e);
(b)--[horizontal,very thick](c);
(d)--[horizontal](f);
(e)--[horizontal,red](c);
(f)--[horizontal,red](h);
(e)--[horizontal,red](r);
};
\end{feynman}
\node[below=0.7cm,scale=0.8]{(7c)};
\end{tikzpicture}
\begin{tikzpicture}[scale=0.6]
\begin{feynman}
\vertex  (a) at (0,0);
\vertex (b) at (-1,1);
\vertex (c) at (1,1);
\vertex (d) at (-1,-1);
\vertex  (e) at (1,-1);
\vertex (g) at (-0.33,-1);
\vertex (h) at (0.33,-1);
\vertex (f) at (0,1);
\vertex (r) at (-0.33,1);
\vertex (q) at (-0.33,0);
\vertex (s) at (1,0);
\diagram{
(d)--[horizontal,very thick](e);
(b)--[horizontal,very thick](c);
(d)--[horizontal,red](r);
(q)--[horizontal,red](s);
(c)--[horizontal,red](e);
(g)--[horizontal,red](r);
};
\end{feynman}
\node[below=0.7cm,scale=0.8]{(8a)};
\end{tikzpicture}
\begin{tikzpicture}[scale=0.6]
\begin{feynman}
\vertex  (a) at (0,0);
\vertex (b) at (-1,1);
\vertex (c) at (1,1);
\vertex (d) at (-1,-1);
\vertex  (e) at (1,-1);
\vertex (g) at (-0.33,-1);
\vertex (h) at (0.33,-1);
\vertex (f) at (0,1);
\vertex (r) at (-0.33,1);
\vertex (q) at (-0.33,0);
\vertex (s) at (1,0);
\diagram{
(d)--[horizontal,very thick](e);
(b)--[horizontal,very thick](c);
(d)--[horizontal,red](r);
(q)--[horizontal,double](s);
(c)--[horizontal,red](e);
(g)--[horizontal,red](r);
};
\end{feynman}
\node[below=0.7cm,scale=0.8]{(8b)};
\end{tikzpicture}
\begin{tikzpicture}[scale=0.6]
\begin{feynman}
\vertex  (a) at (0,0);
\vertex (b) at (-1,1);
\vertex (c) at (1,1);
\vertex (d) at (-1,-1);
\vertex  (e) at (1,-1);
\vertex (g) at (-0.33,-1);
\vertex (h) at (0.33,-1);
\vertex (f) at (0,1);
\vertex (r) at (-0.33,1);
\vertex (q) at (-0.33,0);
\vertex (s) at (1,0);
\diagram{
(d)--[horizontal,very thick](e);
(b)--[horizontal,very thick](c);
(d)--[horizontal,red](r);
(q)--[horizontal,double](s);
(c)--[horizontal](e);
(g)--[horizontal,red](r);
};
\end{feynman}
\node[below=0.7cm,scale=0.8]{(8c)};
\end{tikzpicture}

\begin{tikzpicture}[scale=0.6]
\begin{feynman}
\vertex  (a) at (0,0);
\vertex (b) at (-1,1);
\vertex (c) at (1,1);
\vertex (d) at (-1,-1);
\vertex  (e) at (1,-1);
\vertex (g) at (-0.33,-1);
\vertex (h) at (0.33,-1);
\vertex (f) at (0,1);
\vertex (r) at (0.5,0);
\vertex (s) at (-0.34,0);
\diagram{
(d)--[horizontal,very thick](e);
(b)--[horizontal,very thick](c);
(d)--[horizontal,red](s);
(r)--[horizontal,red](c);
(b)--[horizontal,red](h);
(h)--[horizontal,red](r);
(f)--[horizontal,red](r);
};
\end{feynman}
\node[below=0.7cm,scale=0.8]{(9a)};
\end{tikzpicture}
\begin{tikzpicture}[scale=0.6]
\begin{feynman}
\vertex  (a) at (0,0);
\vertex (b) at (-1,1);
\vertex (c) at (1,1);
\vertex (d) at (-1,-1);
\vertex  (e) at (1,-1);
\vertex (g) at (-0.33,-1);
\vertex (h) at (0.33,-1);
\vertex (f) at (0,1);
\vertex (r) at (-0.5,0);
\vertex (q) at (0,-0.33);
\diagram{
(d)--[horizontal,very thick](e);
(b)--[horizontal,very thick](c);
(d)--[horizontal,red](b);
(d)--[horizontal,red](f);
(e)--[horizontal,red](r);
(g)--[horizontal,red](q);
};
\end{feynman}
\node[below=0.7cm,scale=0.8]{(10a)};
\end{tikzpicture}
\begin{tikzpicture}[scale=0.6]
\begin{feynman}
\vertex  (a) at (0,0);
\vertex (b) at (-1,1);
\vertex (c) at (1,1);
\vertex (d) at (-1,-1);
\vertex  (e) at (1,-1);
\vertex (g) at (-0.33,-1);
\vertex (h) at (0.33,-1);
\vertex (f) at (0,1);
\vertex (r) at (-0.5,0);
\vertex (q) at (0,-0.33);
\diagram{
(d)--[horizontal,very thick](e);
(b)--[horizontal,very thick](c);
(d)--[horizontal,red](b);
(d)--[horizontal,red](f);
(e)--[horizontal,red](q);
(q)--[horizontal,double](r);
(g)--[horizontal,red](q);
};
\end{feynman}
\node[below=0.7cm,scale=0.8]{(10b)};
\end{tikzpicture}
\begin{tikzpicture}[scale=0.6]
\begin{feynman}
\vertex  (a) at (0,0);
\vertex (b) at (-1,1);
\vertex (c) at (1,1);
\vertex (d) at (-1,-1);
\vertex  (e) at (1,-1);
\vertex (g) at (-0.33,-1);
\vertex (h) at (0.33,-1);
\vertex (f) at (0,1);
\vertex (r) at (-0.5,0);
\vertex (q) at (0,-0.33);
\diagram{
(d)--[horizontal,very thick](e);
(b)--[horizontal,very thick](c);
(d)--[horizontal,red](b);
(d)--[horizontal,red](f);
(e)--[horizontal](q);
(q)--[horizontal,double](r);
(g)--[horizontal](q);
};
\end{feynman}
\node[below=0.7cm,scale=0.8]{(10c)};
\end{tikzpicture}
\begin{tikzpicture}[scale=0.6]
\begin{feynman}
\vertex  (a) at (0,0);
\vertex (b) at (-1,1);
\vertex (c) at (1,1);
\vertex (d) at (-1,-1);
\vertex  (e) at (1,-1);
\vertex (g) at (-0.33,-1);
\vertex (h) at (0.33,-1);
\vertex (f) at (0,1);
\vertex (r) at (-0.5,0);
\vertex (q) at (0,-0.33);
\diagram{
(d)--[horizontal,very thick](e);
(b)--[horizontal,very thick](c);
(d)--[horizontal,red](b);
(d)--[horizontal,red](f);
(e)--[horizontal,red](r);
(g)--[horizontal,red](r);
};
\end{feynman}
\node[below=0.7cm,scale=0.8]{(11a)};
\end{tikzpicture}

\begin{tikzpicture}[scale=0.6]
\begin{feynman}
\vertex  (a) at (0,0);
\vertex (b) at (-1,1);
\vertex (c) at (1,1);
\vertex (d) at (-1,-1);
\vertex  (e) at (1,-1);
\vertex (g) at (-0.33,-1);
\vertex (h) at (0.33,-1);
\vertex (f) at (0,1);
\vertex (r) at (-0.33,1);
\vertex (q) at (-0.33,0);
\vertex (s) at (1,0);
\vertex (w) at (0.33,1);
\diagram{
(d)--[horizontal,very thick](e);
(b)--[horizontal,very thick](c);
(d)--[horizontal,red](r);
(q)--[horizontal,red](s);
(c)--[horizontal,red](e);
(q)--[horizontal,red](r);
(q)--[horizontal,red](w);
};
\end{feynman}
\node[below=0.7cm,scale=0.8]{(12a)};
\end{tikzpicture}
\begin{tikzpicture}[scale=0.6]
\begin{feynman}
\vertex  (a) at (0,0);
\vertex (b) at (-1,1);
\vertex (c) at (1,1);
\vertex (d) at (-1,-1);
\vertex  (e) at (1,-1);
\vertex (g) at (-0.33,-1);
\vertex (h) at (0.33,-1);
\vertex (f) at (0,1);
\vertex (r) at (-0.33,1);
\vertex (q) at (-0.33,0);
\vertex (s) at (1,0);
\vertex (w) at (0.33,1);
\diagram{
(d)--[horizontal,very thick](e);
(b)--[horizontal,very thick](c);
(d)--[horizontal,red](r);
(q)--[horizontal,double](s);
(c)--[horizontal,red](e);
(q)--[horizontal,red](r);
(q)--[horizontal,red](w);
};
\end{feynman}
\node[below=0.7cm,scale=0.8]{(12b)};
\end{tikzpicture}
\begin{tikzpicture}[scale=0.6]
\begin{feynman}
\vertex  (a) at (0,0);
\vertex (b) at (-1,1);
\vertex (c) at (1,1);
\vertex (d) at (-1,-1);
\vertex  (e) at (1,-1);
\vertex (g) at (-0.33,-1);
\vertex (h) at (0.33,-1);
\vertex (f) at (0,1);
\vertex (r) at (-0.33,1);
\vertex (q) at (-0.33,0);
\vertex (s) at (1,0);
\vertex (w) at (0.33,1);
\diagram{
(d)--[horizontal,very thick](e);
(b)--[horizontal,very thick](c);
(d)--[horizontal,red](r);
(q)--[horizontal,double](s);
(c)--[horizontal](e);
(q)--[horizontal,red](r);
(q)--[horizontal,red](w);
};
\end{feynman}
\node[below=0.7cm,scale=0.8]{(12c)};
\end{tikzpicture}
\begin{tikzpicture}[scale=0.6]
\begin{feynman}
\vertex  (a) at (0,0);
\vertex (b) at (-1,1);
\vertex (c) at (1,1);
\vertex (d) at (-1,-1);
\vertex  (e) at (1,-1);
\vertex (g) at (-0.33,-1);
\vertex (h) at (0.33,-1);
\vertex (f) at (0,1);
\vertex (r) at (-1,0);
\vertex (s) at (1,0);
\diagram{
(d)--[horizontal,very thick](e);
(b)--[horizontal,very thick](c);
(d)--[horizontal,red](b);
(e)--[horizontal,red](c);
(r)--[horizontal,red](f);
(s)--[horizontal,red](f);
};
\end{feynman}
\node[below=0.7cm,scale=0.8]{(13a)};
\end{tikzpicture}

\caption{Diagrams contributing at order $G^4m_1^2m_2(m_1+m_2)\lambda_2^{(0)}$}
\end{figure}
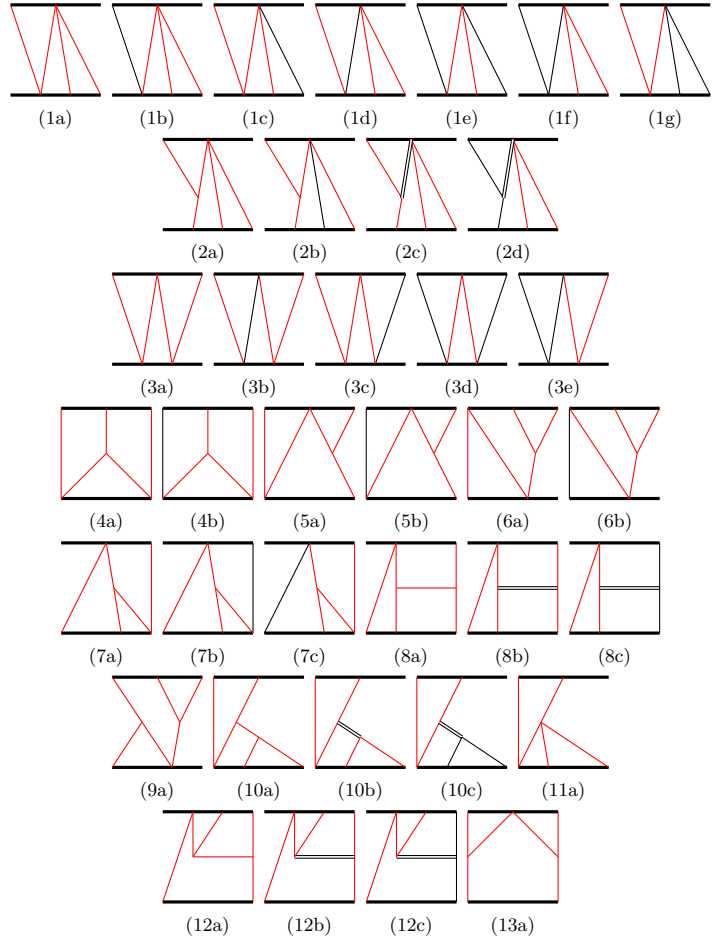

\begin{align}
Fig.(5.1a)=&-12\frac{\tilde{f}_1^{(2)}}{\tilde{f}_0^{(2)}}\tilde{d}_1^{(2)}\tilde{d}_2^{(1)}\nn\\
Fig.(5.1b)=&-6\frac{\tilde{f}_1^{(2)}}{\tilde{f}_0^{(2)}}\tilde{d}_1^{(1)}\nn\\
Fig.(5.1c)=&-12\frac{\tilde{d}_1^{(2)}\tilde{d}_2^{(1)}}{\tilde{d}_1^{(1)}}\nn\\
Fig.(5.1d)=&-6\tilde{d}_1^{(1)}\tilde{d}_1^{(2)}\nn\\
Fig.(5.1e)=&-6\nn\\
Fig.(5.1f)=&-3\nn\\
Fig.(5.2a)=&6\frac{\tilde{f}_1^{(2)}}{\tilde{f}_0^{(2)}}\tilde{d}_1^{(1)}\tilde{d}_1^{(2)}x_1\nn\\
Fig.(5.2b)=&6\tilde{d}_1^{(2)}x_1\nn\\
Fig.(5.2c)=&-8\tilde{d}_1^{(1)}\tilde{d}_1^{(2)}\nn\\
Fig.(5.2d)=&-4\nn\\
Fig.(5.3a)=&-8\tilde{d}_2^{(1)}\tilde{d}_2^{(2)}\nn\\
Fig.(5.3b)=&-4\tilde{d}_1^{(1)}\tilde{d}_1^{(2)}\nn\\
Fig.(5.3c)=&-4\frac{\tilde{d}_1^{(2)}\tilde{d}_2^{(1)}}{\tilde{d}_1^{(1)}}\nn\\
Fig.(5.3e)=&-2\nn\\
Fig.(5.4a)=&4 \tilde{d}_1^{(1)}\tilde{d}_2^{(2)}x_1\nn\\
Fig.(5.4b)=&2\tilde{d}_1^{(2)}x_1\nn\\
Fig.(5.5a)=&4\tilde{d}_1^{(2)}\tilde{d}_2^{(1)}x_1\nn\\
Fig.(5.6a)=&4\tilde{d}_1^{(2)}\tilde{d}_2^{(1)}x_1\nn\\
Fig.(5.6b)=&2\tilde{d}_1^{(1)}x_1\nn\\
Fig.(5.7a)=&8\tilde{d}_1^{(2)}\tilde{d}_2^{(1)}x_1\nn\\
Fig.(5.7b)=&4\tilde{d}_1^{(1)}x_1\nn\\
Fig.(5.8a)=&-6\tilde{d}_1^{(1)}\tilde{d}_1^{(2)}(x_1)^2\nn\\
Fig.(5.8b)=&-24\tilde{d}_1^{(1)}\tilde{d}_1^{(2)}\nn\\
Fig.(5.8c)=&-12\nn\\
Fig.(5.9a)=&-2\tilde{d}_1^{(1)}\tilde{d}_1^{(2)}(x_1)^2\nn\\
\end{align}

\subsection{$G^4m_1^2m_2^2\lambda_{(2)}^0\tilde{f}^{(2)}_n$}

The values of the diagrams are normalized with $\frac{G^4m_1^2m_2^2\lambda_{(2)}^0}{r^6}\tilde{f}_0^{(2)}$. We have to take into account also the mirror images $(1\leftrightarrow 2)$.

\begin{align}
Fig.(5.1a)=&-8\tilde{d}_1^{(1)}(\tilde{d}_1^{(2)})^2\tilde{d}_3^{(1)}\nn\\
Fig.(5.1c)=&-8\tilde{d}_1^{(1)}\tilde{d}_1^{(2)}\tilde{d}_2^{(1)}\nn\\
Fig.(5.1g)=&-2(\tilde{d}_1^{(1)})^2\nn\\
Fig.(5.3a)=&-8(\tilde{d}_1^{(2)})^2(\tilde{d}_2^{(1)})^2\nn\\
Fig.(5.3c)=&-2(\tilde{d}_1^{(1)})^2\nn\\
Fig.(5.3d)=&-8\tilde{d}_1^{(1)}\tilde{d}_1^{(2)}\tilde{d}_2^{(1)}\nn\\
Fig.(5.5a)=&8\tilde{d}_1^{(1)}(\tilde{d}_1^{(2)})^2\tilde{d}_2^{(1)}x_1\nn\\
Fig.(5.5b)=&4(\tilde{d}_1^{(1)})^2\tilde{d}_1^{(2)}x_1\nn\\
Fig.(5.6a)=&4\tilde{d}_1^{(1)}(\tilde{d}_1^{(2)})^2\tilde{d}_2^{(1)}x_1\nn\\
Fig.(5.7a)=&8\tilde{d}_1^{(1)}(\tilde{d}_1^{(2)})^2\tilde{d}_2^{(1)}x_1\nn\\
Fig.(5.7c)=&4(\tilde{d}_1^{(1)})^2\tilde{d}_1^{(2)}x_1\nn\\
Fig.(5.10a)=&-\frac{12}{5}(\tilde{d}_1^{(1)}\tilde{d}_1^{(2)}x_1)^2\nn\\
Fig.(5.10b)=&-\frac{24}{5}(\tilde{d}_1^{(1)}\tilde{d}_1^{(2)})^2\nn\\
Fig.(5.10c)=&-\frac{12}{5}(\tilde{d}_1^{(1)})^2\nn\\
Fig.(5.11a)=&4(\tilde{d}_1^{(1)}\tilde{d}_1^{(2)})^2x_2\nn\\
Fig.(5.12a)=&-\frac{28}{5}(\tilde{d}_1^{(1)}\tilde{d}_1^{(2)}x_1)^2\nn\\
Fig.(5.12b)=&-\frac{16}{5}(\tilde{d}_1^{(1)}\tilde{d}_1^{(2)})^2\nn\\
Fig.(5.12c)=&-\frac{8}{5}\tilde{d}_1^{(1)}\tilde{d}_1^{(2)}\nn\\
Fig.(5.13a)=&-2(\tilde{d}_1^{(1)}\tilde{d}_1^{(2)}x_1)^2\nn\\
\end{align}


